# A Generalized Dispersion Equation Combining Coupled Modes, Parity-Time Symmetry, and Leaky Waves Describing Non-Decaying Leaky Surface Waves

A. Abbaszadeh, J. Budhu, *Member, IEEE*

*Abstract*— A generalized dispersion equation is derived featuring coupled mode theory, parity-time symmetry, and leaky wave antennas of arbitrary periodic modulation. It can be specialized to each of these cases individually or can describe a structure containing all three electromagnetic phenomena simultaneously in a single antenna. This very general dispersion equation is derived using both mode matching and the transverse resonance method, the latter lacking the ability to provide the field descriptions and wave impedances at the cost of computational simplicity. Using the dispersion equation, a sinusoidally modulated reactive sheet (SMRS) supported by an active impedance sheet backed dielectric spacer is designed. The active impedance sheet is designed to compensate the SMRS radiative leakage loss when coupled in close proximity. Due to the coupling, each spatial harmonic generated by the SMRS modulation is characterized by a purely real propagation constant and hence leaky wave radiation is generated from surface waves which do not decay. Due to the non-decaying nature of the surface waves, an input to output surface wave port-to-port $|S_{21}|$ would be unity despite an open far field channel being generated. The non-decaying leaky surface wave antenna is compared to the traditional PEC backed SMRS. The non-decaying leaky surface waves give rise to perfect aperture efficiency and the associated radiation pattern making it far superior to the traditional PEC backed SMRS case. Full-wave simulations corroborate the results. These types of antennas can make sensors probed from the far field where changes in $S_{21}$ amplitude directly correlate to local changes in the waveguiding environment.

## I. Introduction

MODES guided by waveguides are characterized by their modal wave numbers which are usually dispersive. Dispersion analysis can be used to obtain the modal wavenumber and its associated frequency dependance. In dispersion analysis, an equation is derived which relates the wavenumber to the waveguide geometry and its material properties. A particular type of waveguide geometry which couples guided to unguided modes is called a leaky wave antenna [1], [2]. A leaky wave antenna is realized by periodically modulating the waveguides material properties (its surface impedance for example) generating an infinite number of Floquet harmonics [3], [4], [5], [6]. Some of these harmonics fall within the light cone and hence open far field channels. The modulated surface impedance couples the spatial harmonics together [4]. Hence, the fundamental surface wave used to excite the antenna couples to the harmonics that fall within the light cone leading to far field radiation stemming from the initial surface wave. Since the far field beam pointing angle is dependent on the wavenumber of the spatial harmonic within the light cone and this wavenumber is related to the fundamental wavenumber by Floquet's theorem, the design of a leaky wave antenna reduces to determining the fundamental wavenumber given a desired far field beam pointing angle [3], [7]. A dispersion equation is used to find this wavenumber.

A single dispersion equation can model multiple physical phenomena simultaneously. For example, a dispersion equation which models two conjugate impedance sheets models Parity-Time (PT) symmetric waveguides [8], [9]. Parity-Time symmetry refers to a special property that a waveguide may possess where the geometry and fields remain constant when both spatial and temporal inversions are applied [10], [11]. This unique symmetry leads to modes which are characterized by purely real wavenumbers despite the presence of loss and gain. The particular point in parametric space where this occurs is called an exceptional point [12]. If one periodically modulates one of the impedance sheets, a leaky wave antenna which is also PT-symmetric is formed capturing coupled mode, PT-symmetry, and leaky wave phenomena all in a single dispersion equation [13]. If the impedance sheets are defined parametrically, then one may specialize the general dispersion equation to any one of these three cases by proper choice of parameters. In this paper, one such dispersion equation is derived for a leaky waveguide consisting of a periodically modulated impedance sheet

A. Abbaszadeh is with the Bradley Department of Electrical and Computer Engineering at Virginia Tech, Blacksburg, VA 24061 USA (e-mail: afshinabbaszadeh@vt.edu).
J. Budhu is with the Bradley Department of Electrical and Computer Engineering at Virginia Tech, Blacksburg, VA 24061 USA (e-mail: jbudhu@vt.edu).

coupled through a dielectric spacer to an active impedance sheet. At some point in the parametric space, the active sheet gain balances the loss associated with the leaky wave leakage leading to the exceptional point [11]. The unique property of the leaky waveguide operated at this point is that no spatial harmonics decays as they leak energy to the far field. In other words, a leaky wave antenna with a purely real wavenumber results.

Some recent works published in scientific literature investigate similar phenomena. For example, in [14], a coplanar configuration of a periodically modulated transmission line coupled to an active transmission line is shown to exhibit exceptional points. Also, in [15], a leaky wave antenna coupled to a passive waveguide made from two coplanar transmission lines was found to also contain exceptional points. In [12], it was found that the open stopband problem of broadside radiation in periodic leaky wave antennas was related to the coupling of eigenmodes supported by the structure and that an exceptional point perspective can provide insight into ways to mitigate the problem. Each of these referenced works take a circuit modelling perspective. The presented work derives a more rigorous dispersion equation starting from field descriptions and applying boundary conditions. The particular problem being solved is described next, followed by the derivation of the dispersion equation and its application to non-decaying leaky surface waves.

## II. COUPLED LEAKY WAVE ANTENNA GEOMETRY

A sinusoidally modulated impedance sheet placed on a grounded dielectric substrate has previously been shown to generate leaky wave-radiation [3], [7], [16]. Here, a similar configuration is proposed only now the ground plane is replaced by an active impedance sheet (see Fig. 1a). The coupled waveguiding system therefore comprises of a sinusoidally modulated reactive sheet (SMRS) on the top surface and an active impedance sheet described by a constant complex impedance with a negative real part on the bottom surface of a dielectric spacer. The impedance sheets on the top and bottom surfaces can mathematically be described as

$$\eta_{sheet1} = jX_s \left[1 + M \cos\left[\frac{2\pi x}{a}\right]\right] \quad (1)$$
$$\eta_{sheet2} = -R + jX_{AS}$$

where $X_s$, $M$, and $a$ denote the average reactance, modulation index, and period of modulation of the SMRS, and $R$ and $X_{AS}$ represents the constant gain and reactance of the bottom sheet. The dielectric spacer separating the two impedance sheets has a thickness of $d$ and relative permittivity of $\epsilon_r$. The waveguiding structure is assumed infinite and invariant in the $y$ direction reducing the electromagnetics problem to two effective dimensions (2D) with $x$ as the direction of propagation and $z$ normal to this. The SMRS is designed to support $TM^x$ modes. Each of the harmonics, generated by the modulation of the top sheet couple to the bottom sheet forming a coupled waveguide system. Hence the mode power is seen to oscillate between being dominantly in the top sheet to being dominantly in the bottom sheet [17]. Since at least one spatial harmonic is designed to be within the light cone, every spatial harmonic generated by the SMRS is accompanied by an exponential decay factor of the form $e^{-\alpha x}$ arising due to radiative leakage losses. Thus, each harmonic's amplitude decays as it propagates along the SMRS. In this work, the active sheet is designed to compensate the leakage loss such that the exponential decay factor associated with each harmonic disappears. Consequently, the surface waves leak without decaying. The configuration therefore contains balanced gain and loss reminiscent of PT symmetric waveguiding structures only here the loss is due to leaky waves. In the following sections, the dispersion equation required to analyze both the waveguiding and radiative properties of the configuration is derived and solved. The dispersion equation is general and represents leaky waveguides with any periodic modulation (not just sinusoidal) and any value of complex impedance of the bottom sheet including the PEC case. The derived dispersion equation thus describes coupled mode theory, PT symmetry, and leaky wave antennas of arbitrary periodic modulation all in a single equation. It can be specialized to each of these cases individually or can describe a structure containing all these electromagnetic phenomena simultaneously in a single antenna. This very general dispersion equation is derived using both mode matching and the transverse resonance method, the latter lacking the ability to provide the field descriptions and wave impedances at the cost of computational simplicity.

## III. DERIVATION OF THE DISPERSION EQUATION

### A. Modal Expansion and Mode Matching

In this section a general dispersion equation for an arbitrary periodic modulation is derived and subsequently reduced to the sinusoidally modulated case. Since the top sheet has a periodic impedance profile, the electric and magnetic fields in all three regions are expanded into a summation of an infinite number of Floquet harmonics. Since, it is expected to have both $+z$ and $-z$ travelling waves in the dielectric region (region 1), two sets of fields denoted incident $(\vec{E}_1^i, \vec{H}_1^i)$ and reflected $(\vec{E}_1^r, \vec{H}_1^r)$ are assumed. Transmitted waves into the free space regions (regions 2 and 3) are labeled as $(\vec{E}_2^t, \vec{H}_2^t)$ and $(\vec{E}_3^t, \vec{H}_3^t)$. For the $TM^x$ excitation, the suitable choices for the electric and magnetic fields along with their longitudinal and transverse propagation constants are depicted in Fig. 1b for all three regions. By phase matching, the tangential component of the propagation constants should be the same for all three regions (see Fig. 1b) and by the Floquet theorem they are

$$k_{xn} = k + \frac{2n\pi}{a} \quad (2)$$

where $n$, an integer, is denoted as the mode index and $k = \beta - j\alpha$ is the propagation constant of the fundamental mode ($n = 0$).

The total electric and magnetic fields in each region can be written as follows (See appendix part I for more details)

$$H_{1y} = \sum_{n=-\infty}^{n=+\infty} \left( A_n e^{+jk_{zn1}z} - B_n e^{-jk_{zn1}z} \right) e^{-jk_{xn}x}$$

$$E_{x1} = -\frac{1}{j\omega\varepsilon_1} \sum_{n=-\infty}^{n=+\infty} jk_{zn1} \left( A_n e^{+jk_{zn1}z} + B_n e^{-jk_{zn1}z} \right) e^{-jk_{xn}x}$$

$$E_{z1} = \frac{1}{j\omega\varepsilon_1} \sum_{n=-\infty}^{n=+\infty} -jk_{xn} \left( A_n e^{+jk_{zn1}z} - B_n e^{-jk_{zn1}z} \right) e^{-jk_{xn}x}$$

$$H_{2y} = -\sum_{n=-\infty}^{n=+\infty} D_n e^{-jk_{xn}x} e^{-jk_{zn2}z}$$

$$E_{2x} = -\frac{1}{\omega\varepsilon_2} \sum_{n=-\infty}^{n=+\infty} k_{zn2} D_n e^{-jk_{xn}x} e^{-jk_{zn2}z}$$

$$E_{2z} = \frac{1}{\omega\varepsilon_2} \sum_{n=-\infty}^{n=+\infty} k_{xn} D_n e^{-jk_{xn}x} e^{-jk_{zn2}z}$$

$$H_{3y} = \sum_{n=-\infty}^{n=+\infty} C_n e^{-jk_{xn}x} e^{+jk_{zn3}z}$$

$$E_{3x} = -\frac{1}{\omega\varepsilon_3} \sum_{n=-\infty}^{n=+\infty} k_{zn3} C_n e^{-jk_{xn}x} e^{+jk_{zn3}z}$$

$$E_{3z} = -\frac{1}{\omega\varepsilon_3} \sum_{n=-\infty}^{n=+\infty} k_{xn} C_n e^{-jk_{xn}x} e^{+jk_{zn3}z}$$

(3)

where

$$k_{zn1}^2 = k_1^2 - k_{xn}^2 = \omega^2 \mu_0 \varepsilon_1 - k_{xn}^2$$
$$k_{zn2}^2 = k_{zn3}^2 = k_0^2 - k_{xn}^2 = \omega^2 \mu_0 \varepsilon_0 - k_{xn}^2 \quad (4)$$

The field expansions in (3) contain unknown complex constant coefficients $A_n, B_n, C_n,$ and $D_n$ and an unknown fundamental wavenumber $k$. Boundary conditions can be applied to eliminate the field expansion coefficients leaving an expression containing a single unknown $k$ called the dispersion equation. The boundary conditions include the continuity of the tangential components of the total electric fields at the interfaces between regions and the discontinuity of the tangential components of the total magnetic field proportional to the induced surface current density on the region interfaces

$$E_{1x}(z=0) = E_{2x}(z=0)$$
$$E_{1x}(z=-d) = E_{3x}(z=-d)$$
$$H_{1y}(z=0) - H_{2y}(z=0) = J_{x1} = \frac{E_{2x}(z=0)}{\eta_{sheet1}} \quad (5)$$
$$H_{3y}(z=-d) - H_{1y}(z=-d) = J_{x2} = \frac{E_{3x}(z=-d)}{\eta_{sheet2}}$$

At this point, the modulation of the top sheet is assumed a general periodic function with period $a$ and hence $\eta_{sheet1}$ and $\eta_{sheet2}$ in (5) can be written as

$$\eta_{sheet1} = \sum_{m=-\infty}^{m=+\infty} \eta_m e^{-j\frac{2\pi m}{a}x} \quad (6)$$

$$\eta_{sheet2} = -R + jX_{AS}$$

By applying the boundary conditions (5), the field expressions in (3) can be written in the form [16] (See appendix part I for more details)

$$\left[ \bar{\bar{Q}} - \frac{1}{\omega\varepsilon_1} \bar{\bar{I}} \right] \bar{A}_n = 0 \quad (7)$$

where

$$\bar{\bar{Q}} = \begin{bmatrix}
\ddots & \ddots & \vdots & \vdots & \vdots & \vdots & \cdots \\
\ddots & Q_{-2,-2} & Q_{-2,-1} & Q_{-2,0} & Q_{-2,+1} & Q_{-2,+2} & \cdots \\
\ddots & Q_{-1,-2} & Q_{-1,-1} & Q_{-1,0} & Q_{-1,+1} & Q_{-1,+2} & \cdots \\
\cdots & Q_{0,-2} & Q_{0,-1} & Q_{0,0} & Q_{0,+1} & Q_{0,+2} & \cdots \\
\cdots & Q_{+1,-2} & Q_{+1,-1} & Q_{+1,0} & Q_{+1,+1} & Q_{+1,+2} & \ddots \\
\cdots & Q_{+2,-2} & Q_{+2,-1} & Q_{+2,0} & Q_{+2,+1} & Q_{+2,+2} & \ddots \\
\cdots & \vdots & \vdots & \vdots & \vdots & \ddots & \ddots
\end{bmatrix}$$

$$Q_{p,n} = -\frac{\eta_{p-n}}{k_{zp1}} \left[ 1 + \frac{(-R+jX_{AS})\left(1 - \frac{\varepsilon_3}{\varepsilon_1}\frac{k_{zp1}}{k_{zp3}}\right) - \frac{k_{zp1}}{\omega\varepsilon_1}}{\frac{k_{zp1}}{\omega\varepsilon_1} + (-R+jX_{AS})\left(1 + \frac{\varepsilon_3}{\varepsilon_1}\frac{k_{zp1}}{k_{zp3}}\right)} e^{-2jk_{zp1}d} \right]^{-1} \times$$

$$\left[ 1 + \frac{\varepsilon_2 k_{zn1}}{\varepsilon_1 k_{zn2}} + \left[ \frac{\varepsilon_2 k_{zn1}}{\varepsilon_1 k_{zn2}} - 1 \right] \frac{(-R+jX_{AS})\left(1 - \frac{\varepsilon_3 k_{zn1}}{\varepsilon_1 k_{zn3}}\right) - \frac{k_{zn1}}{\omega\varepsilon_1}}{\frac{k_{zn1}}{\omega\varepsilon_1} + (-R+jX_{AS})\left(1 + \frac{\varepsilon_3 k_{zn1}}{\varepsilon_1 k_{zn3}}\right)} e^{-2jk_{zp1}d} \right]$$

(8)

Note the dense nature of $\bar{\bar{Q}}$ with full rows and columns. This indicates that every spatial harmonic couples to every other harmonic. The dispersion equation can be obtained by setting the determinant of the coefficient matrix to zero creating an equation which solves for the wavenumber $k$ leading to a non-trivial null-space

$$\det\left[ \bar{\bar{Q}} - \frac{1}{\omega\varepsilon_1} \bar{\bar{I}} \right] = 0 \quad (9)$$

The value of $k$ that satisfies (9) is the solution of interest. From here, one can compute $k_{xn}$ from (2) then $k_{zn1,2,3}$ from (4) thereby fully determining $\bar{\bar{Q}}$. Once $\bar{\bar{Q}}$ is known, (7) can be solved for its eigenvectors of coefficients leading to determined field components from (3). Hence, the problem has been reduced to solving (9) for the fundamental wavenumber $k$.

For the sinusoidally modulated case, $\eta_{sheet1}$ in (6) reduces to

$$\eta_{sheet1} = jX_s \left[ 1 + M \cos\left[\frac{2\pi x}{a}\right] \right] \quad (10)$$

Limiting the range of $m$ in (6) to $-1$ to $+1$, as per (10), reduces (7) and (8) to the dispersion equation

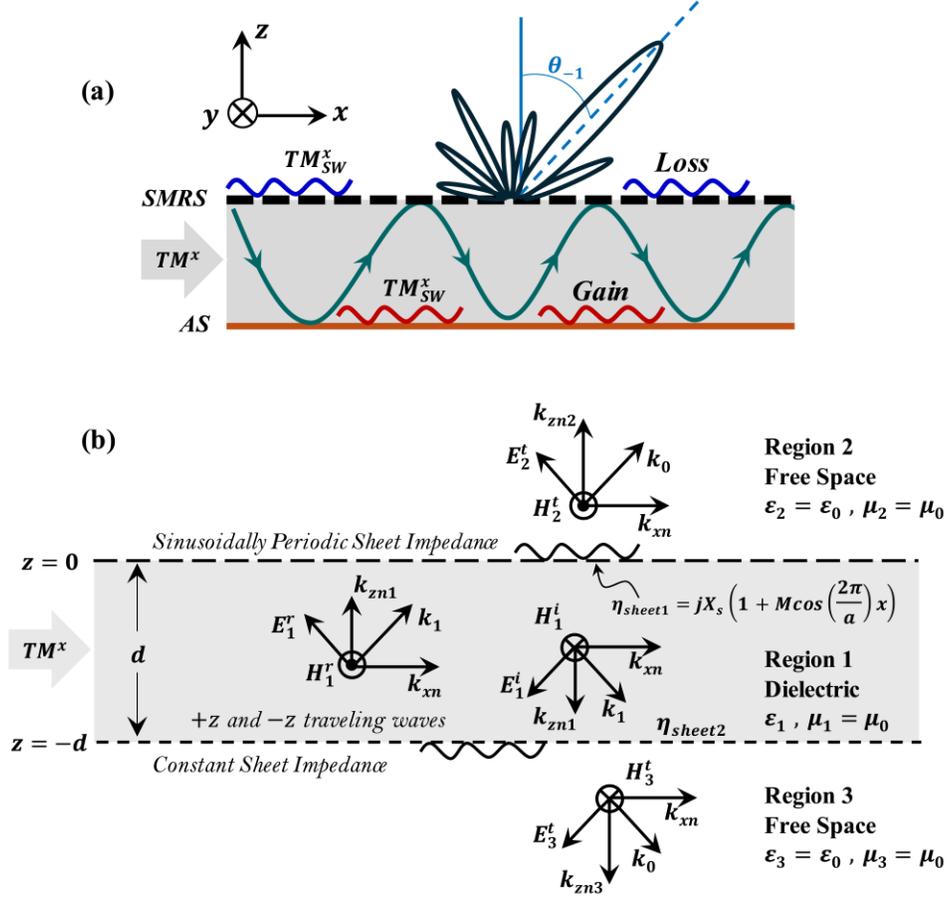

Figure 1: (a) Conceptual framework of the coupled leaky wave antenna geometry, (b) Schematic representation of the designed LWA and the supported Floquet-waves. Note: SMRS: Sinusoidally Modulated Reactive Sheet, and AS: Active Sheet.

$$P_{n-1} A_{n-1} + Q_n A_n + P_{n+1} A_{n+1} = 0 \qquad (11)$$

where

$$P_n = \frac{jX_s M}{2} + \frac{jX_s M}{2} \frac{\varepsilon_2}{\varepsilon_1} \frac{k_{zn1}}{k_{zn2}}$$

$$-\frac{jX_s M}{2} \frac{(-R+jX_{AS})\left(1-\frac{\varepsilon_3}{\varepsilon_1}\frac{k_{zn1}}{k_{zn3}}\right) - \frac{k_{zn1}}{\omega\varepsilon_1}}{\frac{k_{zn1}}{\omega\varepsilon_1} + (-R+jX_{AS})\left(1+\frac{\varepsilon_3}{\varepsilon_1}\frac{k_{zn1}}{k_{zn3}}\right)} e^{-2jk_{zn1}d}$$

$$+\frac{jX_s M}{2}\frac{\varepsilon_2}{\varepsilon_1}\frac{k_{zn1}}{k_{zn2}}\frac{(-R+jX_{AS})\left(1-\frac{\varepsilon_3}{\varepsilon_1}\frac{k_{zn1}}{k_{zn3}}\right) - \frac{k_{zn1}}{\omega\varepsilon_1}}{\frac{k_{zn1}}{\omega\varepsilon_1} + (-R+jX_{AS})\left(1+\frac{\varepsilon_3}{\varepsilon_1}\frac{k_{zn1}}{k_{zn3}}\right)} e^{-2jk_{zn1}d}$$

(12)

$$Q_n = \frac{2}{M} P_n + \frac{k_{zn1}}{\omega\varepsilon_1}$$

$$+\frac{k_{zn1}}{\omega\varepsilon_1}\frac{(-R+jX_{AS})\left(1-\frac{\varepsilon_3}{\varepsilon_1}\frac{k_{zn1}}{k_{zn3}}\right) - \frac{k_{zn1}}{\omega\varepsilon_1}}{\frac{k_{zn1}}{\omega\varepsilon_1} + (-R+jX_{AS})\left(1+\frac{\varepsilon_3}{\varepsilon_1}\frac{k_{zn1}}{k_{zn3}}\right)} e^{-2jk_{zn1}d} \qquad (13)$$

One may also arrive at the same dispersion equation by applying the boundary conditions in (5) with $\eta_{sheet1}$ as defined (10) (See appendix part II for more details). The dispersion equation (11) can be represented in matrix form as

$$\left[\bar{\bar{W}}\right]\bar{A}_n = 0 \qquad (14)$$

or

$$\begin{bmatrix} \ddots & \ddots & \vdots & \vdots & \vdots & \vdots & \cdots \\ \ddots & Q_{n-2} & P_{n-1} & 0 & 0 & 0 & \cdots \\ \ddots & P_{n-2} & Q_{n-1} & P_n & 0 & 0 & \cdots \\ \cdots & 0 & P_{n-1} & Q_n & P_{n+1} & 0 & \cdots \\ \cdots & 0 & 0 & P_n & Q_{n+1} & P_{n+2} & \ddots \\ \cdots & 0 & 0 & 0 & P_{n+1} & Q_{n+2} & \ddots \\ \cdots & \vdots & \vdots & \vdots & \vdots & \ddots & \ddots \end{bmatrix} \begin{bmatrix} \vdots \\ A_{n-2} \\ A_{n-1} \\ A_n \\ A_{n+1} \\ A_{n+2} \\ \vdots \end{bmatrix} = 0$$

Note now, due to the sinusoidal modulation, which can be written as the sum of two exponential functions by limiting $-1 \leq m \leq 1$ in (6), the coefficient matrix, $\bar{\bar{W}}$ in this case, is of tridiagonal format indicating that each spatial harmonic only couples to its nearest neighbors in index [4], [7]. Similarly, nontrivial solutions are obtained by setting the determinant of the $\bar{\bar{W}}$ matrix to zero

$$\det\left[\overline{\overline{W}}\right] = 0 \tag{15}$$

Leading to a solution for $k$. The solution of both (9) and (15) are treated in section III.C.

### B. Transmission Line Model

The dispersion equation can be obtained much more rapidly using the transverse resonance method applied to an equivalent transmission line model for the problem. To begin, mode expansion coefficients, dubbed modal voltages and modal currents, can be defined via the orthogonality of the spatial harmonics

$$V_n = \int_0^a \vec{E}_t(x) \cdot \vec{e}^{\,*}_{x,n}(x) dx$$
$$I_n = \int_0^a \vec{H}_t(x) \cdot \vec{h}^{\,*}_{y,n}(x) dx \tag{16}$$

Where $\vec{e}_{x,n}(x)$ and $\vec{h}_{y,n}(x)$ are the set of orthonormal spatial harmonic functions defined as [4]

$$\vec{e}_{x,n}(x) = \frac{1}{\sqrt{2\pi}} e^{-jk_{x,n}x} \hat{x}$$
$$\vec{h}_{y,n}(x) = -\frac{1}{\sqrt{2\pi}} e^{-jk_{x,n}x} \hat{y} \tag{17}$$

$\vec{E}_t(x)$ represents the tangential component of the total electric field and for penetrable and impenetrable impedance sheet cases can be obtained respectively as

$$\vec{E}_t(x) = Z_s\left[\hat{z} \times \left(\vec{H}_t\big|_{0^+} - \vec{H}_t\big|_{0^-}\right)\right] = Z_s \vec{J}_s$$
$$\vec{E}_t(x) = Z_{op}\left[\hat{z} \times \vec{H}_t\big|_{0^+}\right] = Z_{op} \vec{J}_s \tag{18}$$

where $Z_s = V_n/I_n$.

*Impenetrable case:*

For the impenetrable case, the surface (input) impedance is denoted as

$$Z_{op} = jX_{op}\left[1 + M\cos\left[\frac{2\pi x}{a}\right]\right] \tag{19}$$

Where $X_{op}$ is the average reactance of the impenetrable (opaque) modulated surface impedance. Substituting (17)-(19) back into (16) gives rise to the following dispersion equation (See appendix part V for more details)

$$jX_{op}\frac{M}{2}I_{n-1} + \left(jX_{op} - Z_s\right)I_n + jX_{op}\frac{M}{2}I_{n+1} = 0 \tag{20}$$

By using the transverse resonance condition, $Z_s = -Z_0^{TM}$, the surface impedance $Z_s$ can be eliminated from (20). Note, $Z_0^{TM}$ is defined in (23). In the actual geometry, the top impedance sheet is assumed penetrable, and hence the dispersion equation (20) is approximate as it assumes an impenetrable or opaque sheet. To derive a more accurate dispersion equation in agreement with (15), (20) must be modified as described in the following case.

*Penetrable case:*

Next, the dispersion equation for a penetrable SMRS over a grounded dielectric substrate is derived, followed by the dispersion relation for the case when the ground plane is replaced by the active impedance sheet. In both cases, the sheet impedance modulation function can now be written as

$$Z_s = jX_s\left[1 + M\cos\left[\frac{2\pi x}{a}\right]\right] \tag{21}$$

where the notation $X_s$ replaces $X_{op}$ to indicate the top impedance sheet is penetrable. For the ground plane termination case, the following dispersion equation is obtained by modifying the opaque dispersion equation in (20) to [7] (See appendix part V for more details)

$$jX_s\frac{M}{2}I_{n-1} + \left[jX_s + \frac{1}{Y_0^{TM} + Y_{short}}\right]I_n + jX_s\frac{M}{2}I_{n+1} = 0 \tag{22}$$

where

$$Z_0^{TM} = \eta_0 \frac{k_{z_0,n}}{k_0}$$
$$Z_1^{TM} = \eta_1 \frac{k_{zd,n}}{k_1} \tag{23}$$
$$Z_{short} = jZ_1^{TM}\tan(k_{zd,n}d)$$

In (23), the correct branch of the square root function must be chosen

$$k_{z_0,n} = \begin{cases} \sqrt{k_0^2 - k_{x,n}^2} & \text{if } k_{x,n} < k_0 \\ -j\sqrt{k_{x,n}^2 - k_0^2} & \text{if } k_{x,n} > k_0 \end{cases}$$
$$k_{zd,n} = \begin{cases} \sqrt{k_1^2 - k_{x,n}^2} & \text{if } k_{x,n} < k_1 \\ -j\sqrt{k_{x,n}^2 - k_1^2} & \text{if } k_{x,n} > k_1 \end{cases} \tag{24}$$

An expression relating the two models, namely $X_s$ to $X_{op}$, can be obtained by rearranging the input impedance formula and applying the transverse resonance condition to the transmission line model (see appendix part V) as

$$X_s = X_{op}\left[1 - \frac{\frac{X_{op}}{\eta_0}\varepsilon_r}{\sqrt{\varepsilon_r - 1 - \frac{X_{op}^2}{\eta_0^2}}}\cot\left(k_0 d\sqrt{\varepsilon_r - 1 - \frac{X_{op}^2}{\eta_0^2}}\right)\right]^{-1} \tag{25}$$

where $Z_{op} = jX_{op}$ is the equivalent input impedance looking down into the structure from above the SMRS.

Similarly, for the proposed structure shown in Fig.2a with an active sheet termination (SMRS-AS), the following modified dispersion equation is obtained

$$jX_s\frac{M}{2}I_{n-1} + \left[jX_s + \frac{1}{Y_0^{TM} + Y_{down2}}\right]I_n + jX_s\frac{M}{2}I_{n+1} = 0 \tag{26}$$

where

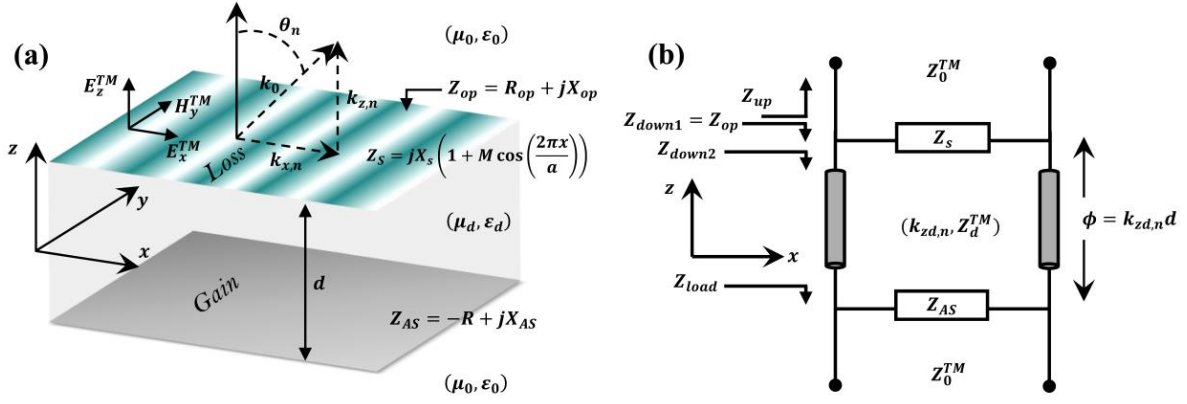

Figure 2: (a) Geometry of an SMRS integrated with an active sheet, (b) Transmission line model (TLM) representation of the structure.

$$Z_{down2} = Z_1^{TM} \frac{Z_0^{TM} Z_{AS} + jZ_1^{TM}(Z_0^{TM} + Z_{AS})\tan(k_{zd,n}d)}{Z_1^{TM}(Z_0^{TM} + Z_{AS}) + jZ_0^{TM} Z_{AS} \tan(k_{zd,n}d)} \quad (27)$$

We can also use matrix notation to write (26) as

$$\left[\bar{\bar{T}}\right]\bar{I}_n = 0 \quad (28)$$

or

$$\begin{bmatrix} \ddots & A & 0 & \cdots & 0 \\ A & B_{n-1} & A & 0 & 0 \\ 0 & A & B_n & A & 0 \\ \vdots & 0 & A & B_{n+1} & A \\ 0 & 0 & 0 & A & \ddots \end{bmatrix} \begin{bmatrix} \vdots \\ I_{n-1} \\ I_n \\ I_{n+1} \\ \vdots \end{bmatrix} = \begin{bmatrix} \vdots \\ 0 \\ 0 \\ 0 \\ \vdots \end{bmatrix}$$

where

$$A = jX_s \frac{M}{2}, \quad B_n = jX_s + \frac{1}{Y_0^{TM} + Y_{down2}} \quad (29)$$

The dispersion equation can now be obtained as

$$\det\left[\bar{\bar{T}}\right] = 0 \quad (30)$$

Given $Y_{down1} = (Z_{down1})^{-1} = Y_{op} = Y_s + Y_{down2}$, we can determine $X_s$ from knowledge of $Y_{down2}$ and $Y_{op}$ by rearranging the input impedance formula and applying the transverse resonance condition $(Y_{op} = -Y_0^{TM})$ to the transmission line model of the proposed structure as

$$X_s = \frac{-jZ_{op}}{1 - \dfrac{Z_{op}}{\dfrac{\eta_0 k_{zd,n}}{\varepsilon_r k_0} \dfrac{\dfrac{j\eta_0 k_{zd,n}(Z_{AS}-Z_{op})}{\varepsilon_r k_0}\tan(k_{zd,n}d) - Z_{op}Z_{AS}}{\dfrac{\eta_0 k_{zd,n}(Z_{AS}-Z_{op})}{\varepsilon_r k_0} - jZ_{op}Z_{AS}\tan(k_{zd,n}d)}}} \quad (31)$$

where

$$k_{zd,n} = k_0 \sqrt{\varepsilon_r - 1 + \frac{1}{\eta_0^2}(Z_{op})^2} \quad (32)$$

See appendix part V for details of the derivation of (31). Finally note that (30) is identical to (15).

### C. Complex Root Finding Algorithm

To solve the dispersion equation the complex roots of (15) or (30) must be obtained. To compute the complex roots of a complex-valued function, a local rational approximation method called type II Padé root-finding algorithm was used as it provides exponential convergence and less dependency on the initial guesses [18]. To use the Padé algorithm for determining the complex roots of the function $f(k)$, the function $f$ is first evaluated at three initial points $(k^{(1)}, k^{(2)}, k^{(3)})$, then the following expression is employed to compute $k^{(i)}$ for $i \geq 4$

$$k^{(i)} = \frac{k^{(i-1)}\dfrac{k^{(i-2)}-k^{(i-3)}}{f(k^{(i-1)})} + k^{(i-2)}\dfrac{k^{(i-3)}-k^{(i-1)}}{f(k^{(i-2)})} + k^{(i-3)}\dfrac{k^{(i-1)}-k^{(i-2)}}{f(k^{(i-3)})}}{\dfrac{k^{(i-2)}-k^{(i-3)}}{f(k^{(i-1)})} + \dfrac{k^{(i-3)}-k^{(i-1)}}{f(k^{(i-2)})} + \dfrac{k^{(i-1)}-k^{(i-2)}}{f(k^{(i-3)})}} \quad (33)$$

After performing a sufficient number of iterations, the algorithm converges to the complex roots of the function $f$. Hence, (33) is used to solve (9), (15), and (30). A choice of $0.75\beta_{x,app}$, $\beta_{x,app}$, and $1.25\beta_{x,app}$, where $\beta_{x,app}$ is defined in (34), were typically used as the three initial points seeding (33).

### IV. COMPENSATING THE LEAKAGE LOSS WITH GAIN

In a leaky wave antenna, the attenuation constant $\alpha$ each spatial harmonic carries, is a function of the modulation index $M$. Thus, in the coupled sheet configuration with $Z_{AS}$ containing a negative real part, a modulation index $M$ can be determined which leads to balanced gain and loss leading to $\alpha = 0$ and hence non-decaying leaky surface waves. To determine the required $M$ to balance the loss (due to $\alpha$) and gain (due to Re$[Z_{AS}]$), the waveguide is first modelled as an opaque SMRS by setting $Z_{op} = Z_{down1}$ (see Fig. 2). Setting $Z_{op}$ equal to a purely reactive value despite the presence of the complex-valued $Z_{AS}$ requires $Z_s$ to be complex-valued with positive real part so that their parallel combination equates to the purely reactive $Z_{op} = Z_{down1}$. This is evident in solving for $Z_s$ using (31) which results in a complex value. The real part of $Z_s$ is synthesized using modulation as $M$ is increased. At some $M$, the $\alpha$ that results in associated with an equivalent resistance of $R_s = \text{Re}[Z_s]$ which balances the active sheet resistance $R$.

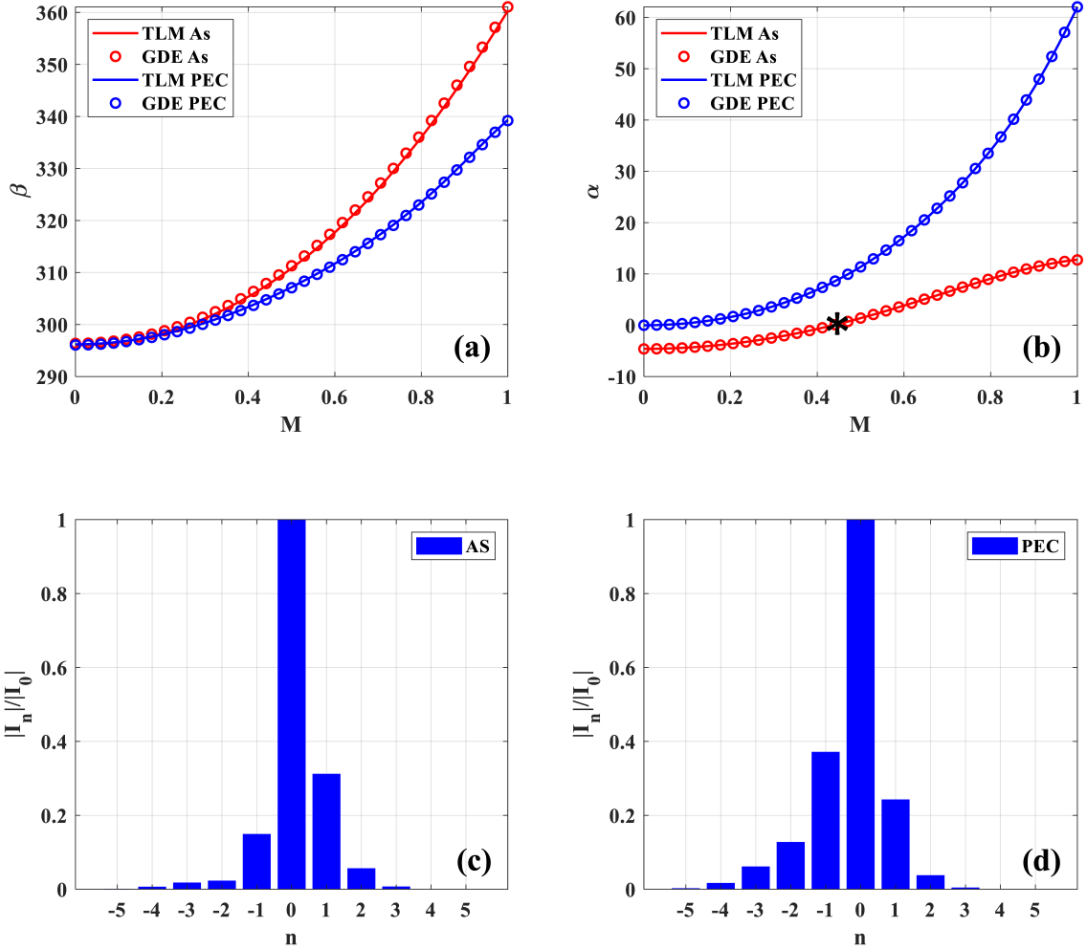

Figure 3: (a) real and (b) imaginary part of the propagation constant of the fundamental mode calculated using Transmission line model (TLM) and general dispersion equation (GDE) for both SMRS-AS and SMRS-PEC cases. (c) computed levels of the modal current amplitudes for SMRS-AS and (d) for SMRS-PEC.

For the SMRS-AS depicted in Fig.1b, the design parameters adapted from [7] are $f = 10\,\text{GHz}$, $\epsilon_r = 6.15$, $Z_{AS} = (-20 + j377)\,\Omega$, $Z_{op} = j377\,\Omega$, $d = 1.905\,\text{mm}$ and $\theta_{-1} = 30°$. Initially treating the SMRS-AS as opaque and assuming no modulation, the fundamental surface wavenumber can be approximated by solving (20) with $M = 0$

$$\beta_{x,app} = k_0\sqrt{1 + \left(X_{op}/\eta_0\right)^2} = 296.1922\,\text{rad/m} \qquad (34)$$

From here, the modulation period can be approximated as

$$a = \frac{2\pi}{\beta_{x,app} - k_0\sin(\theta_{-1})} = 32.8\,\text{mm} \qquad (35)$$

Finally using (31), $X_s = 89.848 - j2.405$, and therefore the SMRS sheet impedance is $Z_s = jX_s = (2.405 + j89.848)\,\Omega$. Retaining the reactive part of $Z_s$, namely $jX_s = +j89.848\,\Omega$, and realizing the loss of $+2.405\,\Omega$ by increasing the modulation index $M$ allows one to design the SMRS. To find the correct $M$, (15) or (30) is solved for a range of $M$ from $M = 0$ to $M = 1$ and $\alpha$ plotted to determine the point at which $\alpha = 0$. The result is shown in Fig. 3b for $N = 5$ (11 spatial harmonics considered). As expected, both (15) and (30) result in a same eigenvalues. When $M = 0$, $\alpha < 0$ indicating gain since the bottom sheet is active while the top sheet does not leak. As $M$ increases, loss due to leakage is introduced as both the $n = -1$ and $n = -2$ spatial harmonics fall within the light cone. At exactly $M = 0.436$, $\alpha = 0$ indicating the loss due to leakage and the active sheet gain are balanced. At this point, the fundamental wavenumber is purely real ($\beta \neq 0, \alpha = 0$) and hence each spatial harmonic does not decay or lose energy as it contributes to leakage to the far field. For $M > 0.436$, the coupled mode leaky wave antenna is dominantly lossy, and $\alpha$ becomes positive. In Fig. 3a, the wavenumber $\beta$ is plotted. As expected, the modulation depth affects the fundamental harmonics wavenumber.

## V. RESULTS AND DISCUSSION

To gain insight into the physical operating principles of the SMRS-AS, its characteristics are compared to a traditional leaky wave antenna made from a SMRS backed by a PEC grounded dielectric substrate. This configuration is dubbed SMRS-PEC and is obtained by setting $Z_{AS} = 0$ with all other parameters the same as the SMRS-AS case. Setting $Z_{AS} = 0$ in (31) leads to a new value for the average SMRS sheet reactance $X_s$. Alternatively, one obtains the same value from (25). The result of either formula is $X_s = -202.91\Omega$. Both (15) and (30) can be solved for case $Z_{AS} = 0$. The result is shown in Fig. 3a and 3b. Since the bottom sheet is now passive and lossless, there is no $\alpha < 0$ region in Fig. 3b, and instead, the attenuation constant starts from a no loss case when there is no modulation ($M = 0$). Since in the SMRS-AS case gain is present across the entire modulation depth range, one can modulate strongly without the high attenuation constant that the SMRS-PEC case presents.

From the solution of the dispersion equations for the fundamental wavenumbers the modal currents $\vec{I}_n$ in (16) can be obtained by solving for the basis vectors of the null space using (28). This can be accomplished by using the MATLAB *null* command. The normalized (to the fundamental) magnitude of the obtained modal currents for both SMRS-AS and SMRS-PEC cases are plotted in Fig. 3c and 3d.

On account of the tridiagonal nature of (14) and (28), the modal currents decay with $|n|$ as each mode couples to only its nearest neighbors with a coupling coefficient proportional to $M$. In general, the decrease in modal current magnitude is monotonic with respect to the fundamental mode. This motivates the choice of symmetric truncation for the number of harmonics about the fundamental considered in (14) and (28). Once the modal currents are obtained, the surface current density can be found from (See appendix part VIII for more details)

$$J_x = -H_y = -e^{-jk_{x,0}x} \sum_{n=-N}^{N} \frac{I_n}{\sqrt{2\pi}} e^{-j\frac{2\pi n}{a}x} \qquad (36)$$

Using (36), the harmonic currents are calculated for both the SMRS-AS and SMRS-PEC cases and their real part plotted in Fig. 4. The non-decaying property of the surface waves in the SMRS-AS case is evident. In contrast, the surface waves in the SMRS-PEC case decay as they leak energy to the far field. The initial amplitudes of the surface waves follow the magnitude plots in Fig. 3c and 3d, and the period follows the wavenumbers in (2). Due to the non-decaying nature of the surface waves in the SMRS-AS case, an input to output port-to-port $|S_{21}|$ would be unity despite an open far field channel being generated. This can be a useful property for far field sensors where amplitude ($|S_{21}|$) changes are measured rather than shifts in wavenumber.

The calculated harmonic currents can be used to formulate the radiation pattern. For $TM^x$ modes, the far electric field component is $\hat{\theta}$-directed (See appendix part VIII for more details) and can be found from

$$E_\theta \approx -jk_0 \frac{e^{-jk_0 r}}{4\pi r} \eta_0 \iint_S J_x \cos(\theta) e^{jk_0 x' \sin(\theta)} dS' \qquad (37)$$

Substituting (36) gives

$$E_\theta \approx jk_0\eta_0 \frac{e^{-jk_0 r}}{4\pi r} \frac{L_y \cos(\theta)}{\sqrt{2\pi}} \sum_{n=-N}^{N} \left[ I_n \int_0^{L_x} e^{-j\left(k_{x,0}+\frac{2n\pi}{a}-k_0\sin(\theta)\right)x'} dx' \right] \qquad (38)$$

where $L_x, L_y$ denotes the size of the assumed aperture antenna. The integral in (38) can be analytically solved as

$$E_\theta \approx jk_0\eta_0 \frac{e^{-jk_0 r}}{4\pi r} \frac{L_y \cos(\theta)}{\sqrt{2\pi}} \times$$

$$\sum_{n=-N}^{N} \left[ \frac{jI_n}{k_{x,0}+\frac{2n\pi}{a}-k_0\sin(\theta)} \left[ e^{-j\left(k_{x,0}+\frac{2n\pi}{a}-k_0\sin(\theta)\right)L_x} -1 \right] \right] \qquad (39)$$

Using the harmonic current coefficients for both SMRS-AS and SMRS-PEC cases for $M = 0.436$, the far fields patterns are plotted in Fig. 5b. As expected, for both cases, a main beam at $\theta \approx 30°$ (around 3° offset is observed) is obtained as per the design procedure. For the design parameters assumed for the SMRS-AS, the fundamental wavenumber computes to $k_{x0} = 307.211$ rad/m. From (2), for the harmonics inside the light cone, their radiation angle can be found from

$$\theta_n = \sin^{-1}\left[\frac{k_{x,n}}{k_0}\right] = \sin^{-1}\left[\frac{k_{x,0}}{k_0}+\frac{2\pi n}{ak_0}\right] \qquad (40)$$

Using (40), it is determined that the $n = -2$ harmonic is also inside the light cone and radiates at $\theta_{-2} = -21.2510°$. In Fig. 5b, an additional lobe is seen at this angle.

The same calculations can be applied to the PEC case, resulting in a $n = -2$ harmonic inside the light cone that radiates at $\theta_{-2} = -21.989°$ as can be seen in Fig. 5b.

As illustrated in Fig. 5b, the proposed SMRS-AS produces a more directive radiation pattern compared to the PEC counterpart. Additionally, the SMRS-AS demonstrates sharper and deeper nulls than the PEC case. Furthermore, the proposed structure exhibits a lower level of undesired higher-order harmonic radiations (in this case: $n = -2$) as its modal current is much lower (see Fig. 3c and 3d).

To validate the performance of the designed SMRS-AS, a 2D full-wave Finite Element Method (FEM) simulation is conducted using COMSOL Multiphysics. The simulated and analytical far field radiation patterns are plotted in Fig. 5a. A good agreement between the theoretical and simulated far-field patterns validates the design approach through full-wave simulation. The simulation setup that was used is shown in Fig. 6a. A numeric port, combined with a boundary mode analysis study step, is employed to effectively excite the proposed structure. At the end of the leaky waveguide, a Perfectly Matched Layer (PML) is used to absorb the generated harmonics, thereby preventing the reflection of surface waves from the structure's end.

Surface current densities are defined based on equation (18) to model the impedance of the sheets, while PML or scattering boundary conditions are used to model the criteria for free space propagation within a bounded simulation domain.

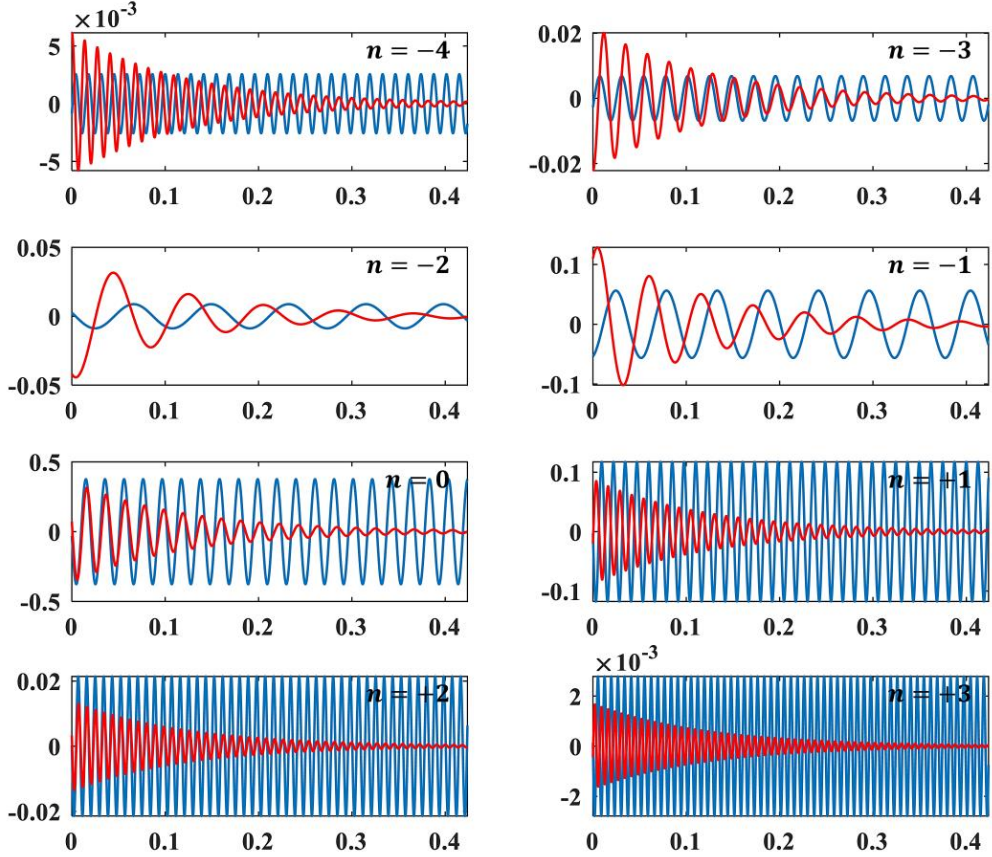

Figure 4: Real part of generated harmonic currents along the LWAs for both SMRS-AS (blue, non-decaying) and SMRS-PEC (red, decaying).

The real part of the $\hat{x}$-component of the electric field $\text{Re}[\vec{E} \cdot \hat{x}]$ is plotted in Figs. 6b and 6c for the SMRS-AS and SMRS-PEC cases. From the plotted fields, it's clear that both structures generate a radiated field at $\theta_{-1} \approx 30°$. As anticipated, the surface wave field in the SMRS-AS maintains a non-decaying property, whereas in the SMRS-PEC counterpart, a noticeable decay in amplitude is observed as the wave propagates along the waveguide. Again, these FEM based simulations validate the proposed design approach, thereby demonstrating the benefits of the designed leaky wave antenna operating in a balanced gain and loss mode through simulation.

## VI. CONCLUSION

This paper presents the derivation of a new multi-phenomena dispersion equation which captures the physics of coupled modes, PT symmetry, and leaky wave antennas in a single expression. The expression models a periodically modulated passive impedance sheet coupled through a dielectric spacer to a uniform active sheet backing it. The expression is derived using both modal expansion/mode matching and via the transverse resonance method. Both approaches result in identical expressions. The dispersion equations are solved using a numerical root finding procedure. The dispersion equation is used to analyze a waveguide which supports leaky wave radiation from surface waves which are described by purely real wavenumbers. Hence, each spatial harmonic shows a constant amplitude despite it leaking energy to the far field. The far fields are calculated from the modal currents and compared to full-wave COMSOL Multiphysics verification. The performance of the PT symmetric leaky waveguide is compared to its classical counterpart containing a periodically modulated impedance sheet backed by a grounded dielectric substrate. It is shown that the presented design is far superior to the classical leaky waveguide in both its radiation pattern and its near fields. The presented leaky waveguide may be realized using patterned metallic claddings and travelling wave amplifiers and may find applications in far field probed sensors with high sensitivity to local changes in the waveguiding environment.

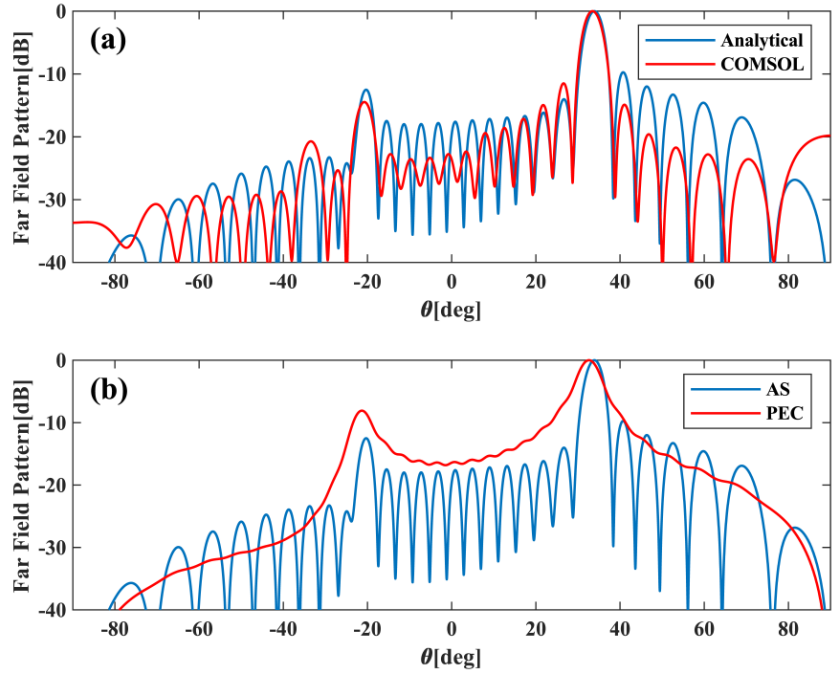

Figure 5: (a) Theoretical and simulated far field radiation patterns for the designed SMRS-AS leaky wave antenna at $f = 10 GHz$, and for the following design parameters: $\epsilon_r = 6.15$, $Z_{AS} = (-20 + j377)\Omega$, $Z_{op} = j377\Omega$, $d = 1.905mm$, $a = 32.8mm$, $M = 0.436$, $L_x = 12a$ and $\theta_{-1} = 30°$, (b) Theoretical radiation patterns for both the designed SMRS-AS and SMRS-PEC cases.

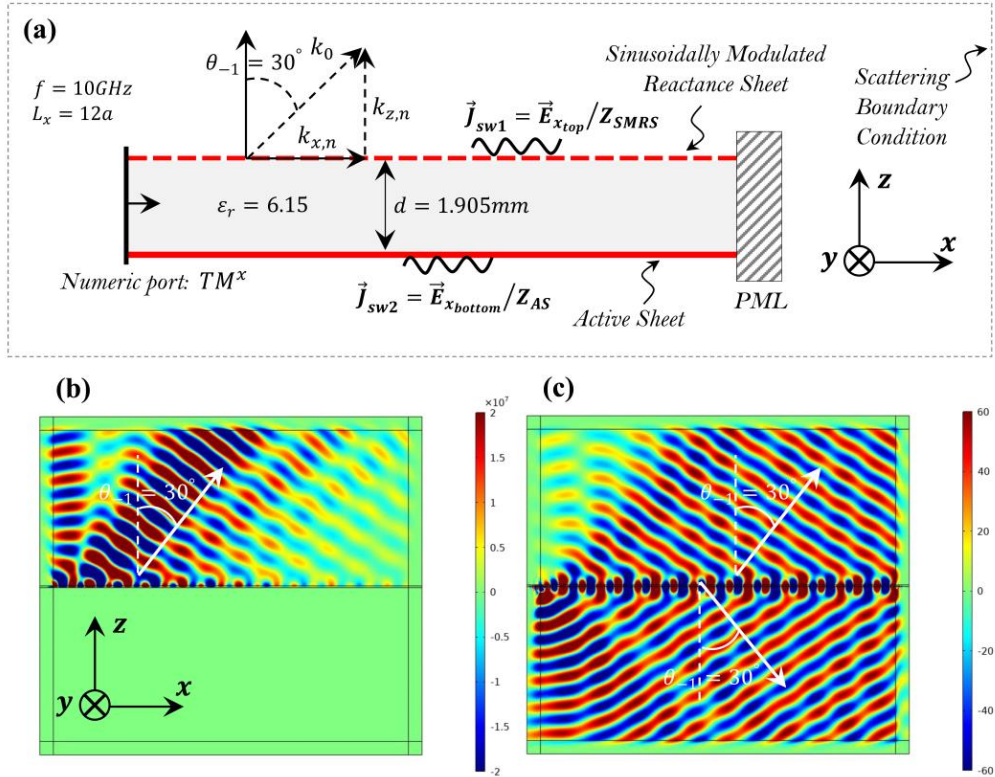

Figure 6: (a) Schematic representation of the COMSOL simulation setup for the far field radiation pattern calculation for the designed SMRS-AS LWA. (b) Real part of the dominant radiating field component (the electric field along the $x$ direction) for the designed SMRS-PEC LWA and (c) for the designed SMRS-AS LWA.

## REFERENCES

[1] D. R. Jackson, C. Caloz, and T. Itoh, "Leaky-Wave Antennas," *Proceedings of the IEEE*, vol. 100, no. 7, pp. 2194–2206, Jul. 2012, doi: 10.1109/JPROC.2012.2187410.

[2] R. Collin and F. Zucker, *Antenna Theory*, vol. 2. United Kingdom: McGraw-Hill, 1969.

[3] A. M. Patel and A. Grbic, "A Printed Leaky-Wave Antenna Based on a Sinusoidally-Modulated Reactance Surface," *IEEE Trans Antennas Propag*, vol. 59, no. 6, pp. 2087–2096, Jun. 2011, doi: 10.1109/TAP.2011.2143668.

[4] A. Oliner and A. Hessel, "Guided waves on sinusoidally-modulated reactance surfaces," *IRE Transactions on Antennas and Propagation*, vol. 7, no. 5, pp. 201–208, Dec. 1959, doi: 10.1109/TAP.1959.1144771.

[5] G. Minatti, F. Caminita, E. Martini, and S. Maci, "Flat Optics for Leaky-Waves on Modulated Metasurfaces: Adiabatic Floquet-Wave Analysis," *IEEE Trans Antennas Propag*, vol. 64, no. 9, pp. 3896–3906, Sep. 2016, doi: 10.1109/TAP.2016.2590559.

[6] G. Minatti *et al.*, "Modulated Metasurface Antennas for Space: Synthesis, Analysis and Realizations," *IEEE Trans Antennas Propag*, vol. 63, no. 4, pp. 1288–1300, Apr. 2015, doi: 10.1109/TAP.2014.2377718.

[7] J. Ruiz-Garcia, "Multi-beam Modulated Metasurface Antennas for Unmanned Aerial Vehicles (UAVs)," University of Rennes, 2021.

[8] A. Abbaszadeh and J. Budhu, "Observation of Exceptional Points in Parity-Time Symmetric Coupled Impedance Sheets," in *2024 18th European Conference on Antennas and Propagation (EuCAP)*, IEEE, Mar. 2024, pp. 1–5. doi: 10.23919/EuCAP60739.2024.10501448.

[9] X. Ma, M. S. Mirmoosa, and S. A. Tretyakov, "Parallel-Plate Waveguides Formed by Penetrable Metasurfaces," *IEEE Trans Antennas Propag*, vol. 68, no. 3, 2020, doi: 10.1109/TAP.2019.2934580.

[10] C. M. Bender and S. Boettcher, "Real Spectra in Non-Hermitian Hamiltonians Having PT Symmetry," *Phys Rev Lett*, vol. 80, no. 24, pp. 5243–5246, Jun. 1998, doi: 10.1103/PhysRevLett.80.5243.

[11] A. Krasnok, N. Nefedkin, and A. Alu, "Parity-Time Symmetry and Exceptional Points [Electromagnetic Perspectives]," *IEEE Antennas Propag Mag*, vol. 63, no. 6, pp. 110–121, Dec. 2021, doi: 10.1109/MAP.2021.3115766.

[12] A. Al-Bassam, D. Heberling, and C. Caloz, "Exceptional Point Perspective of Periodic Leaky-Wave Antennas," in *IEEE Antennas and Propagation Society, AP-S International Symposium (Digest)*, 2023. doi: 10.1109/USNC-URSI52151.2023.10237628.

[13] A. Abbaszadeh and J. Budhu, "Non-Decaying Leaky Surface Waves," in *2024 Eighteenth International Congress on Artificial Materials for Novel Wave Phenomena (Metamaterials)*, Crete, Greece, Sep. 2024, pp. 1–3.

[14] M. A. K. Othman and F. Capolino, "Theory of Exceptional Points of Degeneracy in Uniform Coupled Waveguides and Balance of Gain and Loss," *IEEE Trans Antennas Propag*, vol. 65, no. 10, pp. 5289–5302, Oct. 2017, doi: 10.1109/TAP.2017.2738063.

[15] N. Furman, A. Herrero-Parareda, and F. Capolino, "A Leaky Wave Antenna made of Two TLs with Exceptional Point Degeneracy," in *17th European Conference on Antennas and Propagation, EuCAP 2023*, 2023. doi: 10.23919/EuCAP57121.2023.10132925.

[16] A. M. Patel, "Controlling Electromagnetic Surface Waves with Scalar and Tensor Impedance Surfaces," Doctoral Thesis, University of Michigan, 2013.

[17] W.-P. Huang, "Coupled-mode theory for optical waveguides: an overview," *Journal of the Optical Society of America A*, vol. 11, no. 3, p. 963, Mar. 1994, doi: 10.1364/JOSAA.11.000963.

[18] V. Galdi and I. M. Pinto, "A simple algorithm for accurate location of leaky-wave poles for grounded inhomogeneous dielectric slabs," *Microw Opt Technol Lett*, vol. 24, no. 2, pp. 135–140, Jan. 2000, doi: 10.1002/(SICI)1098-2760(20000120)24:2<135::AID-MOP17>3.0.CO;2-P.

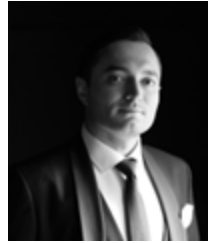

**Afshin Abbaszadeh** earned his B.Sc. degree in electrical engineering from K. N. Toosi University of Technology, Tehran, Iran, in September 2016. Following that, he pursued his M.Sc. degree in electrical engineering at Sharif University of Technology, Tehran, Iran, and graduated in December 2018. Currently, Afshin is a dedicated Ph.D. student within the Bradley Department of Electrical & Computer Engineering at Virginia Tech. From 2018 to 2020, he served as a researcher in the THz and Far-Infrared Laboratory in the Department of Electrical Engineering at Sharif University of Technology, Tehran, Iran. In 2020 and 2021, he continued his research as a researcher at the Optical Network Research Laboratory (ONRL), also in the Department of Electrical Engineering at Sharif University of Technology, Tehran, Iran. His academic journey took a new direction in the fall of 2022 when he joined the Electromagnetics Metamaterials and Antenna Laboratory within the Bradley Department of Electrical & Computer Engineering at Virginia Tech. In this capacity, he is actively engaged in cutting-edge research under the guidance and supervision of Prof. Jordan Budhu. Afshin's research interests span a wide range of areas, including Metamaterials and Metasurfaces, Parity-Time Symmetry, Nano-photonics, integrated optics, and antenna theory. His dedication and hard work have resulted in numerous high-impact papers published in prestigious journals, showcasing his contributions to the field.

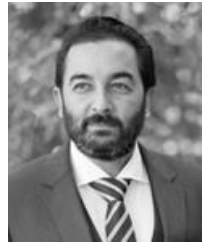

**Jordan Budhu** (Member, IEEE) received the M.S. degree in electrical engineering from California State University, Northridge, CA, USA, in 2010, and the Ph.D. degree in electrical engineering from the University of California, Los Angeles, CA, USA, in 2018. He was hired as an Assistant Professor with Virginia Tech, Blacksburg, VA, USA, in 2022, where he is currently the Steven O. Lane Junior Faculty Fellow of electrical and computer engineering with the Bradley Department of Electrical and Computer Engineering. He was a Postdoctoral Research Fellow in the Radiation Laboratory and a Lecturer in the Department of Electrical Engineering and Computer Science at the University of Michigan, Ann Arbor, MI, USA, from 2019 to 2022. In 2011 and 2012, he was a Graduate Student Intern at the NASA Jet Propulsion Laboratory. In 2017, he was named a Teaching Fellow at the University of California, Los Angeles. His research interests are in metamaterials and metasurfaces, computational electromagnetics algorithms for metamaterial and metasurface design, conformal beamforming antennas, nanophotonics and metamaterials for the infrared, 3-D printed inhomogeneous lens design, CubeSat antennas, reflectarray antennas, and antenna theory. Dr. Budhu's awards and honors include the 2010 Eugene Cota Robles Fellowship from UCLA, the 2012 Best Poster award at the IEEE Coastal Los Angeles Class-Tech Annual Meeting, the 2018 UCLA Henry Samueli School of Engineering and Applied Science Excellence in Teaching Award, the first-place award for the 2019 USNC-URSI Ernst K. Smith Student Paper Competition at the 2019 Boulder National Radio Science Meeting, and the Steven O. Lane Junior Faculty Fellowship of Electrical and Computer Engineering in the Bradley Department of Electrical and Computer Engineering, Virginia Tech.

# APPENDIX

## I. Derivation of the dispersion equation for an arbitrary periodic sheet impedance coupled with a uniform active sheet impedance.

The problem under investigation has been depicted in the figure below:

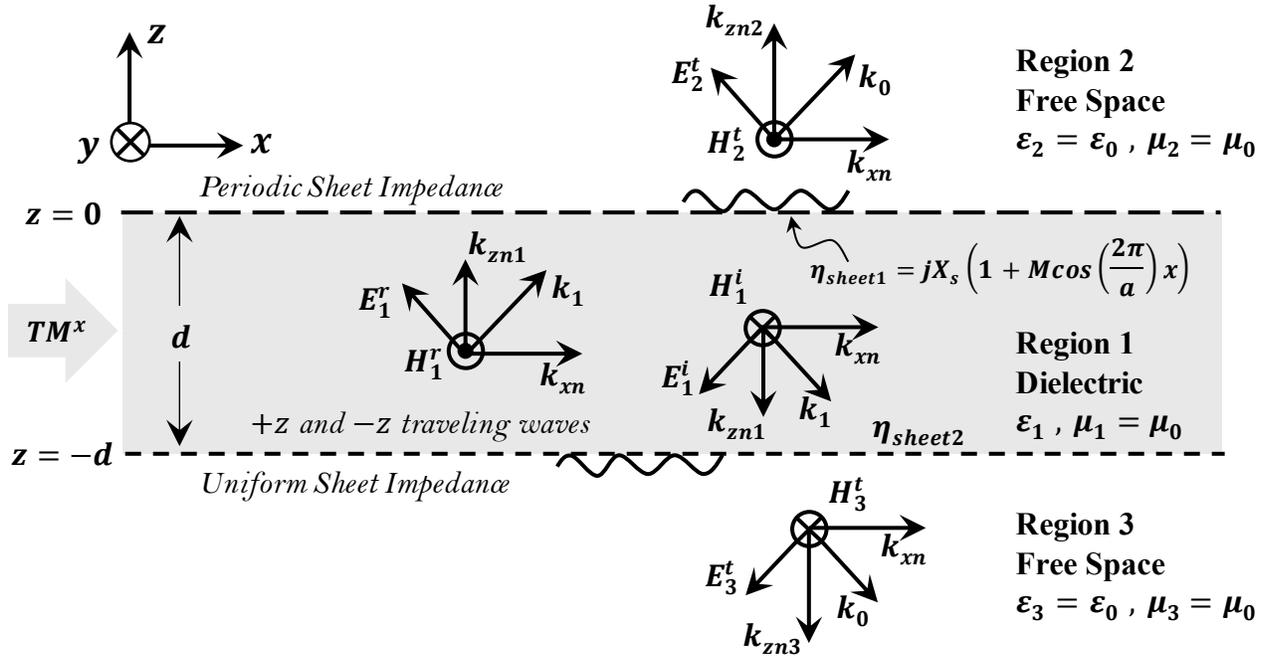

Figure 1: Schematic representation of the designed LWA (an arbitrary periodic sheet impedance coupled with a uniform active sheet which are integrated on the top and bottom of a dielectric slab) and the supported Floquet-waves.

Based on separation of variables method in solving Maxwell's wave equation, the longitudinal and transverse propagation constants are defined as follows:

$$k_{zn1}^2 = k_1^2 - k_{xn}^2 = \omega^2 \mu_0 \varepsilon_1 - k_{xn}^2$$

$$k_{zn2}^2 = k_2^2 - k_{xn}^2 = \omega^2 \mu_0 \varepsilon_0 - k_{xn}^2$$

$$k_{zn3}^2 = k_3^2 - k_{xn}^2 = \omega^2 \mu_0 \varepsilon_0 - k_{xn}^2$$

$$k_{xn} = k + 2n\pi/a$$

$$k = \beta - j\alpha$$

$a$: *period of modulation*

Since a periodic impedance sheet is integrated in to the waveguiding structure, based on Floquet theorem, we can expand the electric and magnetic fields in terms of infinite number of harmonics which are called Floquet waves as follows:

$$\vec{H}_1^i = \sum_{n=-\infty}^{+\infty} A_n e^{-jk_{xn}x} e^{+jk_{zn1}z} \hat{y}$$

$$\vec{H}_1^r = \sum_{n=-\infty}^{+\infty} B_n e^{-jk_{xn}x} e^{-jk_{zn1}z} (-\hat{y})$$

$$\vec{H}_1 = \vec{H}_1^i + \vec{H}_1^r = \sum_{n=-\infty}^{+\infty} A_n e^{-jk_{xn}x} e^{+jk_{zn1}z} \hat{y} + \sum_{n=-\infty}^{+\infty} -B_n e^{-jk_{xn}x} e^{-jk_{zn1}z} \hat{y}$$

$$H_{1y} = \sum_{n=-\infty}^{+\infty} A_n e^{-jk_{xn}x} e^{+jk_{zn1}z} + \sum_{n=-\infty}^{+\infty} -B_n e^{-jk_{xn}x} e^{-jk_{zn1}z}$$

$$H_{1y} = \sum_{n=-\infty}^{+\infty} \left( A_n e^{+jk_{zn1}z} - B_n e^{-jk_{zn1}z} \right) e^{-jk_{xn}x}$$

$$\nabla \times \vec{H}_1 = j\omega\varepsilon_1 \vec{E}_1 \quad \rightarrow \quad \vec{E}_1 = \frac{\nabla \times \vec{H}_1}{j\omega\varepsilon_1} = \frac{1}{j\omega\varepsilon_1} \begin{vmatrix} \hat{x} & \hat{y} & \hat{z} \\ \frac{\partial}{\partial x} & \frac{\partial}{\partial y} & \frac{\partial}{\partial z} \\ H_{1x} & H_{1y} & H_{1z} \end{vmatrix} = \frac{1}{j\omega\varepsilon_1} \begin{vmatrix} \hat{x} & \hat{y} & \hat{z} \\ \frac{\partial}{\partial x} & \frac{\partial}{\partial y} & \frac{\partial}{\partial z} \\ 0 & H_{1y} & 0 \end{vmatrix}$$

$$= \frac{1}{j\omega\varepsilon_1} \left( -\hat{x} \frac{\partial H_{1y}}{\partial z} + \hat{z} \frac{\partial H_{1y}}{\partial x} \right)$$

$$E_{1x} = \frac{-1}{j\omega\varepsilon_1} \frac{\partial H_{1y}}{\partial z} = \frac{-1}{j\omega\varepsilon_1} \frac{\partial}{\partial z} \left( \sum_{n=-\infty}^{+\infty} \left( A_n e^{+jk_{zn1}z} - B_n e^{-jk_{zn1}z} \right) e^{-jk_{xn}x} \right)$$

$$= \frac{-1}{j\omega\varepsilon_1} \sum_{n=-\infty}^{+\infty} jk_{zn1} \left( A_n e^{+jk_{zn1}z} + B_n e^{-jk_{zn1}z} \right) e^{-jk_{xn}x}$$

$$E_{1z} = \frac{1}{j\omega\varepsilon_1} \frac{\partial H_{1y}}{\partial x} = \frac{1}{j\omega\varepsilon_1} \frac{\partial}{\partial x} \left( \sum_{n=-\infty}^{+\infty} \left( A_n e^{+jk_{zn1}z} - B_n e^{-jk_{zn1}z} \right) e^{-jk_{xn}x} \right)$$

$$= \frac{1}{j\omega\varepsilon_1} \sum_{n=-\infty}^{+\infty} -jk_{xn} \left( A_n e^{+jk_{zn1}z} - B_n e^{-jk_{zn1}z} \right) e^{-jk_{xn}x}$$

$$\vec{H}_3^t = \sum_{n=-\infty}^{+\infty} C_n e^{-jk_{xn}x} e^{+jk_{zn3}z} \hat{y}$$

$$E_{3x} = \frac{-1}{j\omega\varepsilon_3} \frac{\partial H_{3y}}{\partial z} = \frac{-1}{j\omega\varepsilon_3} \frac{\partial}{\partial z} \left( \sum_{n=-\infty}^{+\infty} C_n e^{-jk_{xn}x} e^{+jk_{zn3}z} \right) = \frac{-1}{\omega\varepsilon_3} \sum_{n=-\infty}^{+\infty} k_{zn3} C_n e^{-jk_{xn}x} e^{+jk_{zn3}z}$$

$$E_{3z} = \frac{1}{j\omega\varepsilon_3}\frac{\partial H_{3y}}{\partial x} = \frac{1}{j\omega\varepsilon_3}\frac{\partial}{\partial x}\left(\sum_{n=-\infty}^{+\infty} C_n e^{-jk_{xn}x} e^{+jk_{zn3}z}\right) = \frac{-1}{\omega\varepsilon_3}\sum_{n=-\infty}^{+\infty} k_{xn} C_n e^{-jk_{xn}x} e^{+jk_{zn3}z}$$

$$\vec{H}_2^t = \sum_{n=-\infty}^{+\infty} D_n e^{-jk_{xn}x} e^{-jk_{zn2}z}(-\hat{y})$$

$$E_{2x} = \frac{-1}{j\omega\varepsilon_2}\frac{\partial H_{2y}}{\partial z} = \frac{-1}{j\omega\varepsilon_2}\frac{\partial}{\partial z}\left(-\sum_{n=-\infty}^{+\infty} D_n e^{-jk_{xn}x} e^{-jk_{zn2}z}\right) = \frac{-1}{\omega\varepsilon_2}\sum_{n=-\infty}^{+\infty} k_{zn2} D_n e^{-jk_{xn}x} e^{-jk_{zn2}z}$$

$$E_{2z} = \frac{1}{j\omega\varepsilon_2}\frac{\partial H_{2y}}{\partial x} = \frac{1}{j\omega\varepsilon_2}\frac{\partial}{\partial x}\left(-\sum_{n=-\infty}^{+\infty} D_n e^{-jk_{xn}x} e^{-jk_{zn2}z}\right) = \frac{1}{\omega\varepsilon_2}\sum_{n=-\infty}^{+\infty} k_{xn} D_n e^{-jk_{xn}x} e^{-jk_{zn2}z}$$

Therefore, the electric and magnetic fields for a $TM^x$ excitation are summarized and listed below:

$$H_{1y} = \sum_{n=-\infty}^{+\infty} \left(A_n e^{+jk_{zn1}z} - B_n e^{-jk_{zn1}z}\right) e^{-jk_{xn}x}$$

$$E_{1x} = \frac{-1}{j\omega\varepsilon_1}\sum_{n=-\infty}^{+\infty} jk_{zn1}\left(A_n e^{+jk_{zn1}z} + B_n e^{-jk_{zn1}z}\right) e^{-jk_{xn}x}$$

$$E_{1z} = \frac{1}{j\omega\varepsilon_1}\sum_{n=-\infty}^{+\infty} -jk_{xn}\left(A_n e^{+jk_{zn1}z} - B_n e^{-jk_{zn1}z}\right) e^{-jk_{xn}x}$$

$$H_{3y} = \sum_{n=-\infty}^{+\infty} C_n e^{-jk_{xn}x} e^{+jk_{zn3}z}$$

$$E_{3x} = \frac{-1}{\omega\varepsilon_3}\sum_{n=-\infty}^{+\infty} k_{zn3} C_n e^{-jk_{xn}x} e^{+jk_{zn3}z}$$

$$E_{3z} = \frac{-1}{\omega\varepsilon_3}\sum_{n=-\infty}^{+\infty} k_{xn} C_n e^{-jk_{xn}x} e^{+jk_{zn3}z}$$

$$H_{2y} = -\sum_{n=-\infty}^{+\infty} D_n e^{-jk_{xn}x} e^{-jk_{zn2}z}$$

$$E_{2x} = \frac{-1}{\omega\varepsilon_2}\sum_{n=-\infty}^{+\infty} k_{zn2} D_n e^{-jk_{xn}x} e^{-jk_{zn2}z}$$

$$E_{2z} = \frac{1}{\omega\varepsilon_2}\sum_{n=-\infty}^{+\infty} k_{xn} D_n e^{-jk_{xn}x} e^{-jk_{zn2}z}$$

Boundary conditions imply that the tangential component of the total electric fields must be continuous along the interfaces.

Furthermore, the discontinuities of the tangential component of the total magnetic fields are responsible to the generation of surface current densities, therefore:

$$E_{1x}(z=0) = E_{2x}(z=0) \ , \ E_{1x}(z=-d) = E_{3x}(z=-d)$$

$$H_{1y}(z=0) - H_{2y}(z=0) = J_{x1} = \frac{E_{2x}(z=0)}{\eta_{sheet1}}$$

$$H_{3y}(z=-d) - H_{1y}(z=-d) = J_{x2} = \frac{E_{3x}(z=-d)}{\eta_{sheet2}}$$

$$E_{1x}(z=0) = E_{2x}(z=0)$$

$$\frac{1}{\varepsilon_1} \sum_{n=-\infty}^{+\infty} k_{zn1}(A_n + B_n) e^{-j\left(k+\frac{2\pi n}{a}\right)x} = \frac{1}{\varepsilon_2} \sum_{n=-\infty}^{+\infty} k_{zn2} D_n e^{-j\left(k+\frac{2\pi n}{a}\right)x}$$

$$H_{1y}(z=0) - H_{2y}(z=0) = J_{x1} = \frac{E_{2x}(z=0)}{\eta_{sheet1}}$$

$$\sum_{n=-\infty}^{+\infty}(A_n - B_n) e^{-j\left(k+\frac{2\pi n}{a}\right)x} + \sum_{n=-\infty}^{+\infty} D_n e^{-j\left(k+\frac{2\pi n}{a}\right)x} = \frac{1}{\eta_{sheet1}}\left(\frac{-1}{\omega\varepsilon_2} \sum_{n=-\infty}^{+\infty} k_{zn2} D_n e^{-j\left(k+\frac{2\pi n}{a}\right)x}\right)$$

$$E_{1x}(z=-d) = E_{3x}(z=-d)$$

$$\frac{1}{\varepsilon_1} \sum_{n=-\infty}^{+\infty} k_{zn1}\left(A_n e^{-jk_{zn1}d} + B_n e^{+jk_{zn1}d}\right) e^{-j\left(k+\frac{2\pi n}{a}\right)x} = \frac{1}{\varepsilon_3} \sum_{n=-\infty}^{+\infty} k_{zn3} C_n e^{-j\left(k+\frac{2\pi n}{a}\right)x} e^{-jk_{zn3}d}$$

$$H_{3y}(z=-d) - H_{1y}(z=-d) = J_{x2} = \frac{E_{3x}(z=-d)}{\eta_{sheet2}}$$

$$\sum_{n=-\infty}^{+\infty} C_n e^{-j\left(k+\frac{2\pi n}{a}\right)x} e^{-jk_{zn3}d} - \sum_{n=-\infty}^{+\infty} \left(A_n e^{-jk_{zn1}d} - B_n e^{+jk_{zn1}d}\right) e^{-j\left(k+\frac{2\pi n}{a}\right)x}$$
$$= \frac{1}{\eta_{sheet2}}\left(\frac{-1}{\omega\varepsilon_3} \sum_{n=-\infty}^{+\infty} k_{zn3} C_n e^{-j\left(k+\frac{2\pi n}{a}\right)x} e^{-jk_{zn3}d}\right)$$

In general, we assumed that the impedance of the top sheet to be an arbitrary periodic function with the periodicity of $a$, which can be represented as a Fourier series as follows:

$$\eta_{sheet1}(x) = \sum_{m=-\infty}^{+\infty} \eta_m e^{-j\frac{2\pi m}{a}x}$$

And also, the bottom sheet is considered to support the coupled TM harmonics as surface waves with an intrinsic uniform gain as below:

$$\eta_{sheet2} = -R + jX_{AS}$$

Therefore, by substituting the defined impedance sheets, the equations obtained from boundary conditions, take the form of:

$$\frac{1}{\varepsilon_1}\sum_{n=-\infty}^{+\infty}k_{zn1}(A_n+B_n)e^{-j\left(k+\frac{2\pi n}{a}\right)x}=\frac{1}{\varepsilon_2}\sum_{n=-\infty}^{+\infty}k_{zn2}D_ne^{-j\left(k+\frac{2\pi n}{a}\right)x}$$

$$\sum_{m=-\infty}^{+\infty}\eta_m e^{-j\frac{2\pi m}{a}x}\left(\sum_{n=-\infty}^{+\infty}(A_n-B_n)e^{-j\left(k+\frac{2\pi n}{a}\right)x}+\sum_{n=-\infty}^{+\infty}D_ne^{-j\left(k+\frac{2\pi n}{a}\right)x}\right)=\frac{-1}{\omega\varepsilon_2}\sum_{n=-\infty}^{+\infty}k_{zn2}D_ne^{-j\left(k+\frac{2\pi n}{a}\right)x}$$

$$\frac{1}{\varepsilon_1}\sum_{n=-\infty}^{+\infty}k_{zn1}(A_ne^{-jk_{zn1}d}+B_ne^{+jk_{zn1}d})e^{-j\left(k+\frac{2\pi n}{a}\right)x}=\frac{1}{\varepsilon_3}\sum_{n=-\infty}^{+\infty}k_{zn3}C_ne^{-j\left(k+\frac{2\pi n}{a}\right)x}e^{-jk_{zn3}d}$$

$$(-R+jX_{AS})\left(\sum_{n=-\infty}^{+\infty}C_ne^{-j\left(k+\frac{2\pi n}{a}\right)x}e^{-jk_{zn3}d}-\sum_{n=-\infty}^{+\infty}(A_ne^{-jk_{zn1}d}-B_ne^{+jk_{zn1}d})e^{-j\left(k+\frac{2\pi n}{a}\right)x}\right)$$
$$=\frac{-1}{\omega\varepsilon_3}\sum_{n=-\infty}^{+\infty}k_{zn3}C_ne^{-j\left(k+\frac{2\pi n}{a}\right)x}e^{-jk_{zn3}d}$$

We have 4 equations, and 4 unknown parameters $A_n, B_n, C_n, D_n$.

Therefore, we can solve the derived set of equations for the only unknown parameter which is the propagation constant of the fundamental mode $k=\beta-j\alpha$.

First equation:

$$\frac{1}{\varepsilon_1}\sum_{n=-\infty}^{+\infty}k_{zn1}(A_n+B_n)e^{-j\left(k+\frac{2\pi n}{a}\right)x}=\frac{1}{\varepsilon_2}\sum_{n=-\infty}^{+\infty}k_{zn2}D_ne^{-j\left(k+\frac{2\pi n}{a}\right)x}$$

$$\int_0^a\frac{1}{\varepsilon_1}\sum_{n=-\infty}^{+\infty}k_{zn1}(A_n+B_n)e^{-j\left(k+\frac{2\pi n}{a}\right)x}e^{+j\left(k+\frac{2\pi g}{a}\right)x}dx=\int_0^a\frac{1}{\varepsilon_2}\sum_{n=-\infty}^{+\infty}k_{zn2}D_ne^{-j\left(k+\frac{2\pi n}{a}\right)x}e^{+j\left(k+\frac{2\pi g}{a}\right)x}dx$$

$$\frac{1}{\varepsilon_1}\sum_{n=-\infty}^{+\infty}k_{zn1}(A_n+B_n)\int_0^a e^{-j\left(k+\frac{2\pi n}{a}\right)x}e^{+j\left(k+\frac{2\pi g}{a}\right)x}dx=\frac{1}{\varepsilon_2}\sum_{n=-\infty}^{+\infty}k_{zn2}D_n\int_0^a e^{-j\left(k+\frac{2\pi n}{a}\right)x}e^{+j\left(k+\frac{2\pi g}{a}\right)x}dx$$

$$\frac{1}{\varepsilon_1}\sum_{n=-\infty}^{+\infty}k_{zn1}(A_n+B_n)\int_0^a e^{+j\frac{2\pi(g-n)}{a}x}dx=\frac{1}{\varepsilon_2}\sum_{n=-\infty}^{+\infty}k_{zn2}D_n\int_0^a e^{+j\frac{2\pi(g-n)}{a}x}dx$$

Orthogonality yields:

$$\int_0^a e^{+j\frac{2\pi(g-n)}{a}x}dx=\begin{cases}a & n=g\\ 0 & otherwise\end{cases}$$

$$\frac{k_{zn1}}{\varepsilon_1}(A_n+B_n)=\frac{1}{\varepsilon_2}k_{zn2}D_n \quad \Rightarrow \quad D_n=\frac{\varepsilon_2}{\varepsilon_1}\frac{k_{zn1}}{k_{zn2}}(A_n+B_n)$$

Third equation:

$$\frac{1}{\varepsilon_1}\sum_{n=-\infty}^{+\infty}k_{zn1}\left(A_n e^{-jk_{zn1}d}+B_n e^{+jk_{zn1}d}\right)e^{-j\left(k+\frac{2\pi n}{a}\right)x}=\frac{1}{\varepsilon_3}\sum_{n=-\infty}^{+\infty}k_{zn3}C_n e^{-j\left(k+\frac{2\pi n}{a}\right)x}e^{-jk_{zn3}d}$$

$$\int_0^a \frac{1}{\varepsilon_1}\sum_{n=-\infty}^{+\infty}k_{zn1}\left(A_n e^{-jk_{zn1}d}+B_n e^{+jk_{zn1}d}\right)e^{-j\left(k+\frac{2\pi n}{a}\right)x}e^{+j\left(k+\frac{2\pi g}{a}\right)x}dx$$

$$=\int_0^a \frac{1}{\varepsilon_3}\sum_{n=-\infty}^{+\infty}k_{zn3}C_n e^{-j\left(k+\frac{2\pi n}{a}\right)x}e^{+j\left(k+\frac{2\pi g}{a}\right)x}e^{-jk_{zn3}d}dx$$

$$\frac{1}{\varepsilon_1}\sum_{n=-\infty}^{+\infty}k_{zn1}\left(A_n e^{-jk_{zn1}d}+B_n e^{+jk_{zn1}d}\right)\int_0^a e^{-j\left(k+\frac{2\pi n}{a}\right)x}e^{+j\left(k+\frac{2\pi g}{a}\right)x}dx$$

$$=\frac{1}{\varepsilon_3}\sum_{n=-\infty}^{+\infty}k_{zn3}C_n e^{-jk_{zn3}d}\int_0^a e^{-j\left(k+\frac{2\pi n}{a}\right)x}e^{+j\left(k+\frac{2\pi g}{a}\right)x}dx$$

$$\frac{1}{\varepsilon_1}\sum_{n=-\infty}^{+\infty}k_{zn1}\left(A_n e^{-jk_{zn1}d}+B_n e^{+jk_{zn1}d}\right)\int_0^a e^{+j\frac{2\pi(g-n)}{a}x}dx=\frac{1}{\varepsilon_3}\sum_{n=-\infty}^{+\infty}k_{zn3}C_n e^{-jk_{zn3}d}\int_0^a e^{+j\frac{2\pi(g-n)}{a}x}dx$$

Orthogonality yields:

$$\int_0^a e^{+j\frac{2\pi(g-n)}{a}x}dx=\begin{cases}a & n=g\\ 0 & otherwise\end{cases}$$

$$\frac{1}{\varepsilon_1}k_{zn1}\left(A_n e^{-jk_{zn1}d}+B_n e^{+jk_{zn1}d}\right)=\frac{1}{\varepsilon_3}k_{zn3}C_n e^{-jk_{zn3}d}$$

$$\Rightarrow\quad C_n=\frac{\varepsilon_3}{\varepsilon_1}\frac{k_{zn1}}{k_{zn3}}e^{+jk_{zn3}d}\left(A_n e^{-jk_{zn1}d}+B_n e^{+jk_{zn1}d}\right)$$

Therefore, working on the first and third equations, resulted in the following relations:

$$D_n=\frac{\varepsilon_2}{\varepsilon_1}\frac{k_{zn1}}{k_{zn2}}(A_n+B_n)$$

$$C_n=\frac{\varepsilon_3}{\varepsilon_1}\frac{k_{zn1}}{k_{zn3}}e^{+jk_{zn3}d}\left(A_n e^{-jk_{zn1}d}+B_n e^{+jk_{zn1}d}\right)$$

By substituting the results obtained from the first and third equations, into the second and forth expressions we will have:

The second and forth equations are:

$$\sum_{m=-\infty}^{+\infty}\eta_m e^{-j\frac{2\pi m}{a}x}\left(\sum_{n=-\infty}^{+\infty}(A_n-B_n)e^{-j\left(k+\frac{2\pi n}{a}\right)x}+\sum_{n=-\infty}^{+\infty}D_n e^{-j\left(k+\frac{2\pi n}{a}\right)x}\right)=\frac{-1}{\omega\varepsilon_2}\sum_{n=-\infty}^{+\infty}k_{zn2}D_n e^{-j\left(k+\frac{2\pi n}{a}\right)x}$$

$$(-R + jX_{AS})\left(\sum_{n=-\infty}^{+\infty} C_n e^{-j\left(k+\frac{2\pi n}{a}\right)x} e^{-jk_{zn3}d} - \sum_{n=-\infty}^{+\infty} \left(A_n e^{-jk_{zn1}d} - B_n e^{+jk_{zn1}d}\right) e^{-j\left(k+\frac{2\pi n}{a}\right)x}\right)$$
$$= \frac{-1}{\omega\varepsilon_3} \sum_{n=-\infty}^{+\infty} k_{zn3} C_n e^{-j\left(k+\frac{2\pi n}{a}\right)x} e^{-jk_{zn3}d}$$

Substitutions yield:

$$\sum_{m=-\infty}^{+\infty} \eta_m e^{-j\frac{2\pi m}{a}x} \left( \sum_{n=-\infty}^{+\infty} (A_n - B_n) e^{-j\left(k+\frac{2\pi n}{a}\right)x} + \sum_{n=-\infty}^{+\infty} \left(\frac{\varepsilon_2}{\varepsilon_1}\frac{k_{zn1}}{k_{zn2}}(A_n + B_n)\right) e^{-j\left(k+\frac{2\pi n}{a}\right)x} \right)$$
$$= \frac{-1}{\omega\varepsilon_2} \sum_{n=-\infty}^{+\infty} k_{zn2} \left(\frac{\varepsilon_2}{\varepsilon_1}\frac{k_{zn1}}{k_{zn2}}(A_n + B_n)\right) e^{-j\left(k+\frac{2\pi n}{a}\right)x}$$

$$(-R + jX_{AS})\left( \sum_{n=-\infty}^{+\infty} \left(\frac{\varepsilon_3}{\varepsilon_1}\frac{k_{zn1}}{k_{zn3}} e^{+jk_{zn3}d}\left(A_n e^{-jk_{zn1}d} + B_n e^{+jk_{zn1}d}\right)\right) e^{-j\left(k+\frac{2\pi n}{a}\right)x} e^{-jk_{zn3}d} \right.$$
$$\left. - \sum_{n=-\infty}^{+\infty} \left(A_n e^{-jk_{zn1}d} - B_n e^{+jk_{zn1}d}\right) e^{-j\left(k+\frac{2\pi n}{a}\right)x} \right)$$
$$= \frac{-1}{\omega\varepsilon_3} \sum_{n=-\infty}^{+\infty} k_{zn3} \left(\frac{\varepsilon_3}{\varepsilon_1}\frac{k_{zn1}}{k_{zn3}} e^{+jk_{zn3}d}\left(A_n e^{-jk_{zn1}d} + B_n e^{+jk_{zn1}d}\right)\right) e^{-j\left(k+\frac{2\pi n}{a}\right)x} e^{-jk_{zn3}d}$$

Now, let's work on the first expression obtained above:

$$\sum_{m=-\infty}^{+\infty} \eta_m e^{-j\frac{2\pi m}{a}x} \left( \sum_{n=-\infty}^{+\infty} (A_n - B_n) e^{-j\left(k+\frac{2\pi n}{a}\right)x} + \sum_{n=-\infty}^{+\infty} \left(\frac{\varepsilon_2}{\varepsilon_1}\frac{k_{zn1}}{k_{zn2}}(A_n + B_n)\right) e^{-j\left(k+\frac{2\pi n}{a}\right)x} \right)$$
$$= \frac{-1}{\omega\varepsilon_2} \sum_{n=-\infty}^{+\infty} k_{zn2} \left(\frac{\varepsilon_2}{\varepsilon_1}\frac{k_{zn1}}{k_{zn2}}(A_n + B_n)\right) e^{-j\left(k+\frac{2\pi n}{a}\right)x}$$

$$\sum_{m=-\infty}^{+\infty} \eta_m e^{-j\frac{2\pi m}{a}x} \left( \sum_{n=-\infty}^{+\infty} \left(A_n - B_n + \frac{\varepsilon_2}{\varepsilon_1}\frac{k_{zn1}}{k_{zn2}}(A_n + B_n)\right) e^{-j\left(k+\frac{2\pi n}{a}\right)x} \right)$$
$$= \frac{-1}{\omega\varepsilon_2} \sum_{n=-\infty}^{+\infty} \left(\frac{\varepsilon_2}{\varepsilon_1} k_{zn1}(A_n + B_n)\right) e^{-j\left(k+\frac{2\pi n}{a}\right)x}$$

$$\sum_{n=-\infty}^{+\infty} \sum_{m=-\infty}^{+\infty} \left(\eta_m e^{-j\frac{2\pi m}{a}x}\right)\left(A_n - B_n + \frac{\varepsilon_2}{\varepsilon_1}\frac{k_{zn1}}{k_{zn2}}(A_n + B_n)\right) e^{-j\left(k+\frac{2\pi n}{a}\right)x}$$
$$= \frac{-1}{\omega\varepsilon_2} \sum_{n=-\infty}^{+\infty} \left(\frac{\varepsilon_2}{\varepsilon_1} k_{zn1}(A_n + B_n)\right) e^{-j\left(k+\frac{2\pi n}{a}\right)x}$$

$$\sum_{n=-\infty}^{+\infty}\sum_{m=-\infty}^{+\infty}\eta_m\left(A_n-B_n+\frac{\varepsilon_2}{\varepsilon_1}\frac{k_{zn1}}{k_{zn2}}(A_n+B_n)\right)e^{-j\left(k+\frac{2\pi(n+m)}{a}\right)x}$$

$$=\frac{-1}{\omega\varepsilon_2}\sum_{n=-\infty}^{+\infty}\left(\frac{\varepsilon_2}{\varepsilon_1}k_{zn1}(A_n+B_n)\right)e^{-j\left(k+\frac{2\pi n}{a}\right)x}$$

$$\int_0^a\sum_{n=-\infty}^{+\infty}\sum_{m=-\infty}^{+\infty}\eta_m\left(A_n-B_n+\frac{\varepsilon_2}{\varepsilon_1}\frac{k_{zn1}}{k_{zn2}}(A_n+B_n)\right)e^{-j\left(k+\frac{2\pi(n+m)}{a}\right)x}e^{+j\left(k+\frac{2\pi p}{a}\right)x}dx$$

$$=\int_0^a\frac{-1}{\omega\varepsilon_2}\sum_{n=-\infty}^{+\infty}\left(\frac{\varepsilon_2}{\varepsilon_1}k_{zn1}(A_n+B_n)\right)e^{-j\left(k+\frac{2\pi n}{a}\right)x}e^{+j\left(k+\frac{2\pi p}{a}\right)x}dx$$

$$\sum_{n=-\infty}^{+\infty}\sum_{m=-\infty}^{+\infty}\eta_m\left(A_n-B_n+\frac{\varepsilon_2}{\varepsilon_1}\frac{k_{zn1}}{k_{zn2}}(A_n+B_n)\right)\int_0^a e^{+j\frac{2\pi(p-n-m)}{a}x}dx$$

$$=\frac{-1}{\omega\varepsilon_2}\sum_{n=-\infty}^{+\infty}\left(\frac{\varepsilon_2}{\varepsilon_1}k_{zn1}(A_n+B_n)\right)\int_0^a e^{+j\frac{2\pi(p-n)}{a}x}dx$$

$$\sum_{n=-\infty}^{+\infty}\eta_{p-n}\left(A_n-B_n+\frac{\varepsilon_2}{\varepsilon_1}\frac{k_{zn1}}{k_{zn2}}(A_n+B_n)\right)=\frac{-k_{zp1}}{\omega\varepsilon_1}(A_p+B_p)$$

Also, in the following we worked on the second expression:

$$(-R+jX_{AS})\left(\sum_{n=-\infty}^{+\infty}\left(\frac{\varepsilon_3}{\varepsilon_1}\frac{k_{zn1}}{k_{zn3}}e^{+jk_{zn3}d}(A_ne^{-jk_{zn1}d}+B_ne^{+jk_{zn1}d})\right)e^{-j\left(k+\frac{2\pi n}{a}\right)x}e^{-jk_{zn3}d}\right.$$

$$\left.-\sum_{n=-\infty}^{+\infty}(A_ne^{-jk_{zn1}d}-B_ne^{+jk_{zn1}d})e^{-j\left(k+\frac{2\pi n}{a}\right)x}\right)$$

$$=\frac{-1}{\omega\varepsilon_3}\sum_{n=-\infty}^{+\infty}k_{zn3}\left(\frac{\varepsilon_3}{\varepsilon_1}\frac{k_{zn1}}{k_{zn3}}e^{+jk_{zn3}d}(A_ne^{-jk_{zn1}d}+B_ne^{+jk_{zn1}d})\right)e^{-j\left(k+\frac{2\pi n}{a}\right)x}e^{-jk_{zn3}d}$$

$$(-R+jX_{AS})\left(\sum_{n=-\infty}^{+\infty}\left(\frac{\varepsilon_3}{\varepsilon_1}\frac{k_{zn1}}{k_{zn3}}(A_ne^{-jk_{zn1}d}+B_ne^{+jk_{zn1}d})-A_ne^{-jk_{zn1}d}+B_ne^{+jk_{zn1}d}\right)e^{-j\left(k+\frac{2\pi n}{a}\right)x}\right)$$

$$=\frac{-1}{\omega\varepsilon_3}\sum_{n=-\infty}^{+\infty}\frac{\varepsilon_3}{\varepsilon_1}k_{zn1}(A_ne^{-jk_{zn1}d}+B_ne^{+jk_{zn1}d})e^{-j\left(k+\frac{2\pi n}{a}\right)x}$$

$$\int_0^a(-R+jX_{AS})\left(\sum_{n=-\infty}^{+\infty}\left(\frac{\varepsilon_3}{\varepsilon_1}\frac{k_{zn1}}{k_{zn3}}(A_ne^{-jk_{zn1}d}+B_ne^{+jk_{zn1}d})-A_ne^{-jk_{zn1}d}\right.\right.$$

$$\left.\left.+B_ne^{+jk_{zn1}d}\right)e^{-j\left(k+\frac{2\pi n}{a}\right)x}e^{+j\left(k+\frac{2\pi q}{a}\right)x}\right)dx$$

$$=\int_0^a\frac{-1}{\omega\varepsilon_3}\sum_{n=-\infty}^{+\infty}\frac{\varepsilon_3}{\varepsilon_1}k_{zn1}(A_ne^{-jk_{zn1}d}+B_ne^{+jk_{zn1}d})e^{-j\left(k+\frac{2\pi n}{a}\right)x}e^{+j\left(k+\frac{2\pi q}{a}\right)x}dx$$

$$(-R + jX_{AS})\left(\sum_{n=-\infty}^{+\infty}\left(\frac{\varepsilon_3}{\varepsilon_1}\frac{k_{zn1}}{k_{zn3}}(A_n e^{-jk_{zn1}d} + B_n e^{+jk_{zn1}d}) - A_n e^{-jk_{zn1}d}\right.\right.$$
$$\left.\left. + B_n e^{+jk_{zn1}d}\right)\int_0^a e^{-j\left(k+\frac{2\pi n}{a}\right)x} e^{+j\left(k+\frac{2\pi q}{a}\right)x}\,dx\right)$$
$$= \frac{-1}{\omega \varepsilon_3}\sum_{n=-\infty}^{+\infty}\frac{\varepsilon_3}{\varepsilon_1}k_{zn1}(A_n e^{-jk_{zn1}d} + B_n e^{+jk_{zn1}d})\int_0^a e^{-j\left(k+\frac{2\pi n}{a}\right)x} e^{+j\left(k+\frac{2\pi q}{a}\right)x}\,dx$$

$$\Rightarrow (-R + jX_{AS})\left(\frac{\varepsilon_3}{\varepsilon_1}\frac{k_{zn1}}{k_{zn3}}(A_n e^{-jk_{zn1}d} + B_n e^{+jk_{zn1}d}) - A_n e^{-jk_{zn1}d} + B_n e^{+jk_{zn1}d}\right)$$
$$= \frac{-k_{zn1}}{\omega \varepsilon_1}(A_n e^{-jk_{zn1}d} + B_n e^{+jk_{zn1}d})$$

Using the above expression, we can now find $B_n$ in terms of $A_n$.

$$(-R + jX_{AS})\left(\frac{\varepsilon_3}{\varepsilon_1}\frac{k_{zn1}}{k_{zn3}}(A_n + B_n e^{+2jk_{zn1}d}) - A_n + B_n e^{+2jk_{zn1}d}\right) = \frac{-k_{zn1}}{\omega \varepsilon_1}(A_n + B_n e^{+2jk_{zn1}d})$$

$$(-R + jX_{AS})\left(\frac{\varepsilon_3}{\varepsilon_1}\frac{k_{zn1}}{k_{zn3}} - 1\right)A_n + (-R + jX_{AS})\left(\frac{\varepsilon_3}{\varepsilon_1}\frac{k_{zn1}}{k_{zn3}} + 1\right)e^{+2jk_{zn1}d}B_n = \frac{-k_{zn1}}{\omega \varepsilon_1}A_n - \frac{k_{zn1}}{\omega \varepsilon_1}e^{+2jk_{zn1}d}B_n$$

$$(-R + jX_{AS})\left(\frac{\varepsilon_3}{\varepsilon_1}\frac{k_{zn1}}{k_{zn3}} - 1\right)A_n + \frac{k_{zn1}}{\omega \varepsilon_1}A_n = -\frac{k_{zn1}}{\omega \varepsilon_1}e^{+2jk_{zn1}d}B_n - (-R + jX_{AS})\left(\frac{\varepsilon_3}{\varepsilon_1}\frac{k_{zn1}}{k_{zn3}} + 1\right)e^{+2jk_{zn1}d}B_n$$

$$A_n\left((-R + jX_{AS})\left(\frac{\varepsilon_3}{\varepsilon_1}\frac{k_{zn1}}{k_{zn3}} - 1\right) + \frac{k_{zn1}}{\omega \varepsilon_1}\right) = B_n\left(-\frac{k_{zn1}}{\omega \varepsilon_1}e^{+2jk_{zn1}d} - (-R + jX_{AS})\left(\frac{\varepsilon_3}{\varepsilon_1}\frac{k_{zn1}}{k_{zn3}} + 1\right)e^{+2jk_{zn1}d}\right)$$

$$B_n = \frac{(-R + jX_{AS})\left(1 - \frac{\varepsilon_3}{\varepsilon_1}\frac{k_{zn1}}{k_{zn3}}\right) - \frac{k_{zn1}}{\omega \varepsilon_1}}{\frac{k_{zn1}}{\omega \varepsilon_1} + (-R + jX_{AS})\left(1 + \frac{\varepsilon_3}{\varepsilon_1}\frac{k_{zn1}}{k_{zn3}}\right)} e^{-2jk_{zn1}d} A_n$$

Now that we obtained $B_n$ in terms of $A_n$, we can readily substitute it into the following expression that we have previously derived:

Previously we have obtained the following equation:

$$\sum_{n=-\infty}^{+\infty}\eta_{p-n}\left(A_n - B_n + \frac{\varepsilon_2}{\varepsilon_1}\frac{k_{zn1}}{k_{zn2}}(A_n + B_n)\right) = \frac{-k_{zp1}}{\omega \varepsilon_1}(A_p + B_p)$$

$$\to \sum_{n=-\infty}^{+\infty}\eta_{p-n}\left(A_n\left(1 + \frac{\varepsilon_2}{\varepsilon_1}\frac{k_{zn1}}{k_{zn2}}\right) + \left(\frac{\varepsilon_2}{\varepsilon_1}\frac{k_{zn1}}{k_{zn2}} - 1\right)B_n\right) = \frac{-k_{zp1}}{\omega \varepsilon_1}(A_p + B_p)$$

Substitution yields:

$$\sum_{n=-\infty}^{+\infty} \eta_{p-n} \left\{ \left(1 + \frac{\varepsilon_2}{\varepsilon_1}\frac{k_{zn1}}{k_{zn2}}\right) + \left(\frac{\varepsilon_2}{\varepsilon_1}\frac{k_{zn1}}{k_{zn2}} - 1\right) \frac{(-R+jX_{AS})\left(1 - \frac{\varepsilon_3}{\varepsilon_1}\frac{k_{zn1}}{k_{zn3}}\right) - \frac{k_{zn1}}{\omega\varepsilon_1}}{\frac{k_{zn1}}{\omega\varepsilon_1} + (-R+jX_{AS})\left(1 + \frac{\varepsilon_3}{\varepsilon_1}\frac{k_{zn1}}{k_{zn3}}\right)} e^{-2jk_{zn1}d} \right\} A_n$$

$$= \frac{-k_{zp1}}{\omega\varepsilon_1} \left\{ 1 + \frac{(-R+jX_{AS})\left(1 - \frac{\varepsilon_3}{\varepsilon_1}\frac{k_{zp1}}{k_{zp3}}\right) - \frac{k_{zp1}}{\omega\varepsilon_1}}{\frac{k_{zp1}}{\omega\varepsilon_1} + (-R+jX_{AS})\left(1 + \frac{\varepsilon_3}{\varepsilon_1}\frac{k_{zp1}}{k_{zp3}}\right)} e^{-2jk_{zp1}d} \right\} A_p$$

$$\sum_{n=-\infty}^{+\infty} \frac{-1}{k_{zp1}} \left\{ 1 + \frac{(-R+jX_{AS})\left(1 - \frac{\varepsilon_3}{\varepsilon_1}\frac{k_{zp1}}{k_{zp3}}\right) - \frac{k_{zp1}}{\omega\varepsilon_1}}{\frac{k_{zp1}}{\omega\varepsilon_1} + (-R+jX_{AS})\left(1 + \frac{\varepsilon_3}{\varepsilon_1}\frac{k_{zp1}}{k_{zp3}}\right)} e^{-2jk_{zp1}d} \right\}^{-1} \left\{ \left(1 + \frac{\varepsilon_2}{\varepsilon_1}\frac{k_{zn1}}{k_{zn2}}\right) \right.$$

$$\left. + \left(\frac{\varepsilon_2}{\varepsilon_1}\frac{k_{zn1}}{k_{zn2}} - 1\right) \frac{(-R+jX_{AS})\left(1 - \frac{\varepsilon_3}{\varepsilon_1}\frac{k_{zn1}}{k_{zn3}}\right) - \frac{k_{zn1}}{\omega\varepsilon_1}}{\frac{k_{zn1}}{\omega\varepsilon_1} + (-R+jX_{AS})\left(1 + \frac{\varepsilon_3}{\varepsilon_1}\frac{k_{zn1}}{k_{zn3}}\right)} e^{-2jk_{zn1}d} \right\} \eta_{p-n} A_n = \frac{1}{\omega\varepsilon_1} A_p$$

$$\Rightarrow \quad \bar{\bar{Q}} \bar{A}_n = \frac{1}{\omega\varepsilon_1} \bar{A}_p$$

Where:

$$Q_{p,n} = \frac{-1}{k_{zp1}} \left\{ 1 + \frac{(-R+jX_{AS})\left(1 - \frac{\varepsilon_3}{\varepsilon_1}\frac{k_{zp1}}{k_{zp3}}\right) - \frac{k_{zp1}}{\omega\varepsilon_1}}{\frac{k_{zp1}}{\omega\varepsilon_1} + (-R+jX_{AS})\left(1 + \frac{\varepsilon_3}{\varepsilon_1}\frac{k_{zp1}}{k_{zp3}}\right)} e^{-2jk_{zp1}d} \right\}^{-1} \left\{ \left(1 + \frac{\varepsilon_2}{\varepsilon_1}\frac{k_{zn1}}{k_{zn2}}\right) \right.$$

$$\left. + \left(\frac{\varepsilon_2}{\varepsilon_1}\frac{k_{zn1}}{k_{zn2}} - 1\right) \frac{(-R+jX_{AS})\left(1 - \frac{\varepsilon_3}{\varepsilon_1}\frac{k_{zn1}}{k_{zn3}}\right) - \frac{k_{zn1}}{\omega\varepsilon_1}}{\frac{k_{zn1}}{\omega\varepsilon_1} + (-R+jX_{AS})\left(1 + \frac{\varepsilon_3}{\varepsilon_1}\frac{k_{zn1}}{k_{zn3}}\right)} e^{-2jk_{zn1}d} \right\} \eta_{p-n}$$

By truncating the indices $n$ and $p$ symmetrically about zero and to identical ranges (for example, $n = -2$ to $2$ and $p = -2$ to $2$) the above equation can be rewritten as:

$$\bar{\bar{Q}} \bar{A}_n = \frac{1}{\omega\varepsilon_1} \bar{A}_n \quad \rightarrow \quad \bar{\bar{Q}} \bar{A}_n - \frac{1}{\omega\varepsilon_1} \bar{A}_n = 0 \quad \rightarrow \quad \left( \bar{\bar{Q}} - \frac{1}{\omega\varepsilon_1} \bar{\bar{I}} \right) \bar{A}_n = 0$$

$$\rightarrow \quad \bar{\bar{\bar{Q}}} \bar{A}_n = 0 \quad ; \quad \bar{\bar{\bar{Q}}} \triangleq \bar{\bar{Q}} - \frac{1}{\omega\varepsilon_1} \bar{\bar{I}}$$

Using matrix notation, we can represent the dispersion equation in a compact form as follows:

$$Q_{p,n} = \frac{-1}{k_{zp1}} \left\{ 1 + \frac{(-R+jX_{AS})\left(1 - \frac{\varepsilon_3}{\varepsilon_1}\frac{k_{zp1}}{k_{zp3}}\right) - \frac{k_{zp1}}{\omega\varepsilon_1}}{\frac{k_{zp1}}{\omega\varepsilon_1} + (-R+jX_{AS})\left(1 + \frac{\varepsilon_3}{\varepsilon_1}\frac{k_{zp1}}{k_{zp3}}\right)} e^{-2jk_{zp1}d} \right\}^{-1} \left\{ \left(1 + \frac{\varepsilon_2}{\varepsilon_1}\frac{k_{zn1}}{k_{zn2}}\right) \right.$$

$$\left. + \left(\frac{\varepsilon_2}{\varepsilon_1}\frac{k_{zn1}}{k_{zn2}} - 1\right) \frac{(-R+jX_{AS})\left(1 - \frac{\varepsilon_3}{\varepsilon_1}\frac{k_{zn1}}{k_{zn3}}\right) - \frac{k_{zn1}}{\omega\varepsilon_1}}{\frac{k_{zn1}}{\omega\varepsilon_1} + (-R+jX_{AS})\left(1 + \frac{\varepsilon_3}{\varepsilon_1}\frac{k_{zn1}}{k_{zn3}}\right)} e^{-2jk_{zn1}d} \right\} \eta_{p-n}$$

$$\bar{\bar{Q}} = \begin{bmatrix} Q_{-2,-2} - \dfrac{1}{\omega\varepsilon_1} & Q_{-2,-1} & Q_{-2,0} & Q_{-2,+1} & Q_{-2,+2} \\ Q_{-1,-2} & Q_{-1,-1} - \dfrac{1}{\omega\varepsilon_1} & Q_{-1,0} & Q_{-1,+1} & Q_{-1,+2} \\ Q_{0,-2} & Q_{0,-1} & Q_{0,0} - \dfrac{1}{\omega\varepsilon_1} & Q_{0,+1} & Q_{0,+2} \\ Q_{+1,-2} & Q_{+1,-1} & Q_{+1,0} & Q_{+1,+1} - \dfrac{1}{\omega\varepsilon_1} & Q_{+1,+2} \\ Q_{+2,-2} & Q_{+2,-1} & Q_{+2,0} & Q_{+2,+1} & Q_{+2,+2} - \dfrac{1}{\omega\varepsilon_1} \end{bmatrix}$$

The determinant of $\bar{\bar{Q}}$ must be zero for non-trivial solutions:

$$\bar{\bar{Q}}\bar{A}_n = 0 \implies det(\bar{\bar{Q}}) = 0$$

The value of $k$ that satisfies this zero determinant condition will be the solution of interest.

## II. Dispersion equation for a sinusoidally periodic sheet impedance coupled with a uniform active sheet impedance

The formulation discussed in the preceding section is applicable to any periodic variation of $\eta_{sheet}$. In the specific scenario where $\eta_{sheet}$ exhibits a sinusoidal variation, we have:

$$\eta_{sheet1} = jX_s\left(1 + M\cos\left(\frac{2\pi}{a}\right)x\right) = jX_s + \frac{jX_sM}{2}\left(e^{+j\frac{2\pi}{a}x} + e^{-j\frac{2\pi}{a}x}\right)$$

$X_s$: average sheet reactance, $M$: Modulation factor

In general, for an arbitrary periodic sheet impedance coupled with a constant sheet impedance, by using Maxwell's equations and applying boundary conditions we obtained:

$$\frac{1}{\varepsilon_1}\sum_{n=-\infty}^{+\infty} k_{zn1}(A_n + B_n)e^{-j\left(k+\frac{2\pi n}{a}\right)x} = \frac{1}{\varepsilon_2}\sum_{n=-\infty}^{+\infty} k_{zn2}D_n e^{-j\left(k+\frac{2\pi n}{a}\right)x}$$

$$\sum_{n=-\infty}^{+\infty}(A_n - B_n)e^{-j\left(k+\frac{2\pi n}{a}\right)x} + \sum_{n=-\infty}^{+\infty} D_n e^{-j\left(k+\frac{2\pi n}{a}\right)x} = \frac{1}{\eta_{sheet1}}\left(\frac{-1}{\omega\varepsilon_2}\sum_{n=-\infty}^{+\infty} k_{zn2}D_n e^{-j\left(k+\frac{2\pi n}{a}\right)x}\right)$$

$$\frac{1}{\varepsilon_1}\sum_{n=-\infty}^{+\infty} k_{zn1}\left(A_n e^{-jk_{zn1}d} + B_n e^{+jk_{zn1}d}\right)e^{-j\left(k+\frac{2\pi n}{a}\right)x} = \frac{1}{\varepsilon_3}\sum_{n=-\infty}^{+\infty} k_{zn3}C_n e^{-j\left(k+\frac{2\pi n}{a}\right)x}e^{-jk_{zn3}d}$$

$$\sum_{n=-\infty}^{+\infty} C_n e^{-j\left(k+\frac{2\pi n}{a}\right)x}e^{-jk_{zn3}d} - \sum_{n=-\infty}^{+\infty}\left(A_n e^{-jk_{zn1}d} - B_n e^{+jk_{zn1}d}\right)e^{-j\left(k+\frac{2\pi n}{a}\right)x}$$
$$= \frac{1}{\eta_{sheet2}}\left(\frac{-1}{\omega\varepsilon_3}\sum_{n=-\infty}^{+\infty} k_{zn3}C_n e^{-j\left(k+\frac{2\pi n}{a}\right)x}e^{-jk_{zn3}d}\right)$$

Since, here we only specified $\eta_{sheet1}$ as sinusoidal modulation compared to the general case, only the second equation needs to be worked out, and we can use the final results obtained for the rest of the equations as follows:

$$D_n = \frac{\varepsilon_2}{\varepsilon_1}\frac{k_{zn1}}{k_{zn2}}(A_n + B_n)$$

$$C_n = \frac{\varepsilon_3}{\varepsilon_1}\frac{k_{zn1}}{k_{zn3}}e^{+jk_{zn3}d}\left(A_n e^{-jk_{zn1}d} + B_n e^{+jk_{zn1}d}\right)$$

$$B_n = \frac{(-R+jX_{AS})\left(1-\frac{\varepsilon_3}{\varepsilon_1}\frac{k_{zn1}}{k_{zn3}}\right)-\frac{k_{zn1}}{\omega\varepsilon_1}}{\frac{k_{zn1}}{\omega\varepsilon_1}+(-R+jX_{AS})\left(1+\frac{\varepsilon_3}{\varepsilon_1}\frac{k_{zn1}}{k_{zn3}}\right)}e^{-2jk_{zn1}d}A_n$$

By substituting $\eta_{sheet1} = jX_s\left(1+M\cos\left(\frac{2\pi}{a}\right)x\right)$ in the second equation we can write:

$$\sum_{n=-\infty}^{+\infty}(A_n - B_n)e^{-j\left(k+\frac{2\pi n}{a}\right)x} + \sum_{n=-\infty}^{+\infty}D_n e^{-j\left(k+\frac{2\pi n}{a}\right)x} = \frac{1}{\eta_{sheet1}}\left(\frac{-1}{\omega\varepsilon_2}\sum_{n=-\infty}^{+\infty}k_{zn2}D_n e^{-j\left(k+\frac{2\pi n}{a}\right)x}\right)$$

$$\eta_{sheet1} = jX_s\left(1+M\cos\left(\frac{2\pi}{a}\right)x\right) = jX_s + \frac{jX_s M}{2}\left(e^{+j\frac{2\pi}{a}x}+e^{-j\frac{2\pi}{a}x}\right)$$

$$\Rightarrow \left(jX_s + \frac{jX_s M}{2}\left(e^{+j\frac{2\pi}{a}x}+e^{-j\frac{2\pi}{a}x}\right)\right)\left\{\sum_{n=-\infty}^{+\infty}(A_n - B_n)e^{-j\left(k+\frac{2\pi n}{a}\right)x} + \sum_{n=-\infty}^{+\infty}D_n e^{-j\left(k+\frac{2\pi n}{a}\right)x}\right\}$$

$$= \left(\frac{-1}{\omega\varepsilon_2}\sum_{n=-\infty}^{+\infty}k_{zn2}D_n e^{-j\left(k+\frac{2\pi n}{a}\right)x}\right)$$

Substituting $D_n$ in terms of $A_n$ and $B_n$ gives:

$$\Rightarrow \left(jX_s + \frac{jX_s M}{2}\left(e^{+j\frac{2\pi}{a}x}+e^{-j\frac{2\pi}{a}x}\right)\right)\left\{\sum_{n=-\infty}^{+\infty}(A_n - B_n)e^{-j\left(k+\frac{2\pi n}{a}\right)x} + \sum_{n=-\infty}^{+\infty}\frac{\varepsilon_2}{\varepsilon_1}\frac{k_{zn1}}{k_{zn2}}(A_n + B_n)e^{-j\left(k+\frac{2\pi n}{a}\right)x}\right\}$$

$$= \frac{-1}{\omega\varepsilon_2}\sum_{n=-\infty}^{+\infty}k_{zn2}\frac{\varepsilon_2}{\varepsilon_1}\frac{k_{zn1}}{k_{zn2}}(A_n + B_n)e^{-j\left(k+\frac{2\pi n}{a}\right)x}$$

Next, substituting $B_n$ in terms of $A_n$ gives:

$$\Rightarrow \left(jX_s + \frac{jX_s M}{2}\left(e^{+j\frac{2\pi}{a}x}+e^{-j\frac{2\pi}{a}x}\right)\right)$$

$$\times \left\{\sum_{n=-\infty}^{+\infty}A_n\left(1-\frac{(-R+jX_{AS})\left(1-\frac{\varepsilon_3}{\varepsilon_1}\frac{k_{zn1}}{k_{zn3}}\right)-\frac{k_{zn1}}{\omega\varepsilon_1}}{\frac{k_{zn1}}{\omega\varepsilon_1}+(-R+jX_{AS})\left(1+\frac{\varepsilon_3}{\varepsilon_1}\frac{k_{zn1}}{k_{zn3}}\right)}e^{-2jk_{zn1}d}\right)e^{-j\left(k+\frac{2\pi n}{a}\right)x}\right.$$

$$\left. + \sum_{n=-\infty}^{+\infty}\frac{\varepsilon_2}{\varepsilon_1}\frac{k_{zn1}}{k_{zn2}}A_n\left(1+\frac{(-R+jX_{AS})\left(1-\frac{\varepsilon_3}{\varepsilon_1}\frac{k_{zn1}}{k_{zn3}}\right)-\frac{k_{zn1}}{\omega\varepsilon_1}}{\frac{k_{zn1}}{\omega\varepsilon_1}+(-R+jX_{AS})\left(1+\frac{\varepsilon_3}{\varepsilon_1}\frac{k_{zn1}}{k_{zn3}}\right)}e^{-2jk_{zn1}d}\right)e^{-j\left(k+\frac{2\pi n}{a}\right)x}\right\}$$

$$= \frac{-1}{\omega\varepsilon_2}\sum_{n=-\infty}^{+\infty}k_{zn2}\frac{\varepsilon_2}{\varepsilon_1}\frac{k_{zn1}}{k_{zn2}}A_n\left(1+\frac{(-R+jX_{AS})\left(1-\frac{\varepsilon_3}{\varepsilon_1}\frac{k_{zn1}}{k_{zn3}}\right)-\frac{k_{zn1}}{\omega\varepsilon_1}}{\frac{k_{zn1}}{\omega\varepsilon_1}+(-R+jX_{AS})\left(1+\frac{\varepsilon_3}{\varepsilon_1}\frac{k_{zn1}}{k_{zn3}}\right)}e^{-2jk_{zn1}d}\right)e^{-j\left(k+\frac{2\pi n}{a}\right)x}$$

We can encapsulate the summations of the left-hand side and factor out the common terms and write the above equation as:

$$\Rightarrow \left( jX_s + \frac{jX_s M}{2}\left( e^{+j\frac{2\pi}{a}x} + e^{-j\frac{2\pi}{a}x} \right) \right)$$

$$\times \sum_{n=-\infty}^{+\infty} A_n \left\{ 1 - \frac{(-R + jX_{AS})\left(1 - \frac{\varepsilon_3}{\varepsilon_1}\frac{k_{zn1}}{k_{zn3}}\right) - \frac{k_{zn1}}{\omega\varepsilon_1}}{\frac{k_{zn1}}{\omega\varepsilon_1} + (-R + jX_{AS})\left(1 + \frac{\varepsilon_3}{\varepsilon_1}\frac{k_{zn1}}{k_{zn3}}\right)} e^{-2jk_{zn1}d} \right.$$

$$\left. + \frac{\varepsilon_2}{\varepsilon_1}\frac{k_{zn1}}{k_{zn2}}\left(1 + \frac{(-R + jX_{AS})\left(1 - \frac{\varepsilon_3}{\varepsilon_1}\frac{k_{zn1}}{k_{zn3}}\right) - \frac{k_{zn1}}{\omega\varepsilon_1}}{\frac{k_{zn1}}{\omega\varepsilon_1} + (-R + jX_{AS})\left(1 + \frac{\varepsilon_3}{\varepsilon_1}\frac{k_{zn1}}{k_{zn3}}\right)} e^{-2jk_{zn1}d}\right) \right\} e^{-j\left(k + \frac{2\pi n}{a}\right)x}$$

$$= \frac{-1}{\omega\varepsilon_2} \sum_{n=-\infty}^{+\infty} k_{zn2} \frac{\varepsilon_2}{\varepsilon_1}\frac{k_{zn1}}{k_{zn2}} A_n \left( 1 + \frac{(-R + jX_{AS})\left(1 - \frac{\varepsilon_3}{\varepsilon_1}\frac{k_{zn1}}{k_{zn3}}\right) - \frac{k_{zn1}}{\omega\varepsilon_1}}{\frac{k_{zn1}}{\omega\varepsilon_1} + (-R + jX_{AS})\left(1 + \frac{\varepsilon_3}{\varepsilon_1}\frac{k_{zn1}}{k_{zn3}}\right)} e^{-2jk_{zn1}d} \right) e^{-j\left(k + \frac{2\pi n}{a}\right)x}$$

By multiplying the $\eta_{sheet1}$ and further simplifying the above equation, we have:

$$jX_s \sum_{n=-\infty}^{+\infty} A_n \left\{ 1 - \frac{(-R + jX_{AS})\left(1 - \frac{\varepsilon_3}{\varepsilon_1}\frac{k_{zn1}}{k_{zn3}}\right) - \frac{k_{zn1}}{\omega\varepsilon_1}}{\frac{k_{zn1}}{\omega\varepsilon_1} + (-R + jX_{AS})\left(1 + \frac{\varepsilon_3}{\varepsilon_1}\frac{k_{zn1}}{k_{zn3}}\right)} e^{-2jk_{zn1}d} \right.$$

$$\left. + \frac{\varepsilon_2}{\varepsilon_1}\frac{k_{zn1}}{k_{zn2}}\left(1 + \frac{(-R + jX_{AS})\left(1 - \frac{\varepsilon_3}{\varepsilon_1}\frac{k_{zn1}}{k_{zn3}}\right) - \frac{k_{zn1}}{\omega\varepsilon_1}}{\frac{k_{zn1}}{\omega\varepsilon_1} + (-R + jX_{AS})\left(1 + \frac{\varepsilon_3}{\varepsilon_1}\frac{k_{zn1}}{k_{zn3}}\right)} e^{-2jk_{zn1}d}\right) \right\} e^{-j\left(k + \frac{2\pi n}{a}\right)x} +$$

$$\frac{jX_s M}{2} \sum_{n=-\infty}^{+\infty} A_n \left\{ 1 - \frac{(-R + jX_{AS})\left(1 - \frac{\varepsilon_3}{\varepsilon_1}\frac{k_{zn1}}{k_{zn3}}\right) - \frac{k_{zn1}}{\omega\varepsilon_1}}{\frac{k_{zn1}}{\omega\varepsilon_1} + (-R + jX_{AS})\left(1 + \frac{\varepsilon_3}{\varepsilon_1}\frac{k_{zn1}}{k_{zn3}}\right)} e^{-2jk_{zn1}d} \right.$$

$$\left. + \frac{\varepsilon_2}{\varepsilon_1}\frac{k_{zn1}}{k_{zn2}}\left(1 + \frac{(-R + jX_{AS})\left(1 - \frac{\varepsilon_3}{\varepsilon_1}\frac{k_{zn1}}{k_{zn3}}\right) - \frac{k_{zn1}}{\omega\varepsilon_1}}{\frac{k_{zn1}}{\omega\varepsilon_1} + (-R + jX_{AS})\left(1 + \frac{\varepsilon_3}{\varepsilon_1}\frac{k_{zn1}}{k_{zn3}}\right)} e^{-2jk_{zn1}d}\right) \right\} e^{-j\left(k + \frac{2\pi (n-1)}{a}\right)x} +$$

$$\frac{jX_s M}{2} \sum_{n=-\infty}^{+\infty} A_n \left\{ 1 - \frac{(-R + jX_{AS})\left(1 - \frac{\varepsilon_3}{\varepsilon_1}\frac{k_{zn1}}{k_{zn3}}\right) - \frac{k_{zn1}}{\omega\varepsilon_1}}{\frac{k_{zn1}}{\omega\varepsilon_1} + (-R + jX_{AS})\left(1 + \frac{\varepsilon_3}{\varepsilon_1}\frac{k_{zn1}}{k_{zn3}}\right)} e^{-2jk_{zn1}d} \right.$$

$$\left. + \frac{\varepsilon_2}{\varepsilon_1}\frac{k_{zn1}}{k_{zn2}}\left(1 + \frac{(-R + jX_{AS})\left(1 - \frac{\varepsilon_3}{\varepsilon_1}\frac{k_{zn1}}{k_{zn3}}\right) - \frac{k_{zn1}}{\omega\varepsilon_1}}{\frac{k_{zn1}}{\omega\varepsilon_1} + (-R + jX_{AS})\left(1 + \frac{\varepsilon_3}{\varepsilon_1}\frac{k_{zn1}}{k_{zn3}}\right)} e^{-2jk_{zn1}d}\right) \right\} e^{-j\left(k + \frac{2\pi (n+1)}{a}\right)x}$$

$$= \frac{-1}{\omega\varepsilon_2} \sum_{n=-\infty}^{+\infty} k_{zn2} \frac{\varepsilon_2}{\varepsilon_1}\frac{k_{zn1}}{k_{zn2}} A_n \left( 1 + \frac{(-R + jX_{AS})\left(1 - \frac{\varepsilon_3}{\varepsilon_1}\frac{k_{zn1}}{k_{zn3}}\right) - \frac{k_{zn1}}{\omega\varepsilon_1}}{\frac{k_{zn1}}{\omega\varepsilon_1} + (-R + jX_{AS})\left(1 + \frac{\varepsilon_3}{\varepsilon_1}\frac{k_{zn1}}{k_{zn3}}\right)} e^{-2jk_{zn1}d} \right) e^{-j\left(k + \frac{2\pi n}{a}\right)x}$$

Now, let's multiply both sides with $e^{+j\left(k+\frac{2\pi m}{a}\right)x}$, and then integrate both sides over one period $\int_0^a \cdots dx$:

$$\int_0^a jX_s \sum_{n=-\infty}^{+\infty} A_n \left\{ 1 - \frac{(-R+jX_{AS})\left(1-\frac{\varepsilon_3}{\varepsilon_1}\frac{k_{zn1}}{k_{zn3}}\right) - \frac{k_{zn1}}{\omega\varepsilon_1}}{\frac{k_{zn1}}{\omega\varepsilon_1} + (-R+jX_{AS})\left(1+\frac{\varepsilon_3}{\varepsilon_1}\frac{k_{zn1}}{k_{zn3}}\right)} e^{-2jk_{zn1}d} \right.$$

$$\left. + \frac{\varepsilon_2}{\varepsilon_1}\frac{k_{zn1}}{k_{zn2}}\left(1 + \frac{(-R+jX_{AS})\left(1-\frac{\varepsilon_3}{\varepsilon_1}\frac{k_{zn1}}{k_{zn3}}\right) - \frac{k_{zn1}}{\omega\varepsilon_1}}{\frac{k_{zn1}}{\omega\varepsilon_1} + (-R+jX_{AS})\left(1+\frac{\varepsilon_3}{\varepsilon_1}\frac{k_{zn1}}{k_{zn3}}\right)} e^{-2jk_{zn1}d}\right)\right\} e^{-j\left(k+\frac{2\pi n}{a}\right)x} e^{+j\left(k+\frac{2\pi m}{a}\right)x} dx +$$

$$\int_0^a \frac{jX_s M}{2} \sum_{n=-\infty}^{+\infty} A_n \left\{ 1 - \frac{(-R+jX_{AS})\left(1-\frac{\varepsilon_3}{\varepsilon_1}\frac{k_{zn1}}{k_{zn3}}\right) - \frac{k_{zn1}}{\omega\varepsilon_1}}{\frac{k_{zn1}}{\omega\varepsilon_1} + (-R+jX_{AS})\left(1+\frac{\varepsilon_3}{\varepsilon_1}\frac{k_{zn1}}{k_{zn3}}\right)} e^{-2jk_{zn1}d} \right.$$

$$\left. + \frac{\varepsilon_2}{\varepsilon_1}\frac{k_{zn1}}{k_{zn2}}\left(1 + \frac{(-R+jX_{AS})\left(1-\frac{\varepsilon_3}{\varepsilon_1}\frac{k_{zn1}}{k_{zn3}}\right) - \frac{k_{zn1}}{\omega\varepsilon_1}}{\frac{k_{zn1}}{\omega\varepsilon_1} + (-R+jX_{AS})\left(1+\frac{\varepsilon_3}{\varepsilon_1}\frac{k_{zn1}}{k_{zn3}}\right)} e^{-2jk_{zn1}d}\right)\right\} e^{-j\left(k+\frac{2\pi(n-1)}{a}\right)x} e^{+j\left(k+\frac{2\pi m}{a}\right)x} dx +$$

$$\int_0^a \frac{jX_s M}{2} \sum_{n=-\infty}^{+\infty} A_n \left\{ 1 - \frac{(-R+jX_{AS})\left(1-\frac{\varepsilon_3}{\varepsilon_1}\frac{k_{zn1}}{k_{zn3}}\right) - \frac{k_{zn1}}{\omega\varepsilon_1}}{\frac{k_{zn1}}{\omega\varepsilon_1} + (-R+jX_{AS})\left(1+\frac{\varepsilon_3}{\varepsilon_1}\frac{k_{zn1}}{k_{zn3}}\right)} e^{-2jk_{zn1}d} \right.$$

$$\left. + \frac{\varepsilon_2}{\varepsilon_1}\frac{k_{zn1}}{k_{zn2}}\left(1 + \frac{(-R+jX_{AS})\left(1-\frac{\varepsilon_3}{\varepsilon_1}\frac{k_{zn1}}{k_{zn3}}\right) - \frac{k_{zn1}}{\omega\varepsilon_1}}{\frac{k_{zn1}}{\omega\varepsilon_1} + (-R+jX_{AS})\left(1+\frac{\varepsilon_3}{\varepsilon_1}\frac{k_{zn1}}{k_{zn3}}\right)} e^{-2jk_{zn1}d}\right)\right\} e^{-j\left(k+\frac{2\pi(n+1)}{a}\right)x} e^{+j\left(k+\frac{2\pi m}{a}\right)x} dx$$

$$= \int_0^a \frac{-1}{\omega\varepsilon_2} \sum_{n=-\infty}^{+\infty} k_{zn2} \frac{\varepsilon_2}{\varepsilon_1}\frac{k_{zn1}}{k_{zn2}} A_n \left( 1 + \frac{(-R+jX_{AS})\left(1-\frac{\varepsilon_3}{\varepsilon_1}\frac{k_{zn1}}{k_{zn3}}\right) - \frac{k_{zn1}}{\omega\varepsilon_1}}{\frac{k_{zn1}}{\omega\varepsilon_1} + (-R+jX_{AS})\left(1+\frac{\varepsilon_3}{\varepsilon_1}\frac{k_{zn1}}{k_{zn3}}\right)} e^{-2jk_{zn1}d} \right) e^{-j\left(k+\frac{2\pi n}{a}\right)x} e^{+j\left(k+\frac{2\pi m}{a}\right)x} dx$$

Next, since the summation is over $n$ and the integral is over $x$, the integral operator only applies to the exponential terms, therefore we have:

$$jX_s \sum_{n=-\infty}^{+\infty} A_n \left\{ 1 - \frac{(-R+jX_{AS})\left(1-\frac{\varepsilon_3}{\varepsilon_1}\frac{k_{zn1}}{k_{zn3}}\right) - \frac{k_{zn1}}{\omega\varepsilon_1}}{\frac{k_{zn1}}{\omega\varepsilon_1} + (-R+jX_{AS})\left(1+\frac{\varepsilon_3}{\varepsilon_1}\frac{k_{zn1}}{k_{zn3}}\right)} e^{-2jk_{zn1}d} \right.$$

$$\left. + \frac{\varepsilon_2}{\varepsilon_1}\frac{k_{zn1}}{k_{zn2}}\left(1 + \frac{(-R+jX_{AS})\left(1-\frac{\varepsilon_3}{\varepsilon_1}\frac{k_{zn1}}{k_{zn3}}\right) - \frac{k_{zn1}}{\omega\varepsilon_1}}{\frac{k_{zn1}}{\omega\varepsilon_1} + (-R+jX_{AS})\left(1+\frac{\varepsilon_3}{\varepsilon_1}\frac{k_{zn1}}{k_{zn3}}\right)} e^{-2jk_{zn1}d}\right)\right\} \int_0^a e^{-j\left(\frac{2\pi(n-m)}{a}\right)x} dx +$$

$$\frac{jX_s M}{2} \sum_{n=-\infty}^{+\infty} A_n \left\{ 1 - \frac{(-R+jX_{AS})\left(1-\frac{\varepsilon_3}{\varepsilon_1}\frac{k_{zn1}}{k_{zn3}}\right) - \frac{k_{zn1}}{\omega\varepsilon_1}}{\frac{k_{zn1}}{\omega\varepsilon_1} + (-R+jX_{AS})\left(1+\frac{\varepsilon_3}{\varepsilon_1}\frac{k_{zn1}}{k_{zn3}}\right)} e^{-2jk_{zn1}d} \right.$$

$$\left. + \frac{\varepsilon_2}{\varepsilon_1}\frac{k_{zn1}}{k_{zn2}}\left(1 + \frac{(-R+jX_{AS})\left(1-\frac{\varepsilon_3}{\varepsilon_1}\frac{k_{zn1}}{k_{zn3}}\right) - \frac{k_{zn1}}{\omega\varepsilon_1}}{\frac{k_{zn1}}{\omega\varepsilon_1} + (-R+jX_{AS})\left(1+\frac{\varepsilon_3}{\varepsilon_1}\frac{k_{zn1}}{k_{zn3}}\right)} e^{-2jk_{zn1}d}\right)\right\} \int_0^a e^{-j\left(+\frac{2\pi(n-1-m)}{a}\right)x} dx +$$

$$\frac{jX_sM}{2}\sum_{n=-\infty}^{+\infty} A_n \left\{ 1 - \frac{(-R+jX_{AS})\left(1-\frac{\varepsilon_3}{\varepsilon_1}\frac{k_{zn1}}{k_{zn3}}\right) - \frac{k_{zn1}}{\omega\varepsilon_1}}{\frac{k_{zn1}}{\omega\varepsilon_1} + (-R+jX_{AS})\left(1+\frac{\varepsilon_3}{\varepsilon_1}\frac{k_{zn1}}{k_{zn3}}\right)} e^{-2jk_{zn1}d} \right.$$

$$\left. + \frac{\varepsilon_2}{\varepsilon_1}\frac{k_{zn1}}{k_{zn2}}\left(1 + \frac{(-R+jX_{AS})\left(1-\frac{\varepsilon_3}{\varepsilon_1}\frac{k_{zn1}}{k_{zn3}}\right) - \frac{k_{zn1}}{\omega\varepsilon_1}}{\frac{k_{zn1}}{\omega\varepsilon_1} + (-R+jX_{AS})\left(1+\frac{\varepsilon_3}{\varepsilon_1}\frac{k_{zn1}}{k_{zn3}}\right)} e^{-2jk_{zn1}d}\right) \right\} \int_0^a e^{-j\left(+\frac{2\pi(n+1-m)}{a}\right)x} dx$$

$$= \frac{-1}{\omega\varepsilon_2}\sum_{n=-\infty}^{+\infty} k_{zn2}\frac{\varepsilon_2}{\varepsilon_1}\frac{k_{zn1}}{k_{zn2}} A_n \left(1 + \frac{(-R+jX_{AS})\left(1-\frac{\varepsilon_3}{\varepsilon_1}\frac{k_{zn1}}{k_{zn3}}\right) - \frac{k_{zn1}}{\omega\varepsilon_1}}{\frac{k_{zn1}}{\omega\varepsilon_1} + (-R+jX_{AS})\left(1+\frac{\varepsilon_3}{\varepsilon_1}\frac{k_{zn1}}{k_{zn3}}\right)} e^{-2jk_{zn1}d}\right) \int_0^a e^{-j\left(\frac{2\pi(n-m)}{a}\right)x} dx$$

And by using the orthogonality principle, the above equation can be written as:

$$jX_s A_m \left\{ 1 - \frac{(-R+jX_{AS})\left(1-\frac{\varepsilon_3}{\varepsilon_1}\frac{k_{zm1}}{k_{zm3}}\right) - \frac{k_{zm1}}{\omega\varepsilon_1}}{\frac{k_{zm1}}{\omega\varepsilon_1} + (-R+jX_{AS})\left(1+\frac{\varepsilon_3}{\varepsilon_1}\frac{k_{zm1}}{k_{zm3}}\right)} e^{-2jk_{zm1}d} \right.$$

$$\left. + \frac{\varepsilon_2}{\varepsilon_1}\frac{k_{zm1}}{k_{zm2}}\left(1 + \frac{(-R+jX_{AS})\left(1-\frac{\varepsilon_3}{\varepsilon_1}\frac{k_{zm1}}{k_{zm3}}\right) - \frac{k_{zm1}}{\omega\varepsilon_1}}{\frac{k_{zm1}}{\omega\varepsilon_1} + (-R+jX_{AS})\left(1+\frac{\varepsilon_3}{\varepsilon_1}\frac{k_{zm1}}{k_{zm3}}\right)} e^{-2jk_{zm1}d}\right) \right\} +$$

$$\frac{jX_sM}{2} A_{(m+1)} \left\{ 1 - \frac{(-R+jX_{AS})\left(1-\frac{\varepsilon_3}{\varepsilon_1}\frac{k_{z(m+1)1}}{k_{z(m+1)3}}\right) - \frac{k_{z(m+1)1}}{\omega\varepsilon_1}}{\frac{k_{z(m+1)1}}{\omega\varepsilon_1} + (-R+jX_{AS})\left(1+\frac{\varepsilon_3}{\varepsilon_1}\frac{k_{z(m+1)1}}{k_{z(m+1)3}}\right)} e^{-2jk_{z(m+1)1}d} \right.$$

$$\left. + \frac{\varepsilon_2}{\varepsilon_1}\frac{k_{z(m+1)1}}{k_{z(m+1)2}}\left\{ 1 + \frac{(-R+jX_{AS})\left(1-\frac{\varepsilon_3}{\varepsilon_1}\frac{k_{z(m+1)1}}{k_{z(m+1)3}}\right) - \frac{k_{z(m+1)1}}{\omega\varepsilon_1}}{\frac{k_{z(m+1)1}}{\omega\varepsilon_1} + (-R+jX_{AS})\left(1+\frac{\varepsilon_3}{\varepsilon_1}\frac{k_{z(m+1)1}}{k_{z(m+1)3}}\right)} e^{-2jk_{z(m+1)1}d} \right\} \right\} +$$

$$\frac{jX_sM}{2} A_{(m-1)} \left\{ 1 - \frac{(-R+jX_{AS})\left(1-\frac{\varepsilon_3}{\varepsilon_1}\frac{k_{z(m-1)1}}{k_{z(m-1)3}}\right) - \frac{k_{z(m-1)1}}{\omega\varepsilon_1}}{\frac{k_{z(m-1)1}}{\omega\varepsilon_1} + (-R+jX_{AS})\left(1+\frac{\varepsilon_3}{\varepsilon_1}\frac{k_{z(m-1)1}}{k_{z(m-1)3}}\right)} e^{-2jk_{z(m-1)1}d} \right.$$

$$\left. + \frac{\varepsilon_2}{\varepsilon_1}\frac{k_{z(m-1)1}}{k_{z(m-1)2}}\left\{ 1 + \frac{(-R+jX_{AS})\left(1-\frac{\varepsilon_3}{\varepsilon_1}\frac{k_{z(m-1)1}}{k_{z(m-1)3}}\right) - \frac{k_{z(m-1)1}}{\omega\varepsilon_1}}{\frac{k_{z(m-1)1}}{\omega\varepsilon_1} + (-R+jX_{AS})\left(1+\frac{\varepsilon_3}{\varepsilon_1}\frac{k_{z(m-1)1}}{k_{z(m-1)3}}\right)} e^{-2jk_{z(m-1)1}d} \right\} \right\}$$

$$= \frac{-1}{\omega\varepsilon_2} k_{zm2} \frac{\varepsilon_2}{\varepsilon_1}\frac{k_{zm1}}{k_{zm2}} A_m \left(1 + \frac{(-R+jX_{AS})\left(1-\frac{\varepsilon_3}{\varepsilon_1}\frac{k_{zm1}}{k_{zm3}}\right) - \frac{k_{zm1}}{\omega\varepsilon_1}}{\frac{k_{zm1}}{\omega\varepsilon_1} + (-R+jX_{AS})\left(1+\frac{\varepsilon_3}{\varepsilon_1}\frac{k_{zm1}}{k_{zm3}}\right)} e^{-2jk_{zm1}d}\right)$$

Further simplifying leads to:

$$\frac{jX_sM}{2}\left\{1-\frac{(-R+jX_{AS})\left(1-\frac{\varepsilon_3}{\varepsilon_1}\frac{k_{z(m-1)1}}{k_{z(m-1)3}}\right)-\frac{k_{z(m-1)1}}{\omega\varepsilon_1}}{\frac{k_{z(m-1)1}}{\omega\varepsilon_1}+(-R+jX_{AS})\left(1+\frac{\varepsilon_3}{\varepsilon_1}\frac{k_{z(m-1)1}}{k_{z(m-1)3}}\right)}e^{-2jk_{z(m-1)1}d}\right.$$

$$\left.+\frac{\varepsilon_2}{\varepsilon_1}\frac{k_{z(m-1)1}}{k_{z(m-1)2}}\left\{1+\frac{(-R+jX_{AS})\left(1-\frac{\varepsilon_3}{\varepsilon_1}\frac{k_{z(m-1)1}}{k_{z(m-1)3}}\right)-\frac{k_{z(m-1)1}}{\omega\varepsilon_1}}{\frac{k_{z(m-1)1}}{\omega\varepsilon_1}+(-R+jX_{AS})\left(1+\frac{\varepsilon_3}{\varepsilon_1}\frac{k_{z(m-1)1}}{k_{z(m-1)3}}\right)}e^{-2jk_{z(m-1)1}d}\right\}\right\}A_{(m-1)}+$$

$$\left\{+jX_s\left\{1-\frac{(-R+jX_{AS})\left(1-\frac{\varepsilon_3}{\varepsilon_1}\frac{k_{zm1}}{k_{zm3}}\right)-\frac{k_{zm1}}{\omega\varepsilon_1}}{\frac{k_{zm1}}{\omega\varepsilon_1}+(-R+jX_{AS})\left(1+\frac{\varepsilon_3}{\varepsilon_1}\frac{k_{zm1}}{k_{zm3}}\right)}e^{-2jk_{zm1}d}\right.\right.$$

$$\left.+\frac{\varepsilon_2}{\varepsilon_1}\frac{k_{zm1}}{k_{zm2}}\left(1+\frac{(-R+jX_{AS})\left(1-\frac{\varepsilon_3}{\varepsilon_1}\frac{k_{zm1}}{k_{zm3}}\right)-\frac{k_{zm1}}{\omega\varepsilon_1}}{\frac{k_{zm1}}{\omega\varepsilon_1}+(-R+jX_{AS})\left(1+\frac{\varepsilon_3}{\varepsilon_1}\frac{k_{zm1}}{k_{zm3}}\right)}e^{-2jk_{zm1}d}\right)\right\}$$

$$\left.+\frac{1}{\omega\varepsilon_2}k_{zm2}\frac{\varepsilon_2}{\varepsilon_1}\frac{k_{zm1}}{k_{zm2}}\left(1+\frac{(-R+jX_{AS})\left(1-\frac{\varepsilon_3}{\varepsilon_1}\frac{k_{zm1}}{k_{zm3}}\right)-\frac{k_{zm1}}{\omega\varepsilon_1}}{\frac{k_{zm1}}{\omega\varepsilon_1}+(-R+jX_{AS})\left(1+\frac{\varepsilon_3}{\varepsilon_1}\frac{k_{zm1}}{k_{zm3}}\right)}e^{-2jk_{zm1}d}\right)\right\}A_m+$$

$$\frac{jX_sM}{2}\left\{1-\frac{(-R+jX_{AS})\left(1-\frac{\varepsilon_3}{\varepsilon_1}\frac{k_{z(m+1)1}}{k_{z(m+1)3}}\right)-\frac{k_{z(m+1)1}}{\omega\varepsilon_1}}{\frac{k_{z(m+1)1}}{\omega\varepsilon_1}+(-R+jX_{AS})\left(1+\frac{\varepsilon_3}{\varepsilon_1}\frac{k_{z(m+1)1}}{k_{z(m+1)3}}\right)}e^{-2jk_{z(m+1)1}d}\right.$$

$$\left.+\frac{\varepsilon_2}{\varepsilon_1}\frac{k_{z(m+1)1}}{k_{z(m+1)2}}\left\{1+\frac{(-R+jX_{AS})\left(1-\frac{\varepsilon_3}{\varepsilon_1}\frac{k_{z(m+1)1}}{k_{z(m+1)3}}\right)-\frac{k_{z(m+1)1}}{\omega\varepsilon_1}}{\frac{k_{z(m+1)1}}{\omega\varepsilon_1}+(-R+jX_{AS})\left(1+\frac{\varepsilon_3}{\varepsilon_1}\frac{k_{z(m+1)1}}{k_{z(m+1)3}}\right)}e^{-2jk_{z(m+1)1}d}\right\}\right\}A_{(m+1)}=0$$

Therefore, by defining $P_m$ and $Q_m$ as follows, we can write the dispersion expression in more elegant and compact form as:

$$P_m=\frac{jX_sM}{2}\left\{1-\frac{(-R+jX_{AS})\left(1-\frac{\varepsilon_3}{\varepsilon_1}\frac{k_{zm1}}{k_{zm3}}\right)-\frac{k_{zm1}}{\omega\varepsilon_1}}{\frac{k_{zm1}}{\omega\varepsilon_1}+(-R+jX_{AS})\left(1+\frac{\varepsilon_3}{\varepsilon_1}\frac{k_{zm1}}{k_{zm3}}\right)}e^{-2jk_{zm1}d}\right.$$

$$\left.+\frac{\varepsilon_2}{\varepsilon_1}\frac{k_{zm1}}{k_{zm2}}\left\{1+\frac{(-R+jX_{AS})\left(1-\frac{\varepsilon_3}{\varepsilon_1}\frac{k_{zm1}}{k_{zm3}}\right)-\frac{k_{zm1}}{\omega\varepsilon_1}}{\frac{k_{zm1}}{\omega\varepsilon_1}+(-R+jX_{AS})\left(1+\frac{\varepsilon_3}{\varepsilon_1}\frac{k_{zm1}}{k_{zm3}}\right)}e^{-2jk_{zm1}d}\right\}\right\}$$

$$Q_m = +jX_s \left\{ 1 - \frac{(-R+jX_{AS})\left(1 - \frac{\varepsilon_3}{\varepsilon_1}\frac{k_{zm1}}{k_{zm3}}\right) - \frac{k_{zm1}}{\omega\varepsilon_1}}{\frac{k_{zm1}}{\omega\varepsilon_1} + (-R+jX_{AS})\left(1 + \frac{\varepsilon_3}{\varepsilon_1}\frac{k_{zm1}}{k_{zm3}}\right)} e^{-2jk_{zm1}d} \right.$$

$$\left. + \frac{\varepsilon_2}{\varepsilon_1}\frac{k_{zm1}}{k_{zm2}}\left( 1 + \frac{(-R+jX_{AS})\left(1 - \frac{\varepsilon_3}{\varepsilon_1}\frac{k_{zm1}}{k_{zm3}}\right) - \frac{k_{zm1}}{\omega\varepsilon_1}}{\frac{k_{zm1}}{\omega\varepsilon_1} + (-R+jX_{AS})\left(1 + \frac{\varepsilon_3}{\varepsilon_1}\frac{k_{zm1}}{k_{zm3}}\right)} e^{-2jk_{zm1}d} \right) \right\}$$

$$+ \frac{1}{\omega\varepsilon_2} k_{zm2} \frac{\varepsilon_2}{\varepsilon_1}\frac{k_{zm1}}{k_{zm2}}\left( 1 + \frac{(-R+jX_{AS})\left(1 - \frac{\varepsilon_3}{\varepsilon_1}\frac{k_{zm1}}{k_{zm3}}\right) - \frac{k_{zm1}}{\omega\varepsilon_1}}{\frac{k_{zm1}}{\omega\varepsilon_1} + (-R+jX_{AS})\left(1 + \frac{\varepsilon_3}{\varepsilon_1}\frac{k_{zm1}}{k_{zm3}}\right)} e^{-2jk_{zm1}d} \right)$$

$$P_{m-1}A_{m-1} + Q_m A_m + P_{m+1}A_{m+1} = 0$$

The above expression can be represented as matrix format as follows:

$$\begin{bmatrix} \ddots & \ddots & \vdots & \vdots & \vdots & \vdots & \cdots \\ \ddots & Q_{m-2} & P_{m-1} & 0 & 0 & 0 & \cdots \\ \ddots & P_{m-2} & Q_{m-1} & P_m & 0 & 0 & \cdots \\ \cdots & 0 & P_{m-1} & Q_m & P_{m+1} & 0 & \cdots \\ \cdots & 0 & 0 & P_m & Q_{m+1} & P_{m+2} & \ddots \\ \cdots & 0 & 0 & 0 & P_{m+1} & Q_{m+2} & \ddots \\ \cdots & \vdots & \vdots & \vdots & \vdots & \ddots & \ddots \end{bmatrix} \begin{bmatrix} \vdots \\ A_{m-2} \\ A_{m-1} \\ A_m \\ A_{m-1} \\ A_{m-2} \\ \vdots \end{bmatrix} = 0$$

Nontrivial solutions are obviously obtained by setting the determinant of the matrix above to zero.

### III. Derivation of the dispersion equation for an arbitrary periodic sheet impedance over a grounded dielectric substrate

The problem under investigation has been depicted in the figure below:

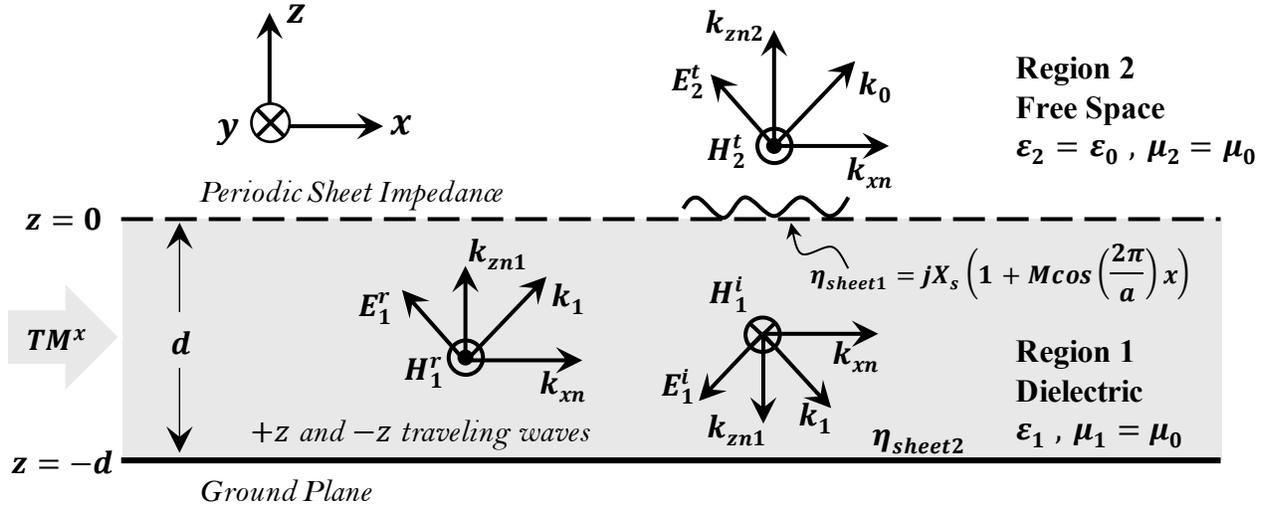

Figure 2: Schematic representation of an arbitrary periodic sheet impedance over a grounded dielectric substrate and the supported Floquet waves.

Similar to section I, we can expand the electric and magnetic fields of both regions as Floquet-waves considering infinite number of harmonics as follows [1,2]:

$$\vec{H}_1^i = \sum_{n=-\infty}^{+\infty} A_n e^{-jk_{xn}x} e^{+jk_{zn1}z} \hat{y}$$

$$\vec{H}_1^r = \sum_{n=-\infty}^{+\infty} -B_n e^{-jk_{xn}x} e^{-jk_{zn1}z} \hat{y}$$

$$\vec{H}_1 = \vec{H}_1^i + \vec{H}_1^r = \sum_{n=-\infty}^{+\infty} A_n e^{-jk_{xn}x} e^{+jk_{zn1}z} \hat{y} + \sum_{n=-\infty}^{+\infty} -B_n e^{-jk_{xn}x} e^{-jk_{zn1}z} \hat{y}$$

$$H_{1y} = \sum_{n=-\infty}^{+\infty} A_n e^{-jk_{xn}x} e^{+jk_{zn1}z} + \sum_{n=-\infty}^{+\infty} -B_n e^{-jk_{xn}x} e^{-jk_{zn1}z} = \sum_{n=-\infty}^{+\infty} \left(A_n e^{+jk_{zn1}z} - B_n e^{-jk_{zn1}z}\right) e^{-jk_{xn}x}$$

$$\nabla \times \vec{H}_1 = j\omega\varepsilon_1 \vec{E}_1 \quad \rightarrow \quad \vec{E}_1 = \frac{\nabla \times \vec{H}_1}{j\omega\varepsilon_1} = \frac{1}{j\omega\varepsilon_1} \begin{vmatrix} \hat{x} & \hat{y} & \hat{z} \\ \frac{\partial}{\partial x} & \frac{\partial}{\partial y} & \frac{\partial}{\partial z} \\ H_{1x} & H_{1y} & H_{1z} \end{vmatrix} = \frac{1}{j\omega\varepsilon_1} \begin{vmatrix} \hat{x} & \hat{y} & \hat{z} \\ \frac{\partial}{\partial x} & \frac{\partial}{\partial y} & \frac{\partial}{\partial z} \\ 0 & H_{1y} & 0 \end{vmatrix}$$

$$= \frac{1}{j\omega\varepsilon_1} \left( -\hat{x} \frac{\partial H_{1y}}{\partial z} + \hat{z} \frac{\partial H_{1y}}{\partial x} \right)$$

$$E_{1x} = \frac{-1}{j\omega\varepsilon_1} \frac{\partial H_{1y}}{\partial z} = \frac{-1}{j\omega\varepsilon_1} \frac{\partial}{\partial z} \left( \sum_{n=-\infty}^{+\infty} \left(A_n e^{+jk_{zn1}z} - B_n e^{-jk_{zn1}z}\right) e^{-jk_{xn}x} \right)$$

$$= \frac{-1}{j\omega\varepsilon_1} \sum_{n=-\infty}^{+\infty} jk_{zn1} \left(A_n e^{+jk_{zn1}z} + B_n e^{-jk_{zn1}z}\right) e^{-jk_{xn}x}$$

$$E_{1z} = \frac{1}{j\omega\varepsilon_1} \frac{\partial H_{1y}}{\partial x} = \frac{1}{j\omega\varepsilon_1} \frac{\partial}{\partial x} \left( \sum_{n=-\infty}^{+\infty} \left(A_n e^{+jk_{zn1}z} - B_n e^{-jk_{zn1}z}\right) e^{-jk_{xn}x} \right)$$

$$= \frac{1}{j\omega\varepsilon_1} \sum_{n=-\infty}^{+\infty} -jk_{xn} \left(A_n e^{+jk_{zn1}z} - B_n e^{-jk_{zn1}z}\right) e^{-jk_{xn}x}$$

$$E_{1x} = \frac{-1}{j\omega\varepsilon_1} \sum_{n=-\infty}^{+\infty} jk_{zn1} \left(A_n e^{+jk_{zn1}z} + B_n e^{-jk_{zn1}z}\right) e^{-jk_{xn}x}$$

Applying the boundary condition for the bottom PEC sheet (the tangential component of the total electric field should be zero at the PEC interface), we have:

$$E_{1x}(z = -d) = \frac{-1}{j\omega\varepsilon_1} \sum_{n=-\infty}^{+\infty} jk_{zn1} \left(A_n e^{-jk_{zn1}d} + B_n e^{+jk_{zn1}d}\right) e^{-jk_{xn}x} = 0$$

$$A_n e^{-jk_{zn1}d} + B_n e^{+jk_{zn1}d} = 0 \quad \rightarrow \quad B_n = -A_n e^{-2jk_{zn1}d}$$

Substituting the derived result back to the $H_{1y}$ gives:

$$H_{1y} = \sum_{n=-\infty}^{+\infty} (A_n e^{+jk_{zn1}z} - B_n e^{-jk_{zn1}z}) e^{-jk_{xn}x} = \sum_{n=-\infty}^{+\infty} (A_n e^{+jk_{zn1}z} + A_n e^{-2jk_{zn1}d} e^{-jk_{zn1}z}) e^{-jk_{xn}x}$$

$$= \sum_{n=-\infty}^{+\infty} A_n (e^{+jk_{zn1}z} + e^{-2jk_{zn1}d} e^{-jk_{zn1}z}) e^{-jk_{xn}x}$$

$$= \sum_{n=-\infty}^{+\infty} A_n \frac{e^{+jk_{zn1}d} e^{+jk_{zn1}z} + e^{-jk_{zn1}d} e^{-jk_{zn1}z}}{e^{+jk_{zn1}d}} e^{-jk_{xn}x} = \sum_{n=-\infty}^{+\infty} \frac{2A_n}{e^{+jk_{zn1}d}} \frac{e^{+jk_{zn1}(z+d)} + e^{-jk_{zn1}(z+d)}}{2} e^{-jk_{xn}x}$$

$$= \sum_{n=-\infty}^{+\infty} \frac{2A_n}{e^{+jk_{zn1}d}} \cos(k_{zn1}(z+d)) e^{-jk_{xn}x}$$

Assuming the following change of parameter gives a final compact representation for $H_{1y}$ as below:

$$C_n = \frac{2A_n}{e^{+jk_{zn1}d}}$$

$$H_{1y} = \sum_{n=-\infty}^{+\infty} C_n \cos(k_{zn1}(z+d)) e^{-jk_{xn}x}$$

Therefore, the electric field in region 1 can be written as:

$$E_{1x} = \frac{-1}{j\omega\varepsilon_1} \frac{\partial H_{1y}}{\partial z} = \frac{-1}{j\omega\varepsilon_1} \frac{\partial}{\partial z} \left( \sum_{n=-\infty}^{+\infty} C_n \cos(k_{zn1}(z+d)) e^{-jk_{xn}x} \right) = \sum_{n=-\infty}^{+\infty} C_n \frac{-jk_{zn1}}{\omega\varepsilon_1} \sin(k_{zn1}(z+d)) e^{-jk_{xn}x}$$

$$E_{1z} = \frac{1}{j\omega\varepsilon_1} \frac{\partial H_{1y}}{\partial x} = \frac{1}{j\omega\varepsilon_1} \frac{\partial}{\partial x} \left( \sum_{n=-\infty}^{+\infty} C_n \cos(k_{zn1}(z+d)) e^{-jk_{xn}x} \right) = \sum_{n=-\infty}^{+\infty} C_n \frac{-k_{xn}}{\omega\varepsilon_1} \cos(k_{zn1}(z+d)) e^{-jk_{xn}x}$$

Similarly for region 2 we have:

$$\vec{H}_2 = \sum_{n=-\infty}^{+\infty} D_n e^{-jk_{xn}x} e^{-jk_{zn2}z} (-\hat{y})$$

$$H_{2y} = -\sum_{n=-\infty}^{+\infty} D_n e^{-jk_{xn}x} e^{-jk_{zn2}z}$$

$$\nabla \times \vec{H}_2 = j\omega\varepsilon_2 \vec{E}_2 \quad \rightarrow \quad \vec{E}_2 = \frac{\nabla \times \vec{H}_2}{j\omega\varepsilon_2} = \frac{1}{j\omega\varepsilon_2} \begin{vmatrix} \hat{x} & \hat{y} & \hat{z} \\ \frac{\partial}{\partial x} & \frac{\partial}{\partial y} & \frac{\partial}{\partial z} \\ H_{2x} & H_{2y} & H_{2z} \end{vmatrix} = \frac{1}{j\omega\varepsilon_2} \begin{vmatrix} \hat{x} & \hat{y} & \hat{z} \\ \frac{\partial}{\partial x} & \frac{\partial}{\partial y} & \frac{\partial}{\partial z} \\ 0 & H_{2y} & 0 \end{vmatrix}$$

$$= \frac{1}{j\omega\varepsilon_2} \left( -\hat{x} \frac{\partial H_{2y}}{\partial z} + \hat{z} \frac{\partial H_{2y}}{\partial x} \right)$$

$$E_{2x} = \frac{-1}{j\omega\varepsilon_2} \frac{\partial H_{2y}}{\partial z} = \frac{-1}{j\omega\varepsilon_2} \frac{\partial}{\partial z} \left( -\sum_{n=-\infty}^{+\infty} D_n e^{-jk_{xn}x} e^{-jk_{zn2}z} \right) = \sum_{n=-\infty}^{+\infty} D_n \frac{-k_{zn2}}{\omega\varepsilon_2} e^{-jk_{xn}x} e^{-jk_{zn2}z}$$

$$E_{2z} = \frac{1}{j\omega\varepsilon_2} \frac{\partial H_{2y}}{\partial x} = \frac{1}{j\omega\varepsilon_2} \frac{\partial}{\partial x}\left(-\sum_{n=-\infty}^{+\infty} D_n e^{-jk_{xn}x} e^{-jk_{zn2}z}\right) = \sum_{n=-\infty}^{+\infty} D_n \frac{k_{xn}}{\omega\varepsilon_2} e^{-jk_{xn}x} e^{-jk_{zn2}z}$$

Assuming the following change of parameter, gives:

$$F_n = -D_n$$

$$H_{2y} = \sum_{n=-\infty}^{+\infty} F_n e^{-jk_{xn}x} e^{-jk_{zn2}z}$$

$$E_{2x} = \sum_{n=-\infty}^{+\infty} F_n \frac{k_{zn2}}{\omega\varepsilon_2} e^{-jk_{xn}x} e^{-jk_{zn2}z}$$

$$E_{2z} = \sum_{n=-\infty}^{+\infty} F_n \frac{-k_{xn}}{\omega\varepsilon_2} e^{-jk_{xn}x} e^{-jk_{zn2}z}$$

Therefore, in summery the electric and magnetic fields for both regions can be written as follows:

$$H_{1y} = \sum_{n=-\infty}^{+\infty} C_n \cos(k_{zn1}(z+d)) e^{-jk_{xn}x}$$

$$E_{1x} = \sum_{n=-\infty}^{+\infty} C_n \frac{-jk_{zn1}}{\omega\varepsilon_1} \sin(k_{zn1}(z+d)) e^{-jk_{xn}x}$$

$$E_{1z} = \sum_{n=-\infty}^{+\infty} C_n \frac{-k_{xn}}{\omega\varepsilon_1} \cos(k_{zn1}(z+d)) e^{-jk_{xn}x}$$

$$H_{2y} = \sum_{n=-\infty}^{+\infty} F_n e^{-jk_{xn}x} e^{-jk_{zn2}z}$$

$$E_{2x} = \sum_{n=-\infty}^{+\infty} F_n \frac{k_{zn2}}{\omega\varepsilon_2} e^{-jk_{xn}x} e^{-jk_{zn2}z}$$

$$E_{2z} = \sum_{n=-\infty}^{+\infty} F_n \frac{-k_{xn}}{\omega\varepsilon_2} e^{-jk_{xn}x} e^{-jk_{zn2}z}$$

$$k_{xn} = k + \frac{2\pi n}{a}$$

In the following, we applied the boundary conditions for the top sheet, which is the continuity of the tangential component of the total electric field and the discontinuity of the tangential component of the total magnetic fields of both regions along the top interface which give rise to the surface current density. As a result:

$$E_{1x}(z=0) = E_{2x}(z=0)$$

$$\sum_{n=-\infty}^{+\infty} C_n \frac{-jk_{zn1}}{\omega\varepsilon_1} \sin(k_{zn1}d) e^{-j\left(k+\frac{2\pi n}{a}\right)x} = \sum_{n=-\infty}^{+\infty} F_n \frac{k_{zn2}}{\omega\varepsilon_2} e^{-j\left(k+\frac{2\pi n}{a}\right)x}$$

$$\int_0^a \sum_{n=-\infty}^{+\infty} C_n \frac{-jk_{zn1}}{\omega\varepsilon_1} \sin(k_{zn1}d) e^{-j\left(k+\frac{2\pi n}{a}\right)x} e^{+j\left(k+\frac{2\pi g}{a}\right)x} dx = \int_0^a \sum_{n=-\infty}^{+\infty} F_n \frac{k_{zn2}}{\omega\varepsilon_2} e^{-j\left(k+\frac{2\pi n}{a}\right)x} e^{+j\left(k+\frac{2\pi g}{a}\right)x} dx$$

$$\sum_{n=-\infty}^{+\infty} C_n \frac{-jk_{zn1}}{\omega\varepsilon_1} \sin(k_{zn1}d) \int_0^a e^{-j\left(k+\frac{2\pi n}{a}\right)x} e^{+j\left(k+\frac{2\pi g}{a}\right)x} dx = \sum_{n=-\infty}^{+\infty} F_n \frac{k_{zn2}}{\omega\varepsilon_2} \int_0^a e^{-j\left(k+\frac{2\pi n}{a}\right)x} e^{+j\left(k+\frac{2\pi g}{a}\right)x} dx$$

$$\sum_{n=-\infty}^{+\infty} C_n \frac{-jk_{zn1}}{\omega\varepsilon_1} \sin(k_{zn1}d) \int_0^a e^{+j\frac{2\pi(g-n)}{a}x} dx = \sum_{n=-\infty}^{+\infty} F_n \frac{k_{zn2}}{\omega\varepsilon_2} \int_0^a e^{+j\frac{2\pi(g-n)}{a}x} dx$$

$$C_n \frac{-jk_{zn1}}{\varepsilon_1} \sin(k_{zn1}d) = F_n \frac{k_{zn2}}{\varepsilon_2} \quad \rightarrow \quad C_n = F_n \frac{k_{zn2}}{\varepsilon_2} \frac{\varepsilon_1}{-jk_{zn1}\sin(k_{zn1}d)}$$

$$H_{1y}(z=0) - H_{2y}(z=0) = J_x = \frac{E_{2x}(z=0)}{\eta_{sheet}}$$

$$\eta_{sheet}(x) \sum_{n=-\infty}^{+\infty} C_n \cos(k_{zn1}d) e^{-j\left(k+\frac{2\pi n}{a}\right)x} - F_n e^{-j\left(k+\frac{2\pi n}{a}\right)x} = \sum_{n=-\infty}^{+\infty} F_n \frac{k_{zn2}}{\omega\varepsilon_2} e^{-j\left(k+\frac{2\pi n}{a}\right)x}$$

Substituting $C_n$ in terms of $F_n$ gives:

$$C_n = F_n \frac{k_{zn2}}{\varepsilon_2} \frac{\varepsilon_1}{-jk_{zn1}\sin(k_{zn1}d)}$$

$$H_{1y}(z=0) - H_{2y}(z=0) = J_x = \frac{E_{2x}(z=0)}{\eta_{sheet}}$$

$$\eta_{sheet}(x) \sum_{n=-\infty}^{+\infty} C_n \cos(k_{zn1}d) e^{-j\left(k+\frac{2\pi n}{a}\right)x} - F_n e^{-j\left(k+\frac{2\pi n}{a}\right)x} = \sum_{n=-\infty}^{+\infty} F_n \frac{k_{zn2}}{\omega\varepsilon_2} e^{-j\left(k+\frac{2\pi n}{a}\right)x}$$

$$\eta_{sheet}(x) \sum_{n=-\infty}^{+\infty} \left(F_n \frac{k_{zn2}}{\varepsilon_2} \frac{\varepsilon_1}{-jk_{zn1}\sin(k_{zn1}d)}\right) \cos(k_{zn1}d) e^{-j\left(k+\frac{2\pi n}{a}\right)x} - F_n e^{-j\left(k+\frac{2\pi n}{a}\right)x}$$

$$= \sum_{n=-\infty}^{+\infty} F_n \frac{k_{zn2}}{\omega\varepsilon_2} e^{-j\left(k+\frac{2\pi n}{a}\right)x}$$

$$\eta_{sheet}(x) \left( \sum_{n=-\infty}^{+\infty} F_n(+j) \frac{k_{zn2}}{k_{zn1}} \frac{\varepsilon_1}{\varepsilon_2} \cot(k_{zn1}d) e^{-j\left(k+\frac{2\pi n}{a}\right)x} - \sum_{n=-\infty}^{+\infty} F_n e^{-j\left(k+\frac{2\pi n}{a}\right)x} \right) = \sum_{n=-\infty}^{+\infty} F_n \frac{k_{zn2}}{\omega\varepsilon_2} e^{-j\left(k+\frac{2\pi n}{a}\right)x}$$

Since $\eta_{sheet}$ is periodic, it can be represented by a Fourier series:

$$\eta_{sheet}(x) = \sum_{m=-\infty}^{+\infty} \eta_m e^{-j\frac{2\pi m}{a}x}$$

Substituting $\eta_{sheet}(x)$ in the obtained equation gives:

$$\sum_{m=-\infty}^{+\infty} \eta_m e^{-j\frac{2\pi m}{a}x} \left( \sum_{n=-\infty}^{+\infty} F_n(+j)\frac{k_{zn2}}{k_{zn1}}\frac{\varepsilon_1}{\varepsilon_2}\cot(k_{zn1}d)e^{-j\left(k+\frac{2\pi n}{a}\right)x} - \sum_{n=-\infty}^{+\infty} F_n e^{-j\left(k+\frac{2\pi n}{a}\right)x} \right)$$
$$= \sum_{n=-\infty}^{+\infty} F_n \frac{k_{zn2}}{\omega\varepsilon_2} e^{-j\left(k+\frac{2\pi n}{a}\right)x}$$

Let's rearrange the above equation as:

$$\sum_{m=-\infty}^{+\infty} \eta_m e^{-j\frac{2\pi m}{a}x} \left( \sum_{n=-\infty}^{+\infty} F_n(+j)\frac{k_{zn2}}{k_{zn1}}\frac{\varepsilon_1}{\varepsilon_2}\cot(k_{zn1}d)e^{-j\left(k+\frac{2\pi n}{a}\right)x} - \sum_{n=-\infty}^{+\infty} F_n e^{-j\left(k+\frac{2\pi n}{a}\right)x} \right)$$
$$= \sum_{n=-\infty}^{+\infty} F_n \frac{k_{zn2}}{\omega\varepsilon_2} e^{-j\left(k+\frac{2\pi n}{a}\right)x}$$

$$\sum_{n=-\infty}^{+\infty}\sum_{m=-\infty}^{+\infty} \eta_m e^{-j\frac{2\pi m}{a}x} \left( F_n(+j)\frac{k_{zn2}}{k_{zn1}}\frac{\varepsilon_1}{\varepsilon_2}\cot(k_{zn1}d)e^{-j\left(k+\frac{2\pi n}{a}\right)x} - F_n e^{-j\left(k+\frac{2\pi n}{a}\right)x} \right)$$
$$= \sum_{n=-\infty}^{+\infty} F_n \frac{k_{zn2}}{\omega\varepsilon_2} e^{-j\left(k+\frac{2\pi n}{a}\right)x}$$

Multiplying both sides with the orthogonal basis function $e^{j\left(k+\frac{2\pi p}{a}\right)x}$ and integrating both sides over one period gives:

$$\int_0^a \sum_{n=-\infty}^{+\infty}\sum_{m=-\infty}^{+\infty} \eta_m e^{-j\frac{2\pi m}{a}x} \left( F_n(+j)\frac{k_{zn2}}{k_{zn1}}\frac{\varepsilon_1}{\varepsilon_2}\cot(k_{zn1}d)e^{-j\left(k+\frac{2\pi n}{a}\right)x} - F_n e^{-j\left(k+\frac{2\pi n}{a}\right)x} \right) e^{j\left(k+\frac{2\pi p}{a}\right)x} dx$$
$$= \int_0^a \sum_{n=-\infty}^{+\infty} F_n \frac{k_{zn2}}{\omega\varepsilon_2} e^{-j\left(k+\frac{2\pi n}{a}\right)x} e^{j\left(k+\frac{2\pi p}{a}\right)x} dx$$

$$\int_0^a \sum_{n=-\infty}^{+\infty}\sum_{m=-\infty}^{+\infty} \eta_m \left( F_n(+j)\frac{k_{zn2}}{k_{zn1}}\frac{\varepsilon_1}{\varepsilon_2}\cot(k_{zn1}d)e^{-j\frac{2\pi(n+m-p)}{a}x} - F_n e^{-j\frac{2\pi(n+m-p)}{a}x} \right) dx$$
$$= \int_0^a \sum_{n=-\infty}^{+\infty} F_n \frac{k_{zn2}}{\omega\varepsilon_2} e^{-j\frac{2\pi(n-p)}{a}x} dx$$

$$\sum_{n=-\infty}^{+\infty}\sum_{m=-\infty}^{+\infty} \eta_m \left( F_n(+j)\frac{k_{zn2}}{k_{zn1}}\frac{\varepsilon_1}{\varepsilon_2}\cot(k_{zn1}d) \int_0^a e^{-j\frac{2\pi(n+m-p)}{a}x} dx - F_n \int_0^a e^{-j\frac{2\pi(n+m-p)}{a}x} dx \right)$$
$$= \sum_{n=-\infty}^{+\infty} F_n \frac{k_{zn2}}{\omega\varepsilon_2} \int_0^a e^{-j\frac{2\pi(n-p)}{a}x} dx$$

Orthogonality yields:

$$\sum_{n=-\infty}^{+\infty} F_n \eta_{p-n} \left( +j\frac{k_{zn2}}{k_{zn1}}\frac{\varepsilon_1}{\varepsilon_2}\cot(k_{zn1}d) - 1 \right) = \frac{1}{\omega\varepsilon_2} F_p k_{zp2}$$

We write the above equation in a matrix form as follows:

$$\sum_{n=-\infty}^{+\infty} \frac{1}{k_{zp2}} \left( +j \frac{k_{zn2}}{k_{zn1}} \frac{\varepsilon_1}{\varepsilon_2} \cot(k_{zn1}d) - 1 \right) \eta_{p-n} F_n = \frac{1}{\omega \varepsilon_2} F_p \quad \rightarrow$$

$$\bar{\bar{Q}} \bar{F}_n = \frac{1}{\omega \varepsilon_2} \bar{F}_p \quad \& \quad Q_{p,n} = \frac{1}{k_{zp2}} \left( +j \frac{k_{zn2}}{k_{zn1}} \frac{\varepsilon_1}{\varepsilon_2} \cot(k_{zn1}d) - 1 \right) \eta_{p-n}$$

By truncating indices $n$ and $p$ symmetrically about zero and to identical ranges (for example, $n = -2$ to 2 and $p = -2$ to 2) the above equation can be rewritten as:

$$\bar{\bar{Q}} \bar{F}_n = \frac{1}{\omega \varepsilon_2} \bar{F}_n \quad \rightarrow \quad \bar{\bar{Q}} \bar{F}_n - \frac{1}{\omega \varepsilon_2} \bar{F}_n = 0 \quad \rightarrow \quad \left( \bar{\bar{Q}} - \frac{1}{\omega \varepsilon_2} \bar{\bar{I}} \right) \bar{F}_n = 0$$

$$\rightarrow \quad \bar{\bar{\tilde{Q}}} \bar{F}_n = 0 \quad ; \quad \bar{\bar{\tilde{Q}}} = \bar{\bar{Q}} - \frac{1}{\omega \varepsilon_2} \bar{\bar{I}}$$

$$\bar{\bar{\tilde{Q}}} = \begin{bmatrix} Q_{-2,-2} - \frac{1}{\omega \varepsilon_2} & Q_{-2,-1} & Q_{-2,0} & Q_{-2,+1} & Q_{-2,+2} \\ Q_{-1,-2} & Q_{-1,-1} - \frac{1}{\omega \varepsilon_2} & Q_{-1,0} & Q_{-1,+1} & Q_{-1,+2} \\ Q_{0,-2} & Q_{0,-1} & Q_{0,0} - \frac{1}{\omega \varepsilon_2} & Q_{0,+1} & Q_{0,+2} \\ Q_{+1,-2} & Q_{+1,-1} & Q_{+1,0} & Q_{+1,+1} - \frac{1}{\omega \varepsilon_2} & Q_{+1,+2} \\ Q_{+2,-2} & Q_{+2,-1} & Q_{+2,0} & Q_{+2,+1} & Q_{+2,+2} - \frac{1}{\omega \varepsilon_2} \end{bmatrix}$$

The determinant of $\bar{\bar{\tilde{Q}}}$ must be zero for non-trivial solutions:

$$\det\left(\bar{\bar{\tilde{Q}}}\right) = 0$$

The value of $k$ that satisfies this zero determinant condition will be the solution of interest.

## IV. Derivation of the dispersion equation for a sinusoidally-modulated sheet impedance over a grounded dielectric substrate

The formulation discussed in the preceding section is applicable to any periodic variation of $\eta_{sheet}$. In the specific scenario where $\eta_{sheet}$ exhibits a sinusoidal variation, we have:

$$\eta_{sheet} = jX_s \left( 1 + M \cos\left(\frac{2\pi}{a}\right) x \right) = jX_s + \frac{jX_s M}{2} \left( e^{+j\frac{2\pi}{a}x} + e^{-j\frac{2\pi}{a}x} \right)$$

$X_s$: average sheet reactance, $M$: Modulation factor

In general, for an arbitrary periodic sheet impedance over a grounded dielectric substrate, by using Maxwell's equations and applying boundary conditions we obtained:

$$\eta_{sheet}(x) \left( \sum_{n=-\infty}^{+\infty} F_n (+j) \frac{k_{zn2}}{k_{zn1}} \frac{\varepsilon_1}{\varepsilon_2} \cot(k_{zn1}d) e^{-j\left(k+\frac{2\pi n}{a}\right)x} - \sum_{n=-\infty}^{+\infty} F_n e^{-j\left(k+\frac{2\pi n}{a}\right)x} \right) = \sum_{n=-\infty}^{+\infty} F_n \frac{k_{zn2}}{\omega \varepsilon_2} e^{-j\left(k+\frac{2\pi n}{a}\right)x}$$

By substituting the sinusoidally varying $\eta_{sheet}$, we have:

$$\left(jX_s + \frac{jX_sM}{2}\left(e^{+j\frac{2\pi}{a}x} + e^{-j\frac{2\pi}{a}x}\right)\right)\left(\sum_{n=-\infty}^{+\infty} F_n(+j)\frac{k_{zn2}\varepsilon_1}{k_{zn1}\varepsilon_2}\cot(k_{zn1}d)e^{-j\left(k+\frac{2\pi n}{a}\right)x} - \sum_{n=-\infty}^{+\infty} F_n e^{-j\left(k+\frac{2\pi n}{a}\right)x}\right)$$
$$= \sum_{n=-\infty}^{+\infty} F_n \frac{k_{zn2}}{\omega\varepsilon_2} e^{-j\left(k+\frac{2\pi n}{a}\right)x}$$

Let's first rearrange the above expression by taking care of the multiplications of the left hand-side and write the equation as:

$$\sum_{n=-\infty}^{+\infty} jX_s F_n(+j)\frac{k_{zn2}\varepsilon_1}{k_{zn1}\varepsilon_2}\cot(k_{zn1}d)e^{-j\left(k+\frac{2\pi n}{a}\right)x} - \sum_{n=-\infty}^{+\infty} jX_s F_n e^{-j\left(k+\frac{2\pi n}{a}\right)x}$$
$$+ \sum_{n=-\infty}^{+\infty} \frac{jX_sM}{2} F_n(+j)\frac{k_{zn2}\varepsilon_1}{k_{zn1}\varepsilon_2}\cot(k_{zn1}d)e^{-j\left(k+\frac{2\pi n}{a}\right)x} e^{+j\frac{2\pi}{a}x}$$
$$- \sum_{n=-\infty}^{+\infty} \frac{jX_sM}{2} F_n e^{-j\left(k+\frac{2\pi n}{a}\right)x} e^{+j\frac{2\pi}{a}x}$$
$$+ \sum_{n=-\infty}^{+\infty} \frac{jX_sM}{2} F_n(+j)\frac{k_{zn2}\varepsilon_1}{k_{zn1}\varepsilon_2}\cot(k_{zn1}d)e^{-j\left(k+\frac{2\pi n}{a}\right)x} e^{-j\frac{2\pi}{a}x}$$
$$- \sum_{n=-\infty}^{+\infty} \frac{jX_sM}{2} F_n e^{-j\left(k+\frac{2\pi n}{a}\right)x} e^{-j\frac{2\pi}{a}x} = \sum_{n=-\infty}^{+\infty} F_n \frac{k_{zn2}}{\omega\varepsilon_2} e^{-j\left(k+\frac{2\pi n}{a}\right)x}$$

Similarly, multiplying both sides with the orthogonal basis function $e^{j\left(k+\frac{2\pi m}{a}\right)x}$ and integrating both sides over one period gives the following:

$$\int_0^a \Biggl\{ \sum_{n=-\infty}^{+\infty} jX_s F_n(+j)\frac{k_{zn2}\varepsilon_1}{k_{zn1}\varepsilon_2}\cot(k_{zn1}d)e^{-j\left(k+\frac{2\pi n}{a}\right)x} - \sum_{n=-\infty}^{+\infty} jX_s F_n e^{-j\left(k+\frac{2\pi n}{a}\right)x}$$
$$+ \sum_{n=-\infty}^{+\infty} \frac{jX_sM}{2} F_n(+j)\frac{k_{zn2}\varepsilon_1}{k_{zn1}\varepsilon_2}\cot(k_{zn1}d)e^{-j\left(k+\frac{2\pi n}{a}\right)x} e^{+j\frac{2\pi}{a}x}$$
$$- \sum_{n=-\infty}^{+\infty} \frac{jX_sM}{2} F_n e^{-j\left(k+\frac{2\pi n}{a}\right)x} e^{+j\frac{2\pi}{a}x}$$
$$+ \sum_{n=-\infty}^{+\infty} \frac{jX_sM}{2} F_n(+j)\frac{k_{zn2}\varepsilon_1}{k_{zn1}\varepsilon_2}\cot(k_{zn1}d)e^{-j\left(k+\frac{2\pi n}{a}\right)x} e^{-j\frac{2\pi}{a}x}$$
$$- \sum_{n=-\infty}^{+\infty} \frac{jX_sM}{2} F_n e^{-j\left(k+\frac{2\pi n}{a}\right)x} e^{-j\frac{2\pi}{a}x} \Biggr\} e^{+j\left(k+\frac{2\pi m}{a}\right)x} dx$$
$$= \int_0^a \Biggl\{ \sum_{n=-\infty}^{+\infty} F_n \frac{k_{zn2}}{\omega\varepsilon_2} e^{-j\left(k+\frac{2\pi n}{a}\right)x} \Biggr\} e^{+j\left(k+\frac{2\pi m}{a}\right)x} dx$$

Distributing the integrations to each term gives:

$$\sum_{n=-\infty}^{+\infty} \int_0^a jX_s F_n(+j)\frac{k_{zn2}}{k_{zn1}}\frac{\varepsilon_1}{\varepsilon_2}\cot(k_{zn1}d)e^{-j\left(k+\frac{2\pi n}{a}\right)x}e^{+j\left(k+\frac{2\pi m}{a}\right)x}dx$$

$$-\sum_{n=-\infty}^{+\infty}\int_0^a jX_s F_n e^{-j\left(k+\frac{2\pi n}{a}\right)x}e^{+j\left(k+\frac{2\pi m}{a}\right)x}dx$$

$$+\sum_{n=-\infty}^{+\infty}\int_0^a \frac{jX_s M}{2}F_n(+j)\frac{k_{zn2}}{k_{zn1}}\frac{\varepsilon_1}{\varepsilon_2}\cot(k_{zn1}d)e^{-j\left(k+\frac{2\pi n}{a}\right)x}e^{+j\frac{2\pi}{a}x}e^{+j\left(k+\frac{2\pi m}{a}\right)x}dx$$

$$-\sum_{n=-\infty}^{+\infty}\int_0^a \frac{jX_s M}{2}F_n e^{-j\left(k+\frac{2\pi n}{a}\right)x}e^{+j\frac{2\pi}{a}x}e^{+j\left(k+\frac{2\pi m}{a}\right)x}dx$$

$$+\sum_{n=-\infty}^{+\infty}\int_0^a \frac{jX_s M}{2}F_n(+j)\frac{k_{zn2}}{k_{zn1}}\frac{\varepsilon_1}{\varepsilon_2}\cot(k_{zn1}d)e^{-j\left(k+\frac{2\pi n}{a}\right)x}e^{-j\frac{2\pi}{a}x}e^{+j\left(k+\frac{2\pi m}{a}\right)x}dx$$

$$-\sum_{n=-\infty}^{+\infty}\int_0^a \frac{jX_s M}{2}F_n e^{-j\left(k+\frac{2\pi n}{a}\right)x}e^{-j\frac{2\pi}{a}x}e^{+j\left(k+\frac{2\pi m}{a}\right)x}dx$$

$$=\sum_{n=-\infty}^{+\infty}\int_0^a F_n\frac{k_{zn2}}{\omega\varepsilon_2}e^{-j\left(k+\frac{2\pi n}{a}\right)x}e^{+j\left(k+\frac{2\pi m}{a}\right)x}dx$$

Now, let's encapsulate the exponential terms and write the above equation in a compact form as:

$$\sum_{n=-\infty}^{+\infty}\int_0^a jX_s F_n(+j)\frac{k_{zn2}}{k_{zn1}}\frac{\varepsilon_1}{\varepsilon_2}\cot(k_{zn1}d)e^{+j\frac{2\pi(m-n)}{a}x}dx - \sum_{n=-\infty}^{+\infty}\int_0^a jX_s F_n e^{+j\frac{2\pi(m-n)}{a}x}dx$$

$$+\sum_{n=-\infty}^{+\infty}\int_0^a \frac{jX_s M}{2}F_n(+j)\frac{k_{zn2}}{k_{zn1}}\frac{\varepsilon_1}{\varepsilon_2}\cot(k_{zn1}d)e^{+j\frac{2\pi(m-n+1)}{a}x}dx$$

$$-\sum_{n=-\infty}^{+\infty}\int_0^a \frac{jX_s M}{2}F_n e^{+j\frac{2\pi(m-n+1)}{a}x}dx$$

$$+\sum_{n=-\infty}^{+\infty}\int_0^a \frac{jX_s M}{2}F_n(+j)\frac{k_{zn2}}{k_{zn1}}\frac{\varepsilon_1}{\varepsilon_2}\cot(k_{zn1}d)e^{+j\frac{2\pi(m-n-1)}{a}x}dx$$

$$-\sum_{n=-\infty}^{+\infty}\int_0^a \frac{jX_s M}{2}F_n e^{+j\frac{2\pi(m-n-1)}{a}x}dx = \sum_{n=-\infty}^{+\infty}\int_0^a F_n\frac{k_{zn2}}{\omega\varepsilon_2}e^{+j\frac{2\pi(m-n)}{a}x}dx$$

Therefore, the orthogonality gives:

$$jX_s F_m(+j)\frac{k_{zm2}}{k_{zm1}}\frac{\varepsilon_1}{\varepsilon_2}\cot(k_{zm1}d) - jX_s F_m + \frac{jX_s M}{2}F_{m+1}(+j)\frac{k_{z(m+1)2}}{k_{z(m+1)1}}\frac{\varepsilon_1}{\varepsilon_2}\cot(k_{z(m+1)1}d) - \frac{jX_s M}{2}F_{m+1}$$

$$+\frac{jX_s M}{2}F_{m-1}(+j)\frac{k_{z(m-1)2}}{k_{z(m-1)1}}\frac{\varepsilon_1}{\varepsilon_2}\cot(k_{z(m-1)1}d) - \frac{jX_s M}{2}F_{m-1} = F_m\frac{k_{zm2}}{\omega\varepsilon_2}$$

Let's rearrange the above equation and write it as:

$$\frac{jX_sM}{2}\left((+j)\frac{k_{z(m-1)2}}{k_{z(m-1)1}}\frac{\varepsilon_1}{\varepsilon_2}\cot(k_{z(m-1)1}d)-1\right)F_{m-1}+\frac{jX_sM}{2}\left((+j)\frac{k_{z(m+1)2}}{k_{z(m+1)1}}\frac{\varepsilon_1}{\varepsilon_2}\cot(k_{z(m+1)1}d)-1\right)F_{m+1}$$
$$+\left(jX_s\left((+j)\frac{k_{zm2}}{k_{zm1}}\frac{\varepsilon_1}{\varepsilon_2}\cot(k_{zm1}d)-1\right)-\frac{k_{zm2}}{\omega\varepsilon_2}\right)F_m=0$$

The above equation can be written in a more elegant format by defining the $P_m$ and $Q_m$ as follows:

$$P_m=\frac{jX_sM}{2}\left((+j)\frac{k_{zm2}}{k_{zm1}}\frac{\varepsilon_1}{\varepsilon_2}\cot(k_{zm1}d)-1\right) \quad \& \quad Q_m=\left(jX_s\left((+j)\frac{k_{zm2}}{k_{zm1}}\frac{\varepsilon_1}{\varepsilon_2}\cot(k_{zm1}d)-1\right)-\frac{k_{zm2}}{\omega\varepsilon_2}\right)$$

Therefore, the dispersion equation is:

$$P_{m-1}F_{m-1}+Q_mF_m+P_{m+1}F_{m+1}=0$$

The equation can be written by matrix notations as follows:

$$\begin{bmatrix} \ddots & \ddots & \vdots & \vdots & \vdots & \cdots \\ \ddots & Q_{m-2} & P_{m-1} & 0 & 0 & 0 & \cdots \\ \ddots & P_{m-2} & Q_{m-1} & P_m & 0 & 0 & \cdots \\ \cdots & 0 & P_{m-1} & Q_m & P_{m+1} & 0 & \cdots \\ \cdots & 0 & 0 & P_m & Q_{m+1} & P_{m+2} & \ddots \\ \cdots & 0 & 0 & 0 & P_{m+1} & Q_{m+2} & \ddots \\ \cdots & \vdots & \vdots & \vdots & \vdots & \ddots & \ddots \end{bmatrix}\begin{bmatrix} \vdots \\ F_{m-2} \\ F_{m-1} \\ F_m \\ F_{m-1} \\ F_{m-2} \\ \vdots \end{bmatrix}=0$$

Nontrivial solutions are obviously obtained by setting the determinant of the matrix above to zero.

## V. Derivation of the dispersion equation for penetrable and impenetrable SMRS

In this section, we will use the transmission line model of the structure and transverse resonance condition to develop the dispersion equations. For simplicity, we will start with the most straightforward problem which is the impenetrable (opaque) case and then we will develop penetrable ones based on the obtained results of this section. The problem under investigation has been represented in the following:

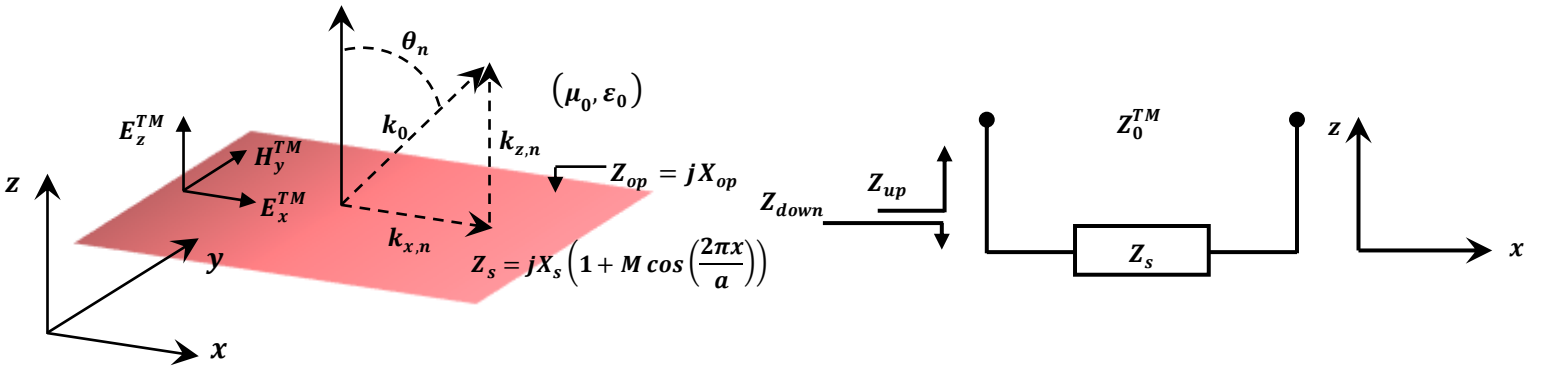

Figure 3: An impenetrable (opaque) sinusoidally modulated sheet impedance supporting $TM^x$ leaky surface waves and its equivalent transmission line model.

The modulation of the sheet $Z_s$ leads to an equivalent modal voltage $V_n$ as follows [1]:

$$V_n = \int_0^a \vec{E}_t(x) \cdot \vec{e}_{x,n}^*(x) dx$$

Based on boundary condition for the penetrable and impenetrable impedance sheets we have the following:

$$\vec{E}_t(x) = Z_s[\hat{z} \times (\vec{H}_t|0^+ - \vec{H}_t|0^-)] = Z_s \vec{J} \quad (penetrable)$$

$$\vec{E}_t(x) = Z_s[\hat{z} \times \vec{H}_t|0^+] = Z_s \vec{J} \quad (impenetrable)$$

Frist we will derive the dispersion relation for impenetrable case and then it can be extended to a general penetrable case. Substituting $\vec{E}_t(x)$ for the impenetrable sheet, gives:

$$V_n = \int_0^a \vec{E}_t(x) \cdot \vec{e}_{x,n}^*(x) dx = \int_0^a Z_s[\hat{z} \times \vec{H}_t|0^+] \cdot \vec{e}_{x,n}^*(x) dx$$

We have the following definitions and relationships for $Z_s$, $\vec{H}_t|0^+$ and $\vec{e}_{x,n}(x)$:

$$Z_s = jX_s\left(1 + M\cos\left(\frac{2\pi x}{a}\right)\right) = jX_s\left(1 + \frac{M}{2}\left(e^{j\phi} + e^{-j\phi}\right)\right), \quad \phi = \frac{2\pi x}{a}$$

$$\vec{H}_t|0^+ = H_y(x)\hat{y} \Rightarrow \hat{z} \times \vec{H}_t|0^+ = \hat{z} \times H_y(x)\hat{y} = -H_y(x)\hat{x}$$

$$\vec{e}_{x,n}(x) = \frac{1}{\sqrt{2\pi}} e^{-jk_{x,n}x}\hat{x} \Rightarrow \vec{e}_{x,n}^*(x) = \frac{1}{\sqrt{2\pi}} e^{+jk_{x,n}x}\hat{x}$$

Let's substitute them back to the expression we derived for $V_n$ in the following:

$$V_n = \int_0^a Z_s[\hat{z} \times \vec{H}_t|0^+] \cdot \vec{e}_{x,n}^*(x) dx = \int_0^a jX_s\left(1 + \frac{M}{2}\left(e^{j\phi} + e^{-j\phi}\right)\right)[-H_y(x)\hat{x}] \cdot \frac{1}{\sqrt{2\pi}} e^{+jk_{x,n}x}\hat{x} dx$$

$$= jX_s \int_0^a \left(1 + \frac{M}{2}\left(e^{j\phi} + e^{-j\phi}\right)\right) H_y(x) \underbrace{\left(\frac{-1}{\sqrt{2\pi}} e^{+jk_{x,n}x}\right)}_{h_{y,n}^*} dx \Rightarrow$$

$$V_n = jX_s \int_0^a \left(1 + \frac{M}{2}\left(e^{j\phi} + e^{-j\phi}\right)\right) H_y(x) h_{y,n}^* dx = jX_s \int_0^a \left(1 + \frac{M}{2}\left(e^{j\phi} + e^{-j\phi}\right)\right) H_y(x) h_{n0} e^{+jk_{x,n}x} dx$$

$$= jX_s \int_0^a H_y(x) h_{n0} \left(e^{+jk_{x,n}x} + \frac{M}{2}\left(e^{+jk_{x,n}x}e^{j\phi} + e^{+jk_{x,n}x}e^{-j\phi}\right)\right) dx$$

$$= jX_s \int_0^a H_y(x) h_{n0} \left(e^{+jk_{x,n}x} + \frac{M}{2}\left(e^{+j(k_{x,n}x+\phi)} + e^{+j(k_{x,n}x-\phi)}\right)\right) dx$$

We can work on the $k_{x,n}x + \phi$ and $k_{x,n}x - \phi$ terms and write them as follows:

$$k_{x,n}x + \phi = k_{x,n}x + \frac{2\pi x}{a} = \left(k_{x,0} + \frac{2\pi n}{a}\right)x + \frac{2\pi x}{a} = k_{x,0}x + \frac{2\pi xn}{a} + \frac{2\pi x}{a} = k_{x,0}x + \frac{2\pi x(n+1)}{a}$$

$$= \underbrace{\left(k_{x,0} + \frac{2\pi(n+1)}{a}\right)}_{k_{x,n+1}}x = k_{x,n+1}x$$

$$k_{x,n}x - \phi = k_{x,n}x - \frac{2\pi x}{a} = \left(k_{x,0} + \frac{2\pi n}{a}\right)x - \frac{2\pi x}{a} = k_{x,0}x + \frac{2\pi xn}{a} - \frac{2\pi x}{a} = k_{x,0}x + \frac{2\pi x(n-1)}{a}$$

$$= \underbrace{\left(k_{x,0} + \frac{2\pi(n-1)}{a}\right)}_{k_{x,n-1}}x = k_{x,n-1}x$$

Substituting them back to the expression we developed for $V_n$ gives:

$$V_n = jX_s \int_0^a H_y(x) h_{n0} \left( e^{+jk_{x,n}x} + \frac{M}{2}\left(e^{+jk_{x,n+1}x} + e^{+jk_{x,n-1}x}\right) \right) dx$$

$$= jX_s \int_0^a H_y(x) h_{n0} e^{+jk_{x,n}x} dx + jX_s \frac{M}{2} \int_0^a H_y(x) h_{n0} e^{+jk_{x,n+1}x} dx + jX_s \frac{M}{2} \int_0^a H_y(x) h_{n0} e^{+jk_{x,n-1}x} dx$$

Equivalently, the modulation of the sheet $Z_s$ leads to an equivalent modal current $I_n$ as follows:

$$I_n = \int_0^a \vec{H}_t(x) \cdot \vec{h}_{y,n}^*(x) dx = \int_0^a H_y(x) h_{n0} e^{+jk_{x,n}x} dx$$

Based on the above definition, $V_n$ can be written in terms of $I_n$ as:

$$V_n = jX_s \underbrace{\int_0^a H_y(x) h_{n0} e^{+jk_{x,n}x} dx}_{I_n} + jX_s \frac{M}{2} \underbrace{\int_0^a H_y(x) h_{n0} e^{+jk_{x,n+1}x} dx}_{I_{n+1}} + jX_s \frac{M}{2} \underbrace{\int_0^a H_y(x) h_{n0} e^{+jk_{x,n-1}x} dx}_{I_{n-1}}$$

$$= jX_s I_n + jX_s \frac{M}{2} I_{n+1} + jX_s \frac{M}{2} I_{n-1}$$

$$\Rightarrow V_n = jX_s I_n + jX_s \frac{M}{2} (I_{n+1} + I_{n-1})$$

Here, we define $I_n$ as the $n$−th mode of the sheet current $I_s$. For each mode (harmonic) $Z_s$ relates $V_n$ and $I_n$ as $V_n = Z_s I_n$. Note that $Z_s$ doesn't depend on mode. So, by substituting $V_n = Z_s I_n$ into the above derived relation, we can write:

$$V_n = jX_s I_n + jX_s \frac{M}{2}(I_{n+1} + I_{n-1})$$

$$\Rightarrow Z_s I_n = jX_s I_n + jX_s \frac{M}{2}(I_{n+1} + I_{n-1})$$

$$\Rightarrow jX_s \frac{M}{2} I_{n-1} + (jX_s - Z_s)I_n + jX_s \frac{M}{2} I_{n+1} = 0$$

In this case (impenetrable or opaque), $X_s = X_{op}$. Therefore:

$$jX_{op}\frac{M}{2}I_{n-1} + (jX_{op} - Z_s)I_n + jX_{op}\frac{M}{2}I_{n+1} = 0 \Rightarrow$$

$$jX_{op}\frac{M}{2}I_{n-1} + \left(jX_{op} - \frac{1}{Y_s}\right)I_n + jX_{op}\frac{M}{2}I_{n+1} = 0$$

$Y_s$ can be obtained using the transverse resonance condition as follows:

$$TRC \Rightarrow Y_{up} + Y_{down} = 0 \Rightarrow \begin{cases} Y_{up} = Y_0^{TM} \\ Y_{down} = Y_s \end{cases} \Rightarrow Y_0^{TM} + Y_s = 0 \Rightarrow Y_s = -Y_0^{TM}$$

Substitution gives:

$$jX_{op}\frac{M}{2}I_{n-1} + \left(jX_{op} + \frac{1}{Y_0^{TM}}\right)I_n + jX_{op}\frac{M}{2}I_{n+1} = 0 \Rightarrow$$

$$\underbrace{jX_{op}\frac{M}{2}}_{A=const.}I_{n-1} + \underbrace{(jX_{op} + Z_0^{TM})}_{B_n=f(k_{x,n})}I_n + \underbrace{jX_{op}\frac{M}{2}}_{A=const.}I_{n+1} = 0$$

The above dispersion relation, can be written as a matrix notation as follows:

$$\begin{bmatrix} \ddots & A & 0 & \cdots & 0 \\ A & B_{n-1} & A & 0 & 0 \\ 0 & A & B_n & A & 0 \\ \vdots & 0 & A & B_{n+1} & A \\ 0 & 0 & 0 & A & \ddots \end{bmatrix} \begin{bmatrix} \vdots \\ I_{n-1} \\ I_n \\ I_{n+1} \\ \vdots \end{bmatrix} = \begin{bmatrix} \vdots \\ 0 \\ 0 \\ 0 \\ \vdots \end{bmatrix}$$

Where:

$$A = jX_{op}\frac{M}{2}, \quad B_n = jX_{op} + Z_0^{TM}$$

And

$$Z_0^{TM} = \eta_0 \frac{k_{z_0,n}}{k_0}$$

Where:

$$k_{z_0,n} = \begin{cases} \sqrt{k_0^2 - k_{x,n}^2} & \text{if } k_{x,n} < k_0 \\ -j\sqrt{k_{x,n}^2 - k_0^2} & \text{if } k_{x,n} > k_0 \end{cases}$$

And

$$k_{x,n} = k_{x,0} + \frac{2n\pi}{a}$$

$$k_{x,0} = k = \beta - j\alpha$$

Nontrivial solutions are obviously obtained by setting the determinant of the matrix above to zero.

Now, let's consider a penetrable sinusoidal modulated reactance surface (SMRS) depicted as follows. The SMRS is integrated at the top of a grounded dielectric slab (SMRS-PEC).

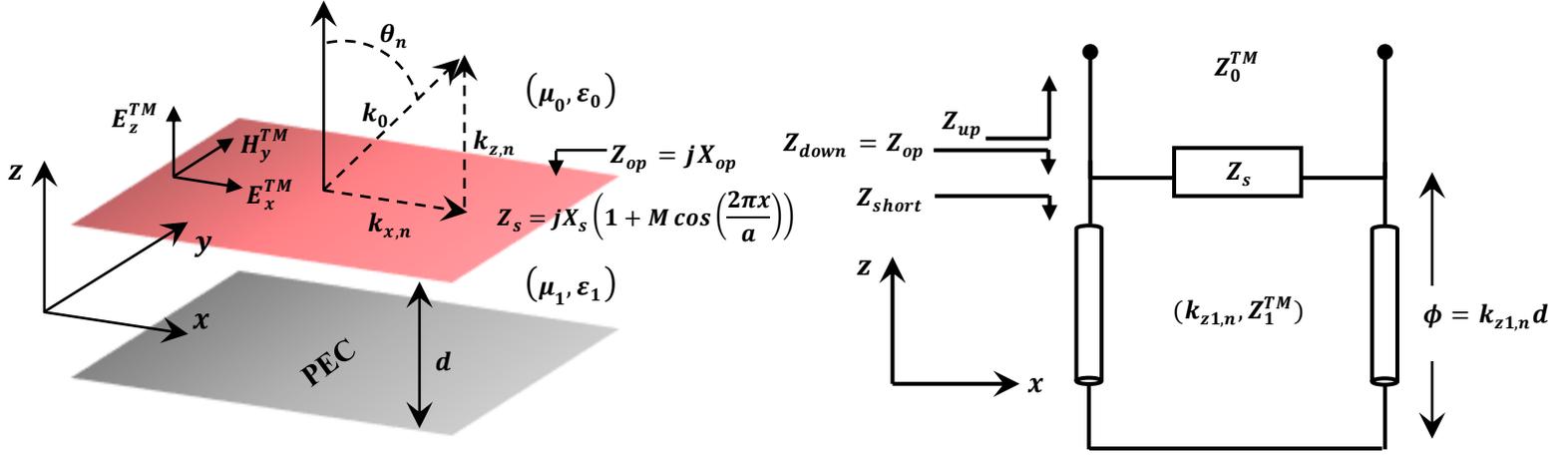

Figure 4: Geometry of a sinusoidally modulated impedance sheet integrated over a grounded dielectric slab (penetrable case) and its equivalent transmission line model.

Note that the impedance looking down to the structure $Z_{op}$ is expected to be purely reactive since a grounded dielectric has a purely reactive impedance and the modulated sheet is also assumed to be reactive. In this case, $X_s \neq X_{op}$ and $Z_s \neq -Z_0^{TM}$ (while for impenetrable case we had: $X_s = X_{op}$ and $Z_s = -Z_0^{TM}$). Therefore, by applying the transverse resonance condition (see the transverse transmission line representation of the structure), we can determine $Z_s$ and $X_s$ and formulate the dispersion equation as follows:

$$jX_s \frac{M}{2} I_{n-1} + (jX_s - Z_s)I_n + jX_s \frac{M}{2} I_{n+1} = 0 \;\Rightarrow\; jX_s \frac{M}{2} I_{n-1} + \left(jX_s - \frac{1}{Y_s}\right)I_n + jX_s \frac{M}{2} I_{n+1} = 0$$

$$TRC \;\Rightarrow\; Y_{up} + Y_{down} = 0$$

$$\Rightarrow \begin{cases} Y_{up} = Y_0^{TM} \\ Y_{down} = Y_s + Y_{short} \end{cases} \Rightarrow\; Y_0^{TM} + Y_s + Y_{short} = 0 \;\Rightarrow\; Y_s = -(Y_0^{TM} + Y_{short})$$

$$\underbrace{jX_s \frac{M}{2}}_{A=const.} I_{n-1} + \underbrace{\left(jX_s + \frac{1}{Y_0^{TM} + Y_{short}}\right)}_{B_n = f(k_{x,n})} I_n + \underbrace{jX_s \frac{M}{2}}_{A=const.} I_{n+1} = 0$$

where:

$$Z_{short} = jZ_1^{TM} \tan(k_{z_1,n} d)$$

$$X_s = X_{op}\left[1 - \frac{X_{op}}{\eta_0} \frac{\varepsilon_r}{\sqrt{\varepsilon_r - 1 - \left(\frac{X_{op}}{\eta_0}\right)^2}} \cot\left(k_0 d \sqrt{\varepsilon_r - 1 - \left(\frac{X_{op}}{\eta_0}\right)^2}\right)\right]^{-1}$$

Note that by definition the modal impedances can be written as:

$$Z_0^{TM} = \eta_0 \frac{k_{z_0,n}}{k_0} \ , \ Z_1^{TM} = \eta_1 \frac{k_{z1,n}}{k_1}$$

The above dispersion relation, can be written as a matrix notation as follows:

$$\begin{bmatrix} \ddots & A & 0 & \cdots & 0 \\ A & B_{n-1} & A & 0 & 0 \\ 0 & A & B_n & A & 0 \\ \vdots & 0 & A & B_{n+1} & A \\ 0 & 0 & 0 & A & \ddots \end{bmatrix} \begin{bmatrix} \vdots \\ I_{n-1} \\ I_n \\ I_{n+1} \\ \vdots \end{bmatrix} = \begin{bmatrix} \vdots \\ 0 \\ 0 \\ 0 \\ \vdots \end{bmatrix}$$

Where:

$$A = jX_s \frac{M}{2} \ , \ B_n = jX_s + \frac{1}{Y_0^{TM} + Y_{short}}$$

Note that:

$$k_{z_0,n} = \begin{cases} \sqrt{k_0^2 - k_{x,n}^2} & \text{if } k_{x,n} < k_0 \\ -j\sqrt{k_{x,n}^2 - k_0^2} & \text{if } k_{x,n} > k_0 \end{cases}$$

$$k_{z1,n} = \begin{cases} \sqrt{k_1^2 - k_{x,n}^2} & \text{if } k_{x,n} < k_1 \\ -j\sqrt{k_{x,n}^2 - k_1^2} & \text{if } k_{x,n} > k_1 \end{cases}$$

And

$$k_{x,n} = k_{x,0} + \frac{2n\pi}{a} \ , \ k_{x,0} = k = \beta - j\alpha$$

Nontrivial solutions are obviously obtained by setting the determinant of the matrix above to zero.

We will obtain the above expression for $X_s$ in the following. Note that for designing using this approach, we first need to stipulate $X_{op}$, $d$, $\varepsilon_r$ and then find the required $X_s$. Then we can solve for supported eigen modes.

➤ Derivation of $X_s$ (average surface reactance in penetrable case) in terms of $X_{op}$ (average surface reactance of the modeled impenetrable equivalent)

$$Y_{down} = Y_{op} = Y_s + Y_{short} \Rightarrow Y_s = Y_{op} - Y_{short} \rightarrow \frac{1}{Z_s} = Y_{op} - Y_{short} \rightarrow Z_s = \frac{1}{Y_{op} - Y_{short}}$$

$$\rightarrow jX_s = \frac{1}{Y_{op} - Y_{short}} \rightarrow X_s = \frac{-j}{Y_{op} - Y_{short}} \rightarrow X_s = \frac{-j}{\frac{1}{Z_{op}} - Y_{short}} = \frac{-j}{\frac{1}{jX_{op}} - \frac{1}{Z_{short}}} \rightarrow$$

$$X_s = \frac{-j}{\frac{1}{jX_{op}} - \frac{1}{Z_{short}}} = \frac{X_{op}}{1 - \frac{jX_{op}}{Z_{short}}} = \frac{X_{op}}{1 - \frac{jX_{op}}{jZ_1^{TM}\tan(k_{z_1,n}d)}} = X_{op}\left[1 - \frac{X_{op}}{Z_1^{TM}\tan(k_{z_1,n}d)}\right]^{-1}$$

$$= X_{op}\left[1 - X_{op}\frac{1}{\eta_1\frac{k_{z_1,n}}{k_1}}\cot(k_{z_1,n}d)\right]^{-1}$$

$$X_s = X_{op}\left[1 - X_{op}\frac{k_1}{\eta_1 k_{z_1,n}}\cot(k_{z_1,n}d)\right]^{-1} = X_{op}\left[1 - X_{op}\frac{\sqrt{\varepsilon_r}k_0}{\frac{\eta_0}{\sqrt{\varepsilon_r}}k_{z_1,n}}\cot(k_{z_1,n}d)\right]^{-1}$$

$$= X_{op}\left[1 - \frac{X_{op}}{\eta_0}\frac{\varepsilon_r k_0}{k_{z_1,n}}\cot(k_{z_1,n}d)\right]^{-1}$$

Let's derive $k_{z1,n}$ in terms of the basic parameters of the problem as follows:

$$k_{z_0,n} = \sqrt{k_0^2 - k_{x,n}^2}, \quad k_{z_1,n} = \sqrt{k_1^2 - k_{x,n}^2}$$

$$k_{z_1,n}^2 - k_{z_0,n}^2 = k_1^2 - k_{x,n}^2 - k_0^2 + k_{x,n}^2 = k_1^2 - k_0^2 \Rightarrow k_{z_1,n} = \sqrt{k_1^2 - k_0^2 + k_{z_0,n}^2}$$

$$k_{z_0,n} = \frac{k_0}{\eta_0}Z_0^{TM}$$

$$k_{z_1,n} = \sqrt{k_1^2 - k_0^2 + \frac{k_0^2}{\eta_0^2}(Z_0^{TM})^2} = \sqrt{\varepsilon_r k_0^2 - k_0^2 + \frac{k_0^2}{\eta_0^2}(Z_0^{TM})^2} = k_0\sqrt{\varepsilon_r - 1 + \frac{1}{\eta_0^2}(Z_0^{TM})^2}$$

$$Z_{up} + Z_{down} = 0 \rightarrow Z_0^{TM} + Z_{op} = 0 \rightarrow Z_0^{TM} + jX_{op} = 0 \rightarrow X_{op} = jZ_0^{TM}$$

$$k_{z_1,n} = k_0\sqrt{\varepsilon_r - 1 - \frac{1}{\eta_0^2}X_{op}^2} = k_0\sqrt{\varepsilon_r - 1 - \left(\frac{X_{op}}{\eta_0}\right)^2}$$

Substituting it back to the expression we derived for $X_s$, gives:

$$X_s = X_{op}\left[1 - \frac{X_{op}}{\eta_0}\frac{\varepsilon_r k_0}{k_{z_1,n}}\cot(k_{z_1,n}d)\right]^{-1} = X_{op}\left[1 - \frac{X_{op}}{\eta_0}\frac{\varepsilon_r k_0}{k_0\sqrt{\varepsilon_r - 1 - \left(\frac{X_{op}}{\eta_0}\right)^2}}\cot\left(k_0\sqrt{\varepsilon_r - 1 - \left(\frac{X_{op}}{\eta_0}\right)^2}d\right)\right]^{-1}$$

$$X_s = X_{op}\left[1 - \frac{X_{op}}{\eta_0}\frac{\varepsilon_r}{\sqrt{\varepsilon_r - 1 - \left(\frac{X_{op}}{\eta_0}\right)^2}}\cot\left(k_0 d\sqrt{\varepsilon_r - 1 - \left(\frac{X_{op}}{\eta_0}\right)^2}\right)\right]^{-1}$$

Next, let's extend the problem to a more general penetrable case, where the PEC sheet is replaced by a sheet with uniform active impedance capable of supporting surface TM waves. The top sheet is assumed to be sinusoidally modulated. Note that the impedance sheets are placed on the top and bottom of a dielectric slab. The following figure shows the geometry of the proposed structure (SMRS-AS) along with its transverse transmission line model. Similar to the SMRS-PEC case, we can use the following dispersion equation but with an updated $X_s$ and $Z_s$ supporting the transverse resonance condition for this structure.

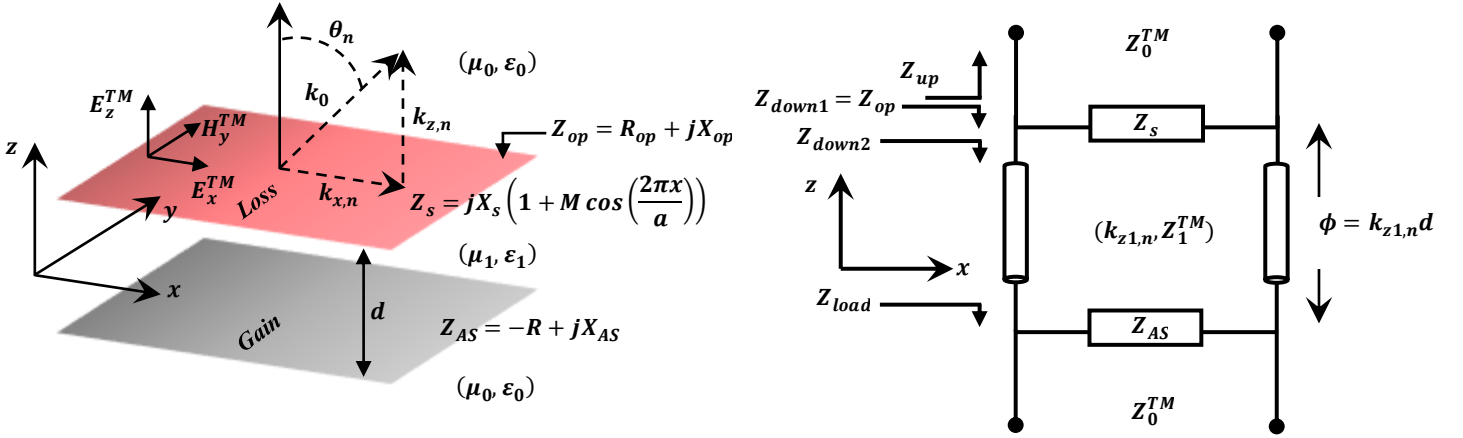

Figure 5: Geometry of the proposed LWA (a sinusoidally modulated impedance sheet coupled with a uniform active sheet which are integrated on both sides of a dielectric slab) and its equivalent transmission line model.

$$jX_s \frac{M}{2} I_{n-1} + (jX_s - Z_s)I_n + jX_s \frac{M}{2} I_{n+1} = 0 \quad \Rightarrow \quad jX_s \frac{M}{2} I_{n-1} + \left(jX_s - \frac{1}{Y_s}\right)I_n + jX_s \frac{M}{2} I_{n+1} = 0$$

$$TRC \quad \Rightarrow \quad Y_{up} + Y_{down1} = 0 \quad \Rightarrow \quad \begin{cases} Y_{up} = Y_0^{TM} \\ Y_{down1} = Y_s + Y_{down2} \end{cases} \quad \Rightarrow \quad Y_0^{TM} + Y_s + Y_{down2} = 0$$

$$\Rightarrow \quad Y_s = -(Y_0^{TM} + Y_{down2})$$

$$\underbrace{jX_s \frac{M}{2}}_{A=const.} I_{n-1} + \underbrace{\left(jX_s + \frac{1}{Y_0^{TM} + Y_{down2}}\right)}_{B_n = f(k_{x,n})} I_n + \underbrace{jX_s \frac{M}{2}}_{A=const.} I_{n+1} = 0$$

Where:

$$Z_{down2} = Z_1^{TM} \left\{ \frac{Z_0^{TM} Z_{AS} + jZ_1^{TM}(Z_0^{TM} + Z_{AS})\tan(k_{z1,n}d)}{Z_1^{TM}(Z_0^{TM} + Z_{AS}) + jZ_0^{TM} Z_{AS}\tan(k_{z1,n}d)} \right\}$$

$$X_s = \frac{-jZ_{op}}{1 - \dfrac{Z_{op}}{\dfrac{\eta_0 k_{z_1,n}}{\varepsilon_r k_0} \left\{ \dfrac{Z_0^{TM} Z_{AS} + j\dfrac{\eta_0 k_{z_1,n}}{\varepsilon_r k_0}(Z_0^{TM} + Z_{AS})\tan(k_{z1,n}d)}{\dfrac{\eta_0 k_{z_1,n}}{\varepsilon_r k_0}(Z_0^{TM} + Z_{AS}) + jZ_0^{TM} Z_{AS}\tan(k_{z1,n}d)} \right\}}}, \quad k_{z_1,n} = k_0 \sqrt{\varepsilon_r - 1 + \frac{Z_{op}^2}{\eta_0^2}}$$

Note that by definition the modal impedances can be written as:

$$Z_0^{TM} = \eta_0 \frac{k_{z_0,n}}{k_0} \;,\; Z_1^{TM} = \eta_1 \frac{k_{z_1,n}}{k_1}$$

The above dispersion relation, can be written as a matrix notation as follows:

$$\begin{bmatrix} \ddots & A & 0 & \cdots & 0 \\ A & B_{n-1} & A & 0 & 0 \\ 0 & A & B_n & A & 0 \\ \vdots & 0 & A & B_{n+1} & A \\ 0 & 0 & 0 & A & \ddots \end{bmatrix} \begin{bmatrix} \vdots \\ I_{n-1} \\ I_n \\ I_{n+1} \\ \vdots \end{bmatrix} = \begin{bmatrix} \vdots \\ 0 \\ 0 \\ 0 \\ \vdots \end{bmatrix}$$

Where:

$$A = jX_s \frac{M}{2} \;,\; B_n = jX_s + \frac{1}{Y_0^{TM} + Y_{down2}}$$

Note that:

$$k_{z_0,n} = \begin{cases} \sqrt{k_0^2 - k_{x,n}^2} & \text{if } k_{x,n} < k_0 \\ -j\sqrt{k_{x,n}^2 - k_0^2} & \text{if } k_{x,n} > k_0 \end{cases}$$

$$k_{z_1,n} = \begin{cases} \sqrt{k_1^2 - k_{x,n}^2} & \text{if } k_{x,n} < k_1 \\ -j\sqrt{k_{x,n}^2 - k_1^2} & \text{if } k_{x,n} > k_1 \end{cases}$$

And

$$k_{x,n} = k_{x,0} + \frac{2n\pi}{a} \;,\; k_{x,0} = k = \beta - j\alpha$$

Nontrivial solutions are obviously obtained by setting the determinant of the matrix above to zero.

> Derivation of $X_s$ (average surface reactance in penetrable case) in terms of $Z_{op}$ (average surface impedance of the modeled impenetrable equivalent)

$$Y_{down1} = Y_{op} = Y_s + Y_{down2} \Rightarrow Y_s = Y_{op} - Y_{down2} \rightarrow \frac{1}{Z_s} = Y_{op} - Y_{down2} \rightarrow Z_s = \frac{1}{Y_{op} - Y_{down2}}$$

$$\rightarrow jX_s = \frac{1}{Y_{op} - Y_{down2}} \rightarrow X_s = \frac{-j}{Y_{op} - Y_{down2}} \rightarrow X_s = \frac{-j}{\frac{1}{Z_{op}} - Y_{down2}} = \frac{-j}{\frac{1}{Z_{op}} - \frac{1}{Z_{down2}}}$$

$$\rightarrow X_s = \frac{-j}{\frac{1}{Z_{op}} - \frac{1}{Z_{down2}}} = \frac{-jZ_{op}}{1 - \frac{Z_{op}}{Z_{down2}}}$$

$$Y_{load} = Y_0^{TM} + Y_{AS} \rightarrow Z_{load} = \frac{1}{Y_{load}} = \frac{1}{Y_0^{TM} + Y_{AS}} = \frac{1}{\frac{1}{Z_0^{TM}} + \frac{1}{Z_{AS}}} = \frac{Z_0^{TM} Z_{AS}}{Z_0^{TM} + Z_{AS}}$$

$$Z_{down2} = Z_1^{TM}\left(\frac{Z_{load} + jZ_1^{TM}\tan(k_{z1,n}d)}{Z_1^{TM} + jZ_{load}\tan(k_{z1,n}d)}\right) = Z_1^{TM}\left\{\frac{\frac{Z_0^{TM}Z_{AS}}{Z_0^{TM}+Z_{AS}} + jZ_1^{TM}\tan(k_{z1,n}d)}{Z_1^{TM} + j\frac{Z_0^{TM}Z_{AS}}{Z_0^{TM}+Z_{AS}}\tan(k_{z1,n}d)}\right\}$$

$$= Z_1^{TM}\left\{\frac{Z_0^{TM}Z_{AS} + jZ_1^{TM}(Z_0^{TM}+Z_{AS})\tan(k_{z1,n}d)}{Z_1^{TM}(Z_0^{TM}+Z_{AS}) + jZ_0^{TM}Z_{AS}\tan(k_{z1,n}d)}\right\}$$

$$\Rightarrow X_s = \frac{-jZ_{op}}{1 - \frac{Z_{op}}{Z_1^{TM}\left\{\frac{Z_0^{TM}Z_{AS} + jZ_1^{TM}(Z_0^{TM}+Z_{AS})\tan(k_{z1,n}d)}{Z_1^{TM}(Z_0^{TM}+Z_{AS}) + jZ_0^{TM}Z_{AS}\tan(k_{z1,n}d)}\right\}}}$$

Using the definition of the $Z_1^{TM}$, we can write it in terms of $k_{z1,n}$ as follows:

$$Z_1^{TM} = \eta_1 \frac{k_{z_1,n}}{k_1}$$

$$k_1 = \sqrt{\varepsilon_r}k_0, \quad \eta_1 = \frac{\eta_0}{\sqrt{\varepsilon_r}}$$

$$\Rightarrow Z_1^{TM} = \frac{\eta_0}{\sqrt{\varepsilon_r}}\frac{k_{z_1,n}}{\sqrt{\varepsilon_r}k_0} = \frac{\eta_0 k_{z_1,n}}{\varepsilon_r k_0}$$

Substituting it back to the expression that we derived for $X_s$ gives:

$$X_s = \frac{-jZ_{op}}{1 - \frac{Z_{op}}{\frac{\eta_0 k_{z_1,n}}{\varepsilon_r k_0}\left\{\frac{Z_0^{TM}Z_{AS} + j\frac{\eta_0 k_{z_1,n}}{\varepsilon_r k_0}(Z_0^{TM}+Z_{AS})\tan(k_{z1,n}d)}{\frac{\eta_0 k_{z_1,n}}{\varepsilon_r k_0}(Z_0^{TM}+Z_{AS}) + jZ_0^{TM}Z_{AS}\tan(k_{z1,n}d)}\right\}}}$$

Let's derive $k_{z1,n}$ in terms of the basic parameters of the problem as follows:

$$k_{z_0,n} = \sqrt{k_0^2 - k_{x,n}^2}, \quad k_{z_1,n} = \sqrt{k_1^2 - k_{x,n}^2}$$

$$k_{z_1,n}^2 - k_{z_0,n}^2 = k_1^2 - k_{x,n}^2 - k_0^2 + k_{x,n}^2 = k_1^2 - k_0^2 \Rightarrow k_{z_1,n} = \sqrt{k_1^2 - k_0^2 + k_{z_0,n}^2}$$

$$k_{z_0,n} = \frac{k_0}{\eta_0}Z_0^{TM}$$

$$k_{z_1,n} = \sqrt{k_1^2 - k_0^2 + \frac{k_0^2}{\eta_0^2}(Z_0^{TM})^2} = \sqrt{\varepsilon_r k_0^2 - k_0^2 + \frac{k_0^2}{\eta_0^2}(Z_0^{TM})^2} = k_0\sqrt{\varepsilon_r - 1 + \frac{1}{\eta_0^2}(Z_0^{TM})^2}$$

$$Z_{up} + Z_{down1} = 0 \quad \rightarrow \quad Z_0^{TM} + Z_{op} = 0 \quad \rightarrow \quad Z_0^{TM} = -Z_{op}$$

$$k_{z_1,n} = k_0\sqrt{\varepsilon_r - 1 + \frac{1}{\eta_0^2}(-Z_{op})^2} = k_0\sqrt{\varepsilon_r - 1 + \frac{1}{\eta_0^2}Z_{op}^2}$$

Therefore $X_s$ can be written as:

$$X_s = \frac{-jZ_{op}}{1 - \dfrac{Z_{op}}{\dfrac{\eta_0 k_{z_1,n}}{\varepsilon_r k_0}\left\{\dfrac{Z_0^{TM}Z_{AS} + j\dfrac{\eta_0 k_{z_1,n}}{\varepsilon_r k_0}(Z_0^{TM} + Z_{AS})\tan(k_{z_1,n}d)}{\dfrac{\eta_0 k_{z_1,n}}{\varepsilon_r k_0}(Z_0^{TM} + Z_{AS}) + jZ_0^{TM}Z_{AS}\tan(k_{z_1,n}d)}\right\}}}, \quad k_{z_1,n} = k_0\sqrt{\varepsilon_r - 1 + \frac{Z_{op}^2}{\eta_0^2}}$$

Note: Using this approach, for design, we first need to stipulate $R_{op}$, $X_{op}$, $d$, $\varepsilon_r$ and then find the required $X_s$. Then we can solve for supported eigen modes.

## VI. Verification of the dispersion equation of an arbitrary periodic sheet impedance coupled with a uniform active sheet impedance

Let's first check with the case that we had an arbitrary periodic sheet impedance over a PEC grounded substrate:

Previously for a PEC grounded case, we obtained the following:

$$\left(\bar{\bar{Q}} - \frac{1}{\omega\varepsilon_2}\bar{\bar{I}}\right)\bar{F}_n = 0 \quad \rightarrow \quad \bar{\bar{\tilde{Q}}}\bar{F}_n = 0 \;;\quad \bar{\bar{\tilde{Q}}} = \bar{\bar{Q}} - \frac{1}{\omega\varepsilon_2}\bar{\bar{I}}$$

$$Q'_{p,n} = \frac{1}{k_{zp2}}\left(+j\frac{k_{zn2}}{k_{zn1}}\frac{\varepsilon_1}{\varepsilon_2}\cot(k_{zn1}d) - 1\right)\eta_{p-n}$$

The following figure, shows the parameters we assumed in the process of solving the two problems.

$$\left(\bar{\bar{\tilde{Q}}} - \frac{1}{\omega\varepsilon_1}\bar{\bar{I}}\right)\bar{A}_n = 0 \qquad\qquad \left(\bar{\bar{\tilde{Q}}} - \frac{1}{\omega\varepsilon_2}\bar{\bar{I}}\right)\bar{F}_n = 0$$

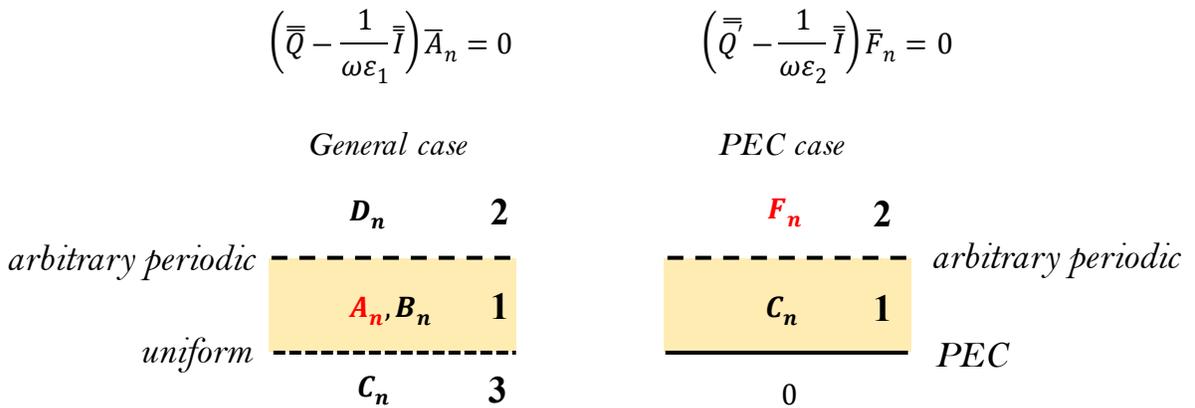

Figure 6: Schematic representation of the LWAs for the general and grounded cases, assumed coefficients for their different regions along with their dispersion equations.

From the boundary conditions for the general case, we have:

$$D_n = \frac{\varepsilon_2}{\varepsilon_1}\frac{k_{zn1}}{k_{zn2}}(A_n + B_n)$$

$$C_n = \frac{\varepsilon_3}{\varepsilon_1}\frac{k_{zn1}}{k_{zn3}}e^{+jk_{zn3}d}\left(A_n e^{-jk_{zn1}d} + B_n e^{+jk_{zn1}d}\right) = 0 \quad \Rightarrow \quad A_n e^{-jk_{zn1}d} + B_n e^{+jk_{zn1}d} = 0$$

$$\Rightarrow \quad B_n = -A_n e^{-2jk_{zn1}d}$$

$$D_n = \frac{\varepsilon_2}{\varepsilon_1}\frac{k_{zn1}}{k_{zn2}}(A_n + B_n) = \frac{\varepsilon_2}{\varepsilon_1}\frac{k_{zn1}}{k_{zn2}}\left(A_n - A_n e^{-2jk_{zn1}d}\right) = \frac{\varepsilon_2}{\varepsilon_1}\frac{k_{zn1}}{k_{zn2}}\left(1 - e^{-2jk_{zn1}d}\right)A_n$$

$$\Rightarrow \quad A_n = \frac{\varepsilon_1}{\varepsilon_2}\frac{k_{zn2}}{k_{zn1}}\left(1 - e^{-2jk_{zn1}d}\right)^{-1}D_n$$

$D_n \equiv F_n$ (according to our definition for two cases in region 2). Therefore:

$$A_n = \frac{\varepsilon_1}{\varepsilon_2}\frac{k_{zn2}}{k_{zn1}}\left(1 - e^{-2jk_{zn1}d}\right)^{-1}F_n$$

In general, we have the following expression for $A_n$ and $A_p$,

$$\sum_{n=-\infty}^{+\infty}\frac{-1}{k_{zp1}}\left\{1 + \frac{(-R+jX_{AS})\left(1-\frac{\varepsilon_3}{\varepsilon_1}\frac{k_{zp1}}{k_{zp3}}\right) - \frac{k_{zp1}}{\omega\varepsilon_1}}{\frac{k_{zp1}}{\omega\varepsilon_1} + (-R+jX_{AS})\left(1+\frac{\varepsilon_3}{\varepsilon_1}\frac{k_{zp1}}{k_{zp3}}\right)}e^{-2jk_{zp1}d}\right\}^{-1}\left\{\left(1+\frac{\varepsilon_2}{\varepsilon_1}\frac{k_{zn1}}{k_{zn2}}\right)\right.$$
$$\left.+\left(\frac{\varepsilon_2}{\varepsilon_1}\frac{k_{zn1}}{k_{zn2}}-1\right)\frac{(-R+jX_{AS})\left(1-\frac{\varepsilon_3}{\varepsilon_1}\frac{k_{zn1}}{k_{zn3}}\right) - \frac{k_{zn1}}{\omega\varepsilon_1}}{\frac{k_{zn1}}{\omega\varepsilon_1} + (-R+jX_{AS})\left(1+\frac{\varepsilon_3}{\varepsilon_1}\frac{k_{zn1}}{k_{zn3}}\right)}e^{-2jk_{zn1}d}\right\}\eta_{p-n}A_n = \frac{1}{\omega\varepsilon_1}A_p$$

Now, for comparison lets write it in terms of $F_n$:

$$\sum_{n=-\infty}^{+\infty}\frac{-1}{k_{zp1}}\left\{1 + \frac{(-R+jX_{AS})\left(1-\frac{\varepsilon_3}{\varepsilon_1}\frac{k_{zp1}}{k_{zp3}}\right) - \frac{k_{zp1}}{\omega\varepsilon_1}}{\frac{k_{zp1}}{\omega\varepsilon_1} + (-R+jX_{AS})\left(1+\frac{\varepsilon_3}{\varepsilon_1}\frac{k_{zp1}}{k_{zp3}}\right)}e^{-2jk_{zp1}d}\right\}^{-1}\left\{\left(1+\frac{\varepsilon_2}{\varepsilon_1}\frac{k_{zn1}}{k_{zn2}}\right)\right.$$
$$\left.+\left(\frac{\varepsilon_2}{\varepsilon_1}\frac{k_{zn1}}{k_{zn2}}-1\right)\frac{(-R+jX_{AS})\left(1-\frac{\varepsilon_3}{\varepsilon_1}\frac{k_{zn1}}{k_{zn3}}\right) - \frac{k_{zn1}}{\omega\varepsilon_1}}{\frac{k_{zn1}}{\omega\varepsilon_1} + (-R+jX_{AS})\left(1+\frac{\varepsilon_3}{\varepsilon_1}\frac{k_{zn1}}{k_{zn3}}\right)}e^{-2jk_{zn1}d}\right\}\eta_{p-n}\frac{\varepsilon_1}{\varepsilon_2}\frac{k_{zn2}}{k_{zn1}}\left(1-e^{-2jk_{zn1}d}\right)^{-1}F_n$$
$$= \frac{1}{\omega\varepsilon_1}\frac{\varepsilon_1}{\varepsilon_2}\frac{k_{zp2}}{k_{zp1}}\left(1-e^{-2jk_{zp1}d}\right)^{-1}F_p$$

Let's rearrange the above equation and write it as:

$$\sum_{n=-\infty}^{+\infty}\frac{-1}{k_{zp1}}\left\{1 + \frac{(-R+jX_{AS})\left(1-\frac{\varepsilon_3}{\varepsilon_1}\frac{k_{zp1}}{k_{zp3}}\right) - \frac{k_{zp1}}{\omega\varepsilon_1}}{\frac{k_{zp1}}{\omega\varepsilon_1} + (-R+jX_{AS})\left(1+\frac{\varepsilon_3}{\varepsilon_1}\frac{k_{zp1}}{k_{zp3}}\right)}e^{-2jk_{zp1}d}\right\}^{-1}\left\{\left(1+\frac{\varepsilon_2}{\varepsilon_1}\frac{k_{zn1}}{k_{zn2}}\right)\right.$$
$$\left.+\left(\frac{\varepsilon_2}{\varepsilon_1}\frac{k_{zn1}}{k_{zn2}}-1\right)\frac{(-R+jX_{AS})\left(1-\frac{\varepsilon_3}{\varepsilon_1}\frac{k_{zn1}}{k_{zn3}}\right) - \frac{k_{zn1}}{\omega\varepsilon_1}}{\frac{k_{zn1}}{\omega\varepsilon_1} + (-R+jX_{AS})\left(1+\frac{\varepsilon_3}{\varepsilon_1}\frac{k_{zn1}}{k_{zn3}}\right)}e^{-2jk_{zn1}d}\right\}\eta_{p-n}\frac{k_{zp1}}{k_{zp2}}\frac{\varepsilon_1}{\varepsilon_2}\frac{k_{zn2}}{k_{zn1}}\left(1\right.$$
$$\left.- e^{-2jk_{zn1}d}\right)^{-1}\left(1-e^{-2jk_{zp1}d}\right)F_n = \frac{1}{\omega\varepsilon_2}F_p$$

Therefore, the $Q_{p,n}$ can be written as:

$$Q_{p,n} = \frac{-1}{k_{zp1}} \left\{ 1 + \frac{(-R + jX_{AS})\left(1 - \frac{\varepsilon_3}{\varepsilon_1}\frac{k_{zp1}}{k_{zp3}}\right) - \frac{k_{zp1}}{\omega\varepsilon_1}}{\frac{k_{zp1}}{\omega\varepsilon_1} + (-R + jX_{AS})\left(1 + \frac{\varepsilon_3}{\varepsilon_1}\frac{k_{zp1}}{k_{zp3}}\right)} e^{-2jk_{zp1}d} \right\}^{-1} \left\{ \left(1 + \frac{\varepsilon_2}{\varepsilon_1}\frac{k_{zn1}}{k_{zn2}}\right) \right.$$
$$\left. + \left(\frac{\varepsilon_2}{\varepsilon_1}\frac{k_{zn1}}{k_{zn2}} - 1\right) \frac{(-R + jX_{AS})\left(1 - \frac{\varepsilon_3}{\varepsilon_1}\frac{k_{zn1}}{k_{zn3}}\right) - \frac{k_{zn1}}{\omega\varepsilon_1}}{\frac{k_{zn1}}{\omega\varepsilon_1} + (-R + jX_{AS})\left(1 + \frac{\varepsilon_3}{\varepsilon_1}\frac{k_{zn1}}{k_{zn3}}\right)} e^{-2jk_{zn1}d} \right\} \eta_{p-n} \frac{k_{zp1}}{k_{zp2}} \frac{\varepsilon_1}{\varepsilon_2} \frac{k_{zn2}}{k_{zn1}} \left(1 - e^{-2jk_{zn1}d}\right)^{-1} \left(1 - e^{-2jk_{zp1}d}\right)$$

Let's set $\eta_{sheet2} = 0 \rightarrow -R + jX_{AS} = 0$, in this case we have:

$$Q_{p,n} = \left(-\left(1 + \frac{\varepsilon_2}{\varepsilon_1}\frac{k_{zn1}}{k_{zn2}}\right) + \left(\frac{\varepsilon_2}{\varepsilon_1}\frac{k_{zn1}}{k_{zn2}} - 1\right)e^{-2jk_{zn1}d}\right) \eta_{p-n} \frac{1}{k_{zp2}}\frac{\varepsilon_1}{\varepsilon_2}\frac{k_{zn2}}{k_{zn1}}\left(1 - e^{-2jk_{zn1}d}\right)^{-1}$$

$$Q_{p,n} = \left(-1 - \frac{\varepsilon_2}{\varepsilon_1}\frac{k_{zn1}}{k_{zn2}} + \frac{\varepsilon_2}{\varepsilon_1}\frac{k_{zn1}}{k_{zn2}}e^{-2jk_{zn1}d} - e^{-2jk_{zn1}d}\right) \eta_{p-n} \frac{1}{k_{zp2}}\frac{\varepsilon_1}{\varepsilon_2}\frac{k_{zn2}}{k_{zn1}}\left(1 - e^{-2jk_{zn1}d}\right)^{-1}$$

$$= \left(\frac{\varepsilon_2}{\varepsilon_1}\frac{k_{zn1}}{k_{zn2}}\left(e^{-2jk_{zn1}d} - 1\right) - \left(e^{-2jk_{zn1}d} + 1\right)\right) \eta_{p-n} \frac{1}{k_{zp2}}\frac{\varepsilon_1}{\varepsilon_2}\frac{k_{zn2}}{k_{zn1}}\left(1 - e^{-2jk_{zn1}d}\right)^{-1}$$

$$Q_{p,n} = \left(\frac{\varepsilon_2}{\varepsilon_1}\frac{k_{zn1}}{k_{zn2}}\left(e^{-2jk_{zn1}d} - 1\right) - \left(e^{-2jk_{zn1}d} + 1\right)\right) \eta_{p-n} \frac{1}{k_{zp2}}\frac{\varepsilon_1}{\varepsilon_2}\frac{k_{zn2}}{k_{zn1}}\left(1 - e^{-2jk_{zn1}d}\right)^{-1}$$

Knowing that: $e^{-2jk_{zn1}d} + 1 = j\left(e^{-2jk_{zn1}d} - 1\right)\cot(k_{zn1}d)$, gives:

$$Q_{p,n} = \left(\frac{\varepsilon_2}{\varepsilon_1}\frac{k_{zn1}}{k_{zn2}}\left(e^{-2jk_{zn1}d} - 1\right) - j\left(e^{-2jk_{zn1}d} - 1\right)\cot(k_{zn1}d)\right) \eta_{p-n} \frac{1}{k_{zp2}}\frac{\varepsilon_1}{\varepsilon_2}\frac{k_{zn2}}{k_{zn1}}\left(1 - e^{-2jk_{zn1}d}\right)^{-1}$$

$$Q_{p,n} = \left(-\frac{\varepsilon_2}{\varepsilon_1}\frac{k_{zn1}}{k_{zn2}}\left(1 - e^{-2jk_{zn1}d}\right) + j\left(1 - e^{-2jk_{zn1}d}\right)\cot(k_{zn1}d)\right) \eta_{p-n} \frac{1}{k_{zp2}}\frac{\varepsilon_1}{\varepsilon_2}\frac{k_{zn2}}{k_{zn1}}\left(1 - e^{-2jk_{zn1}d}\right)^{-1}$$

$$Q_{p,n} = \left(1 - e^{-2jk_{zn1}d}\right)\left(-\frac{\varepsilon_2}{\varepsilon_1}\frac{k_{zn1}}{k_{zn2}} + j\cot(k_{zn1}d)\right) \eta_{p-n} \frac{1}{k_{zp2}}\frac{\varepsilon_1}{\varepsilon_2}\frac{k_{zn2}}{k_{zn1}}\left(1 - e^{-2jk_{zn1}d}\right)^{-1}$$

$$Q_{p,n} = \left(-\frac{\varepsilon_2}{\varepsilon_1}\frac{k_{zn1}}{k_{zn2}} + j\cot(k_{zn1}d)\right) \eta_{p-n} \frac{1}{k_{zp2}}\frac{\varepsilon_1}{\varepsilon_2}\frac{k_{zn2}}{k_{zn1}}$$

$$\Rightarrow Q_{p,n} = \frac{1}{k_{zp2}}\left(+j\frac{k_{zn2}}{k_{zn1}}\frac{\varepsilon_1}{\varepsilon_2}\cot(k_{zn1}d) - 1\right)\eta_{p-n} = Q'_{p,n}$$

The above expression is identical to what we obtained for an arbitrary periodic impedance sheet over a grounded dielectric layer.

This is expected because we solved the dispersion relation for a general case, and by considering the bottom layer as a perfect electric conductor (PEC), it should reduce to the same expression.

Let's check with [3]:

In this paper, we introduced two coupled sheet impedances with balanced gain and loss supporting TM surface waves. Therefore, we need to apply the following uniform impedance sheets for the general dispersion equation:

The general dispersion equation is:

$$\sum_{n=-\infty}^{+\infty} \frac{-1}{k_{zp1}} \left\{ 1 + \frac{(-R+jX_{AS})\left(1-\frac{\varepsilon_3}{\varepsilon_1}\frac{k_{zp1}}{k_{zp3}}\right) - \frac{k_{zp1}}{\omega\varepsilon_1}}{\frac{k_{zp1}}{\omega\varepsilon_1} + (-R+jX_{AS})\left(1+\frac{\varepsilon_3}{\varepsilon_1}\frac{k_{zp1}}{k_{zp3}}\right)} e^{-2jk_{zp1}d} \right\}^{-1} \left\{ \left(1 + \frac{\varepsilon_2}{\varepsilon_1}\frac{k_{zn1}}{k_{zn2}}\right) \right.$$

$$\left. + \left(\frac{\varepsilon_2}{\varepsilon_1}\frac{k_{zn1}}{k_{zn2}} - 1\right)\frac{(-R+jX_{AS})\left(1-\frac{\varepsilon_3}{\varepsilon_1}\frac{k_{zn1}}{k_{zn3}}\right) - \frac{k_{zn1}}{\omega\varepsilon_1}}{\frac{k_{zn1}}{\omega\varepsilon_1} + (-R+jX_{AS})\left(1+\frac{\varepsilon_3}{\varepsilon_1}\frac{k_{zn1}}{k_{zn3}}\right)} e^{-2jk_{zn1}d} \right\} \eta_{p-n}A_n = \frac{1}{\omega\varepsilon_1}A_p$$

The uniform impedance sheets are assumed to be:

$$\eta_{p-n} \equiv +R + jX \quad , \quad \eta_{sheet2} = -R + jX_{AS} = -R + jX$$

This will terminate the summation operator and $n$ and $p$ indices. Therefore:

$$\frac{-1}{k_{z1}}\left\{1 + \frac{(-R+jX)\left(1-\frac{\varepsilon_3}{\varepsilon_1}\frac{k_{z1}}{k_{z3}}\right) - \frac{k_{z1}}{\omega\varepsilon_1}}{\frac{k_{z1}}{\omega\varepsilon_1} + (-R+jX)\left(1+\frac{\varepsilon_3}{\varepsilon_1}\frac{k_{z1}}{k_{z3}}\right)} e^{-2jk_{z1}d}\right\}^{-1} \left\{\left(1+\frac{\varepsilon_2}{\varepsilon_1}\frac{k_{z1}}{k_{z2}}\right)\right.$$

$$\left. + \left(\frac{\varepsilon_2}{\varepsilon_1}\frac{k_{z1}}{k_{z2}} - 1\right)\frac{(-R+jX)\left(1-\frac{\varepsilon_3}{\varepsilon_1}\frac{k_{z1}}{k_{z3}}\right) - \frac{k_{z1}}{\omega\varepsilon_1}}{\frac{k_{z1}}{\omega\varepsilon_1} + (-R+jX)\left(1+\frac{\varepsilon_3}{\varepsilon_1}\frac{k_{z1}}{k_{z3}}\right)} e^{-2jk_{z1}d}\right\} (+R+jX) = \frac{1}{\omega\varepsilon_1}$$

Regions 2 and 3 are identical (free space, air), which gives:

$$\left\{1 + \frac{(-R+jX)\left(1-\frac{\varepsilon_2}{\varepsilon_1}\frac{k_{z1}}{k_{z2}}\right) - \frac{k_{z1}}{\omega\varepsilon_1}}{\frac{k_{z1}}{\omega\varepsilon_1} + (-R+jX)\left(1+\frac{\varepsilon_2}{\varepsilon_1}\frac{k_{z1}}{k_{z2}}\right)} e^{-2jk_{z1}d}\right\}^{-1} \left\{\left(1+\frac{\varepsilon_2}{\varepsilon_1}\frac{k_{z1}}{k_{z2}}\right)\right.$$

$$\left. + \left(\frac{\varepsilon_2}{\varepsilon_1}\frac{k_{z1}}{k_{z2}} - 1\right)\frac{(-R+jX)\left(1-\frac{\varepsilon_2}{\varepsilon_1}\frac{k_{z1}}{k_{z2}}\right) - \frac{k_{z1}}{\omega\varepsilon_1}}{\frac{k_{z1}}{\omega\varepsilon_1} + (-R+jX)\left(1+\frac{\varepsilon_2}{\varepsilon_1}\frac{k_{z1}}{k_{z2}}\right)} e^{-2jk_{z1}d}\right\} (+R+jX) = \frac{-k_{z1}}{\omega\varepsilon_1}$$

Or equivalently we can write the above equation as:

$$\frac{\left(1+\frac{\varepsilon_2}{\varepsilon_1}\frac{k_{z1}}{k_{z2}}\right)\left\{1 + \frac{\left(\frac{\varepsilon_2}{\varepsilon_1}\frac{k_{z1}}{k_{z2}}-1\right)}{\left(1+\frac{\varepsilon_2}{\varepsilon_1}\frac{k_{z1}}{k_{z2}}\right)} \times \frac{(-R+jX)\left(1-\frac{\varepsilon_2}{\varepsilon_1}\frac{k_{z1}}{k_{z2}}\right) - \frac{k_{z1}}{\omega\varepsilon_1}}{\frac{k_{z1}}{\omega\varepsilon_1} + (-R+jX)\left(1+\frac{\varepsilon_2}{\varepsilon_1}\frac{k_{z1}}{k_{z2}}\right)} e^{-2jk_{z1}d}\right\}(+R+jX)}{1 + \frac{(-R+jX)\left(1-\frac{\varepsilon_2}{\varepsilon_1}\frac{k_{z1}}{k_{z2}}\right) - \frac{k_{z1}}{\omega\varepsilon_1}}{\frac{k_{z1}}{\omega\varepsilon_1} + (-R+jX)\left(1+\frac{\varepsilon_2}{\varepsilon_1}\frac{k_{z1}}{k_{z2}}\right)} e^{-2jk_{z1}d}} = \frac{-k_{z1}}{\omega\varepsilon_1}$$

We previously assumed $\varepsilon_{r1} = 1$, which gives:

$$k_{z1} = \sqrt{k_1^2 - k_x^2} = \sqrt{\varepsilon_{r1} k_0^2 - k_x^2} \quad , \quad k_{z2} = \sqrt{k_2^2 - k_x^2} = \sqrt{k_0^2 - k_x^2}$$

$$\frac{\varepsilon_2 k_{z1}}{\varepsilon_1 k_{z2}} = \frac{1}{\varepsilon_{r1}} \frac{\sqrt{\varepsilon_{r1} k_0^2 - k_x^2}}{\sqrt{k_0^2 - k_x^2}} \xrightarrow{\varepsilon_{r1}=1} \frac{\varepsilon_2 k_{z1}}{\varepsilon_1 k_{z2}} = 1$$

Therefore, the dispersion equation reduces to:

$$\frac{2(+R+jX)}{1 + \frac{-\frac{k_{z1}}{\omega \varepsilon_1}}{\frac{k_{z1}}{\omega \varepsilon_1} + 2(-R+jX)} e^{-2jk_{z1}d}} = \frac{-k_{z1}}{\omega \varepsilon_1} \Rightarrow 2(+R+jX) = \frac{-k_{z1}}{\omega \varepsilon_1} \left(1 + \frac{-\frac{k_{z1}}{\omega \varepsilon_1}}{\frac{k_{z1}}{\omega \varepsilon_1} + 2(-R+jX)} e^{-2jk_{z1}d}\right)$$

Let's work on the above expression and write it in a more elegant way:

$$2\frac{\omega \varepsilon_1}{k_{z1}} (+R+jX) = -1 + \frac{\frac{k_{z1}}{\omega \varepsilon_1}}{\frac{k_{z1}}{\omega \varepsilon_1} + 2(-R+jX)} e^{-2jk_{z1}d}$$

$$\frac{\omega \varepsilon_1}{k_{z1}} \left(2 \frac{\omega \varepsilon_1}{k_{z1}} (+R+jX) + 1\right) \left(\frac{k_{z1}}{\omega \varepsilon_1} + 2(-R+jX)\right) = e^{-2jk_{z1}d}$$

$$\left(1 + 2\frac{\omega \varepsilon_1}{k_{z1}} (+R+jX)\right) \left(1 + 2\frac{\omega \varepsilon_1}{k_{z1}} (-R+jX)\right) = e^{-2jk_{z1}d} \Rightarrow$$

$$e^{-jk_{z1}d} = \sqrt{\left(1 + 2\frac{\omega \varepsilon_1}{k_{z1}} (+R+jX)\right) \left(1 + 2\frac{\omega \varepsilon_1}{k_{z1}} (-R+jX)\right)}$$

Now, by using the following definition of the tan function, we can write:

$$\tan(k_{z1}d) = \frac{e^{jk_{z1}d} - e^{-jk_{z1}d}}{j(e^{jk_{z1}d} + e^{-jk_{z1}d})}$$

$$\tan(k_{z1}d) = \frac{1 - \left(1 + 2\frac{\omega \varepsilon_1}{k_{z1}} (+R+jX)\right) \left(1 + 2\frac{\omega \varepsilon_1}{k_{z1}} (-R+jX)\right)}{j\left\{1 + \left(1 + 2\frac{\omega \varepsilon_1}{k_{z1}} (+R+jX)\right) \left(1 + 2\frac{\omega \varepsilon_1}{k_{z1}} (-R+jX)\right)\right\}}$$

Further simplifying the dispersion relation gives:

$$\tan(k_{z1}d) = j \frac{-2R^2 - 2X^2 + 2jX \frac{k_{z1}}{\omega \varepsilon_1}}{-2R^2 - 2X^2 + 2jX \frac{k_{z1}}{\omega \varepsilon_1} + \frac{k_{z1}^2}{\omega^2 \varepsilon_1^2}}$$

Let's use the same definition as [3], for the transverse propagation constant and relative permittivity and rewrite the dispersion equation:

$$k_{z1} = \sqrt{k_1^2 - k_x^2} = \sqrt{\varepsilon_{r1} k_0^2 - k_x^2} = \sqrt{k_0^2 - k_x^2} = \sqrt{k^2 - \beta^2} \quad , \quad \varepsilon_1 = \varepsilon$$

$$\Rightarrow \tan\left(d\sqrt{k^2 - \beta^2}\right) = j \frac{-2R^2 - 2X^2 + 2jX \frac{\sqrt{k^2 - \beta^2}}{\omega \varepsilon}}{-2R^2 - 2X^2 + 2jX \frac{\sqrt{k^2 - \beta^2}}{\omega \varepsilon} + \frac{k^2 - \beta^2}{\omega^2 \varepsilon^2}}$$

The above expression is identical with the reported dispersion equation in [3].

## VII. Verification of the dispersion equation of a sinusoidally periodic sheet impedance coupled with a uniform active sheet impedance

In the following we repeated the final results obtained for a sinusoidally periodic sheet impedance coupled with a uniform active sheet impedance.

$$P_m = \frac{jX_s M}{2} \left\{ 1 - \frac{(-R + jX_{AS})\left(1 - \frac{\varepsilon_3}{\varepsilon_1} \frac{k_{zm1}}{k_{zm3}}\right) - \frac{k_{zm1}}{\omega \varepsilon_1}}{\frac{k_{zm1}}{\omega \varepsilon_1} + (-R + jX_{AS})\left(1 + \frac{\varepsilon_3}{\varepsilon_1} \frac{k_{zm1}}{k_{zm3}}\right)} e^{-2jk_{zm1}d} \right.$$

$$\left. + \frac{\varepsilon_2}{\varepsilon_1} \frac{k_{zm1}}{k_{zm2}} \left\{ 1 + \frac{(-R + jX_{AS})\left(1 - \frac{\varepsilon_3}{\varepsilon_1} \frac{k_{zm1}}{k_{zm3}}\right) - \frac{k_{zm1}}{\omega \varepsilon_1}}{\frac{k_{zm1}}{\omega \varepsilon_1} + (-R + jX_{AS})\left(1 + \frac{\varepsilon_3}{\varepsilon_1} \frac{k_{zm1}}{k_{zm3}}\right)} e^{-2jk_{zm1}d} \right\} \right\}$$

$$Q_m = +jX_s \left\{ 1 - \frac{(-R + jX_{AS})\left(1 - \frac{\varepsilon_3}{\varepsilon_1} \frac{k_{zm1}}{k_{zm3}}\right) - \frac{k_{zm1}}{\omega \varepsilon_1}}{\frac{k_{zm1}}{\omega \varepsilon_1} + (-R + jX_{AS})\left(1 + \frac{\varepsilon_3}{\varepsilon_1} \frac{k_{zm1}}{k_{zm3}}\right)} e^{-2jk_{zm1}d} \right.$$

$$+ \frac{\varepsilon_2}{\varepsilon_1} \frac{k_{zm1}}{k_{zm2}} \left( 1 + \frac{(-R + jX_{AS})\left(1 - \frac{\varepsilon_3}{\varepsilon_1} \frac{k_{zm1}}{k_{zm3}}\right) - \frac{k_{zm1}}{\omega \varepsilon_1}}{\frac{k_{zm1}}{\omega \varepsilon_1} + (-R + jX_{AS})\left(1 + \frac{\varepsilon_3}{\varepsilon_1} \frac{k_{zm1}}{k_{zm3}}\right)} e^{-2jk_{zm1}d} \right) \right\}$$

$$+ \frac{1}{\omega \varepsilon_2} k_{zm2} \frac{\varepsilon_2}{\varepsilon_1} \frac{k_{zm1}}{k_{zm2}} \left( 1 + \frac{(-R + jX_{AS})\left(1 - \frac{\varepsilon_3}{\varepsilon_1} \frac{k_{zm1}}{k_{zm3}}\right) - \frac{k_{zm1}}{\omega \varepsilon_1}}{\frac{k_{zm1}}{\omega \varepsilon_1} + (-R + jX_{AS})\left(1 + \frac{\varepsilon_3}{\varepsilon_1} \frac{k_{zm1}}{k_{zm3}}\right)} e^{-2jk_{zm1}d} \right)$$

Now, let's apply the PEC condition for the bottom sheet $(-R + jX_{AS} = 0)$. The above equations reduce to:

$$P_m = \frac{jX_s M}{2} \left( 1 + e^{-2jk_{zm1}d} + \frac{\varepsilon_2}{\varepsilon_1} \frac{k_{zm1}}{k_{zm2}} \left(1 - e^{-2jk_{zm1}d}\right) \right)$$

$$Q_m = +jX_s \left( 1 + e^{-2jk_{zm1}d} + \frac{\varepsilon_2}{\varepsilon_1} \frac{k_{zm1}}{k_{zm2}} \left(1 - e^{-2jk_{zm1}d}\right) \right) + \frac{1}{\omega \varepsilon_1} k_{zm1} \left(1 - e^{-2jk_{zm1}d}\right)$$

Where the dispersion relation can be written as:

$$P_{m-1} A_{m-1} + Q_m A_m + P_{m+1} A_{m+1} = 0$$

Let's simplify the above expressions using the following mathematical relationship:

$$-j\cot(k_{zm1}d) = \frac{e^{-2jk_{zm1}d} + 1}{1 - e^{-2jk_{zm1}d}} \quad \Rightarrow \quad e^{-2jk_{zm1}d} + 1 = -j\cot(k_{zm1}d)(1 - e^{-2jk_{zm1}d})$$

Therefore, for the $P_m$, we have:

$$P_m = \frac{jX_sM}{2}\left(1 + e^{-2jk_{zm1}d} + \frac{\varepsilon_2}{\varepsilon_1}\frac{k_{zm1}}{k_{zm2}}(1 - e^{-2jk_{zm1}d})\right)$$

$$= \frac{jX_sM}{2}\left(-j\cot(k_{zm1}d)(1 - e^{-2jk_{zm1}d}) + \frac{\varepsilon_2}{\varepsilon_1}\frac{k_{zm1}}{k_{zm2}}(1 - e^{-2jk_{zm1}d})\right)$$

$$= \frac{jX_sM}{2}(1 - e^{-2jk_{zm1}d})\left(-j\cot(k_{zm1}d) + \frac{\varepsilon_2}{\varepsilon_1}\frac{k_{zm1}}{k_{zm2}}\right)$$

$$= \frac{jX_sM}{2}(e^{-2jk_{zm1}d} - 1)\left(+j\cot(k_{zm1}d) - \frac{\varepsilon_2}{\varepsilon_1}\frac{k_{zm1}}{k_{zm2}}\right)$$

$$= (e^{-2jk_{zm1}d} - 1)\frac{\varepsilon_2}{\varepsilon_1}\frac{k_{zm1}}{k_{zm2}}\frac{jX_sM}{2}\left(+j\frac{\varepsilon_1}{\varepsilon_2}\frac{k_{zm2}}{k_{zm1}}\cot(k_{zm1}d) - 1\right)$$

Using $P_m$, we can evaluate $P_{m+1}$ and $P_{m-1}$ as follows:

$$P_{m-1} = (e^{-2jk_{z(m-1)1}d} - 1)\frac{\varepsilon_2}{\varepsilon_1}\frac{k_{z(m-1)1}}{k_{z(m-1)2}}\frac{jX_sM}{2}\left(+j\frac{\varepsilon_1}{\varepsilon_2}\frac{k_{z(m-1)2}}{k_{z(m-1)1}}\cot(k_{z(m-1)1}d) - 1\right)$$

$$P_{m+1} = (e^{-2jk_{z(m+1)1}d} - 1)\frac{\varepsilon_2}{\varepsilon_1}\frac{k_{z(m+1)1}}{k_{z(m+1)2}}\frac{jX_sM}{2}\left(+j\frac{\varepsilon_1}{\varepsilon_2}\frac{k_{z(m+1)2}}{k_{z(m+1)1}}\cot(k_{z(m+1)1}d) - 1\right)$$

Similarly, we can rewrite the $Q_m$ as follows:

$$Q_m = +jX_s\left(1 + e^{-2jk_{zm1}d} + \frac{\varepsilon_2}{\varepsilon_1}\frac{k_{zm1}}{k_{zm2}}(1 - e^{-2jk_{zm1}d})\right) + \frac{1}{\omega\varepsilon_1}k_{zm1}(1 - e^{-2jk_{zm1}d})$$

$$= +jX_s\left(-j\cot(k_{zm1}d)(1 - e^{-2jk_{zm1}d}) + \frac{\varepsilon_2}{\varepsilon_1}\frac{k_{zm1}}{k_{zm2}}(1 - e^{-2jk_{zm1}d})\right) + \frac{1}{\omega\varepsilon_1}k_{zm1}(1 - e^{-2jk_{zm1}d})$$

$$= (1 - e^{-2jk_{zm1}d})\left(+jX_s\left(-j\cot(k_{zm1}d) + \frac{\varepsilon_2}{\varepsilon_1}\frac{k_{zm1}}{k_{zm2}}\right) + \frac{1}{\omega\varepsilon_1}k_{zm1}\right)$$

$$= (e^{-2jk_{zm1}d} - 1)\left(+jX_s\left(+j\cot(k_{zm1}d) - \frac{\varepsilon_2}{\varepsilon_1}\frac{k_{zm1}}{k_{zm2}}\right) - \frac{1}{\omega\varepsilon_1}k_{zm1}\right)$$

$$= (e^{-2jk_{zm1}d} - 1)\frac{\varepsilon_2}{\varepsilon_1}\frac{k_{zm1}}{k_{zm2}}\left(+jX_s\left(+j\frac{\varepsilon_1}{\varepsilon_2}\frac{k_{zm2}}{k_{zm1}}\cot(k_{zm1}d) - 1\right) - \frac{k_{zm2}}{\omega\varepsilon_2}\right)$$

Also, we have obtained the following relationships by applying the boundary conditions to the general structure (arbitrary periodic sheet impedance coupled with a uniform impedance).

Let's again apply the PEC condition for the bottom sheet ($-R + jX_{AS} = 0$).

$$B_n = \frac{(-R + jX_{AS})\left(1 - \frac{\varepsilon_3}{\varepsilon_1}\frac{k_{zn1}}{k_{zn3}}\right) - \frac{k_{zn1}}{\omega\varepsilon_1}}{\frac{k_{zn1}}{\omega\varepsilon_1} + (-R + jX_{AS})\left(1 + \frac{\varepsilon_3}{\varepsilon_1}\frac{k_{zn1}}{k_{zn3}}\right)}e^{-2jk_{zn1}d}A_n \quad \rightarrow \quad B_n = -e^{-2jk_{zn1}d}A_n$$

Substituting the result into the following relationship which is also obtained from the boundary conditions gives:

$$D_n = \frac{\varepsilon_2}{\varepsilon_1}\frac{k_{zn1}}{k_{zn2}}(A_n + B_n)$$

$$D_n = \frac{\varepsilon_2}{\varepsilon_1}\frac{k_{zn1}}{k_{zn2}}(1 - e^{-2jk_{zn1}d})A_n \quad \rightarrow \quad A_n = \frac{\varepsilon_1}{\varepsilon_2}\frac{k_{zn2}}{k_{zn1}}(e^{-2jk_{zn1}d} - 1)^{-1}D_n$$

Also, according to our definition for two cases in region 2 we can write: $D_n \equiv F_n$. Therefore:

$$A_n = \frac{\varepsilon_1}{\varepsilon_2}\frac{k_{zn2}}{k_{zn1}}(e^{-2jk_{zn1}d} - 1)^{-1}F_n$$

Using the above obtained result, we can write the dispersion equation in terms of $F_n$.

$$P_{m-1}A_{m-1} + Q_m A_m + P_{m+1}A_{m+1} = 0$$

$$P_{m-1}\frac{\varepsilon_1}{\varepsilon_2}\frac{k_{z(m-1)2}}{k_{z(m-1)1}}(e^{-2jk_{z(m-1)1}d} - 1)^{-1}F_{m-1} + Q_m\frac{\varepsilon_1}{\varepsilon_2}\frac{k_{zm2}}{k_{zm1}}(e^{-2jk_{zm1}d} - 1)^{-1}F_m$$
$$+ P_{m+1}\frac{\varepsilon_1}{\varepsilon_2}\frac{k_{z(m+1)2}}{k_{z(m+1)1}}(e^{-2jk_{z(m+1)1}d} - 1)^{-1}F_{m+1} = 0$$

Let's substitute $P_{m-1}$, $Q_m$ and $P_{m+1}$ obtained previously, into the above equation:

$$\left((e^{-2jk_{z(m-1)1}d} - 1)\frac{\varepsilon_2}{\varepsilon_1}\frac{k_{z(m-1)1}}{k_{z(m-1)2}}\frac{jX_sM}{2}\left(+j\frac{\varepsilon_1}{\varepsilon_2}\frac{k_{z(m-1)2}}{k_{z(m-1)1}}\cot(k_{z(m-1)1}d) - 1\right)\right)\frac{\varepsilon_1}{\varepsilon_2}\frac{k_{z(m-1)2}}{k_{z(m-1)1}}$$
$$\times (e^{-2jk_{z(m-1)1}d} - 1)^{-1}F_{m-1}$$

$$+\left((e^{-2jk_{zm1}d} - 1)\frac{\varepsilon_2}{\varepsilon_1}\frac{k_{zm1}}{k_{zm2}}\left(+jX_s\left(+j\frac{\varepsilon_1}{\varepsilon_2}\frac{k_{zm2}}{k_{zm1}}\cot(k_{zm1}d) - 1\right) - \frac{k_{zm2}}{\omega\varepsilon_2}\right)\right)\frac{\varepsilon_1}{\varepsilon_2}\frac{k_{zm2}}{k_{zm1}}(e^{-2jk_{zm1}d} - 1)^{-1}F_m$$

$$+\left((e^{-2jk_{z(m+1)1}d} - 1)\frac{\varepsilon_2}{\varepsilon_1}\frac{k_{z(m+1)1}}{k_{z(m+1)2}}\frac{jX_sM}{2}\left(+j\frac{\varepsilon_1}{\varepsilon_2}\frac{k_{z(m+1)2}}{k_{z(m+1)1}}\cot(k_{z(m+1)1}d) - 1\right)\right)\frac{\varepsilon_1}{\varepsilon_2}\frac{k_{z(m+1)2}}{k_{z(m+1)1}}$$
$$\times (e^{-2jk_{z(m+1)1}d} - 1)^{-1}F_{m+1} = 0$$

Therefore, we can simplify the obtained expression as follows:

$$\frac{jX_sM}{2}\left(+j\frac{\varepsilon_1}{\varepsilon_2}\frac{k_{z(m-1)2}}{k_{z(m-1)1}}\cot(k_{z(m-1)1}d) - 1\right)F_{m-1} + \left(+jX_s\left(+j\frac{\varepsilon_1}{\varepsilon_2}\frac{k_{zm2}}{k_{zm1}}\cot(k_{zm1}d) - 1\right) - \frac{k_{zm2}}{\omega\varepsilon_2}\right)F_m$$
$$+ \frac{jX_sM}{2}\left(+j\frac{\varepsilon_1}{\varepsilon_2}\frac{k_{z(m+1)2}}{k_{z(m+1)1}}\cot(k_{z(m+1)1}d) - 1\right)F_{m+1} = 0$$

The above dispersion equation can also be written using the following notation:

$$P'_m = \frac{jX_sM}{2}\left((+j)\frac{k_{zm2}}{k_{zm1}}\frac{\varepsilon_1}{\varepsilon_2}\cot(k_{zm1}d) - 1\right) \quad \& \quad Q'_m = \left(jX_s\left((+j)\frac{k_{zm2}}{k_{zm1}}\frac{\varepsilon_1}{\varepsilon_2}\cot(k_{zm1}d) - 1\right) - \frac{k_{zm2}}{\omega\varepsilon_2}\right)$$

$$P'_{m-1}F_{m-1} + Q'_m F_m + P'_{m+1}F_{m+1} = 0$$

The above expression is identical to what we obtained for a sinusoidally varying impedance sheet over a grounded dielectric layer. This is expected because we solved the dispersion relation for a general case, and by considering the bottom layer as a perfect electric conductor (PEC), it should reduce to the same expression. This verifies that the obtained dispersion expression for the general case is accurate and correct.

## VIII. Far field radiation pattern calculations for the proposed coupled LWA

Here, we will formulate the generated surface currents and consequently the far-field radiation patterns for the proposed coupled LWAs. Note that the following formulations are solely an analytical estimation of the radiated fields [1], and numerical solutions (in this study, finite element method (FEM)-based simulations) can be considered accurate solutions. Both sets of results are compared in the paper. We will first formulate the far-field radiation of an aperture antenna lying on the $xy$ plane and supporting only electric surface current density (assuming that magnetic surface current density is not excited). Then, it will be specified for a spatially modulated impedance sheet along the $x$ direction.

For a rectangular aperture antenna with the size of $L_x$ and $L_y$ in the $xy$ plane, the vector potential can be written as:

$$\vec{A} = \frac{\mu}{4\pi}\iint_S \vec{J}\frac{e^{-jk_0R}}{R}dS' \approx \frac{\mu e^{-jk_0r}}{4\pi r}\vec{N}$$

Where, $S$ is the area of the aperture antenna, $R$ is the distance from the source to the observation point, $r$ is the distance from the origin to the observation point and $\vec{N}$ is called the radiation vector and is expressed as:

$$\vec{N} = \iint_S \vec{J}e^{jk_0r'\cos\psi}dS'$$

Note that:

$$r'\cos\psi = x'\sin\theta\cos\phi + y'\sin\theta\sin\phi$$

Where $\theta$ is the angle with respect to the $z$ axis and $\phi$ is the angle with respect to the $x$ axis. Decomposing $\vec{N}$ in spherical coordinates system gives:

$$N_\theta = \iint_S J_x \cos\theta e^{jk_0(x'\sin\theta\cos\phi + y'\sin\theta\sin\phi)}dS'$$

$$N_\phi = \iint_S J_y e^{jk_0(x'\sin\theta\cos\phi + y'\sin\theta\sin\phi)}dS'$$

A 3D radiation pattern typically is evaluated at 2 planes called E-plane and H-plane. For E-plane $\phi$ is assumed to be $\phi = 0°$ and for H-plane $\phi = 90°$. Note that here, we are especially interested in E-plane radiation patterns, therefore:

$$N_\theta = \iint_S J_x \cos\theta e^{jk_0x'\sin\theta}dS', \quad N_\phi = \iint_S J_y e^{jk_0x'\sin\theta}dS'$$

Since, in our problem only $J_x$ exist and $J_y = 0$, as a result $N_\phi = 0$ and only $N_\theta$ is taking to the consideration.

$J_x$ can be determine using the following boundary condition as:

$$\vec{J} = J_x\hat{x} + J_y\hat{y} = \hat{z} \times \vec{H}_t = \hat{z} \times (H_x\hat{x} + H_y\hat{y}) = H_x\hat{y} - H_y\hat{x} \quad \rightarrow \quad \begin{matrix} J_x = -H_y \\ J_y = H_x \end{matrix}$$

Since we assumed $TM^x$ excitation, $H_x = 0$ and therefore: $J_y = 0$. We already have expanded the electric and magnetic fields in terms of infinite number of spatial harmonics (Floquet-Waves), therefore assuming $2N + 1$ number of harmonics for $H_y$ gives:

$$J_x = -H_y = -e^{-jk_{x,0}x} \sum_{n=-N}^{N} \frac{I_n}{\sqrt{2\pi}} e^{-j\frac{2\pi n}{a}x}$$

Substituting the above $J_x$ into the derived formula for $N_\theta$ gives:

$$N_\theta = \iint_S J_x \cos\theta e^{j\,k_0 x' \sin\theta}\, dS' = \iint_S -e^{-jk_{x,0}x'} \sum_{n=-N}^{N} \frac{I_n}{\sqrt{2\pi}} e^{-j\frac{2\pi n}{a}x'} \cos\theta e^{j\,k_0 x' \sin\theta}\, dS'$$

$$= -L_y \frac{\cos\theta}{\sqrt{2\pi}} \sum_{n=-N}^{N} I_n \int_0^{L_x} e^{-j\left(k_{x,0} + \frac{2\pi n}{a} - k_0\sin\theta\right)x'}\, dx'$$

The integration can be calculated as:

$$\int_0^{L_x} e^{-j\left(k_{x,0} + \frac{2\pi n}{a} - k_0\sin\theta\right)x'}\, dx' = \frac{j}{k_{x,0} + \frac{2\pi n}{a} - k_0\sin\theta}\left(e^{-j\left(k_{x,0} + \frac{2\pi n}{a} - k_0\sin\theta\right)L_x} - 1\right)$$

Substituting back gives:

$$N_\theta = -L_y \frac{\cos\theta}{\sqrt{2\pi}} \sum_{n=-N}^{N} \frac{jI_n}{k_{x,0} + \frac{2\pi n}{a} - k_0\sin\theta}\left(e^{-j\left(k_{x,0} + \frac{2\pi n}{a} - k_0\sin\theta\right)L_x} - 1\right)$$

The electric field components (far field electric components) can be approximated as:

$$E_\theta \approx -jk_0 \frac{e^{-jk_0 r}}{4\pi r} \eta_0 N_\theta, \quad E_\phi \approx -jk_0 \frac{e^{-jk_0 r}}{4\pi r} \eta_0 N_\phi$$

$$N_\phi = 0 \xrightarrow{yields} E_\phi = 0$$

$$E_\theta \approx -jk_0 \frac{e^{-jk_0 r}}{4\pi r} \eta_0 N_\theta$$

$$= -jk_0 \frac{e^{-jk_0 r}}{4\pi r} \eta_0 \left(-L_y \frac{\cos\theta}{\sqrt{2\pi}} \sum_{n=-N}^{N} \frac{jI_n}{k_{x,0} + \frac{2\pi n}{a} - k_0\sin\theta}\left(e^{-j\left(k_{x,0} + \frac{2\pi n}{a} - k_0\sin\theta\right)L_x} - 1\right)\right)$$

$$= jk_0\eta_0 \frac{e^{-jk_0 r}}{4\pi r} \frac{L_y \cos\theta}{\sqrt{2\pi}} \sum_{n=-N}^{N} \frac{jI_n}{k_{x,0} + \frac{2\pi n}{a} - k_0\sin\theta}\left(e^{-j\left(k_{x,0} + \frac{2\pi n}{a} - k_0\sin\theta\right)L_x} - 1\right)$$

The above expression is used as the analytical far field solution in the paper.

## References:

[1] J. Ruiz-García, 'Multi-beam modulated metasurface antennas for unmanned aerial vehicles (UAVs)', Doctoral Thesis, University of Rennes, 2021.
[2] A. M. Patel, Controlling Electromagnetic Surface Waves with Scalar and Tensor Impedance Surfaces, Doctoral Thesis, University of Michigan, 2013.
[3] A. Abbaszadeh and J. Budhu, "Observation of Exceptional Points in Parity-Time Symmetric Coupled Impedance Sheets," 2024 18th European Conference on Antennas and Propagation (EuCAP), Glasgow, Scotland, 2024, pp. 1-5.